\documentclass[aps,prd,twocolumn,superscriptaddress,runinaddress,floatfix,preprintnumbers,showpacs,nofootinbib]{revtex4-1}
\usepackage{amsmath}
\usepackage{amssymb}
\usepackage{amsfonts}
\usepackage{amsthm}

\newtheorem{theorem}{Theorem}[section]
\newtheorem{corollary}[theorem]{Corollary}
\newtheorem{lemma}{Lemma}[section]
\newtheorem{definition}{Definition}[section]
\newtheorem{proposition}{Proposition}[section]
\newtheorem{example}{Example}[section]

\newcommand{\eqn}[1]{ \begin{equation} #1 \end{equation} }

\usepackage{graphicx}
\newcommand{\transpose}{\intercal}

\begin{document}

\preprint{ADP-13-28/T848}

\title{Nucleon Excited State Wave Functions from Lattice QCD}

\author{Dale S.~Roberts}
\author{Waseem Kamleh}
\author{Derek B.~Leinweber}
\affiliation{Special Research Centre for the Subatomic Structure of
  Matter, School of Chemistry and Physics, University of Adelaide, SA,
  5005, Australia} 

\begin{abstract}
We apply the eigenvectors from a variational analysis to successfully
extract the wave functions of even-parity excited states of the
nucleon, including the Roper.  We explore the first four states in the
spectrum excited by the standard nucleon interpolating field.
We find that the states exhibit a structure qualitatively consistent
with a constituent quark model, where the ground, first-,
second- and third-excited states have 0, 1, 2, and 3 nodes in the
radial wave function of the $d$-quark about two $u$ quarks at the
origin.  Moreover the radial amplitude of the probability
distribution is similar to that predicted by constituent quark models.
We present a detailed examination of the quark-mass dependence of the
probability distributions for these states, searching for a nontrivial
role for the multi-particle components mixed in the finite-volume QCD
eigenstates.
Finally we examine the dependence of the $d$-quark probability
distribution on the positions of the two $u$ quarks.  The results are
fascinating, with the underlying $S$-wave orbitals governing the
distributions even at rather large $u$-quark separations.
\end{abstract}

%
%  12.38.Gc Lattice QCD calculations
%  12.39.Jh Nonrelativistic quark model 
%  14.20.Gk Baryon resonances with S=0
%  14.20.Dh Protons and neutrons 

\maketitle

\section{Introduction}

An examination of the wave function of a quark bound within a hadron
provides deep insights into the underlying dynamics of the many-body
theory of QCD.  It enables a few-body projection of the underlying
physics that can be connected with models, shedding light on the
essential effective phenomena emerging from the complex dynamics of
QCD.

The hadron spectrum is the manifestation of the highly complex
dynamics of QCD.  It is an observable that is readily accessible in
collider experiments.  While the quantum numbers of the states can be
ascertained, properties providing more insight into the structure of
the resonances often remain elusive to experiment.  We aim to provide
some insight into the underlying dynamics governing the structure of
these states.

In quantum field theory, a Schr\"{o}dinger-like probability
distribution can be constructed for bound states by taking a
simplified view of the full quantum field theory wave functional in
the form of the Bethe-Salpeter wave function \cite{Salpeter:1951sz},
herein referred to as simply the `wave function'.  Recent advances in
the isolation of nucleon excited states through correlation-matrix
based variational techniques in lattice QCD now enable the exploration
of the structure of these states and how these properties emerge from
the fundamental interactions of QCD.

In this paper, we extend earlier results \cite{Roberts:2013ipa}
focusing on the wave function of the Roper excitation
\cite{Roper:1964zza} to the four lowest-lying even-parity states
excited by the standard $\chi_1$ interpolating field which
incorporates a scalar diquark construction.  We examine the quark mass
dependence of the probability distributions for these states.  Here we
search for a signature of multi-particle components mixed in the
finite-volume QCD eigenstates at the two largest quark masses where
the states sit close to the multi-particle thresholds.  We also explore
the dependence of the $d$-quark probability distribution on the
positions of the two $u$ quarks along an axis through the centre of
the distribution.

In presenting our results we make extensive use of isovolume and
surface plots of the probability distributions for the quarks.  Such
visualizations have already been used to illustrate physical effects
such as Lorentz contraction \cite{Chu:1990ps,Gupta:1993vp}, the effect
of external magnetic fields \cite{Roberts:2010cz} and finite volume
effects \cite{Roberts:2013ipa,Alexandrou:2008ru}, for example.

Early explorations of these states were based on non-relativistic
constituent quark models.  The probability distributions of quarks
within hadrons were determined using a one-gluon-exchange potential
augmented with a confining form \cite{DeRujula:1975ge,Bhaduri:1980fd}.
These models have been the cornerstone of intuition of hadronic
probability distributions for many decades.  In this investigation, we
will confront these early predictions for quark probability
distributions in excited states directly via Lattice QCD.

In a relativistic gauge theory the concept of a hadronic wave function
is not unique and the Bethe-Salpeter wave function underlying the
probability distributions can be defined in several different forms.
For example, the gauge-invariant Bethe-Salpeter amplitude exploits a
string of flux to connect the quarks annihilated at different spatial
positions in a gauge invariant manner.  As this leads to an explicit
path dependence, an average over the paths is desirable.  Another
approach considers Bethe-Salpeter amplitudes in which the gauge degree
of freedom is fixed to a specific gauge.  In lattice field theory,
Coulomb and Landau gauges are most common due to their local gauge
fixing procedure.  Landau gauge provides distributions that compare
favorably with constituent quark model predictions \cite{Roberts:2013ipa}
and therefore we select Landau gauge herein.

\section{Lattice Techniques}

Robust methods have been developed that allow the isolation and study
of the states associated with these resonances in Lattice QCD
\cite{Leinweber:1994nm,Gockeler:2001db,Zanotti:2001yb,Sasaki:2001nf,Melnitchouk:2002eg,%
  Lee:2002gn,Leinweber:2004it,Basak:2007kj,Bulava:2010yg,Mahbub:2010rm,%
  Mahbub:2011zz,Roberts:2013ipa}.  In this study, we apply the variational
method \cite{Michael:1985ne,Luscher:1990ck} to extract the ground
state and first three $P_{11}$ excited states of the proton associated
with the Roper \cite{Roper:1964zza} and other higher-energy $P_{11}$
states.  We then combine this with lattice wave function techniques to
calculate the probability distributions of these states at several
quark masses and quark positions.  We use the $2+1$ flavour
$32^3\times 64$ PACS-CS configurations \cite{Aoki:2008sm} at a pion
mass as low as 156 MeV.

The wave function of a hadron is proportional to the parity-projected
\cite{Lee:1998cx} two point Green's function,
\begin{eqnarray}
G_{ij}^\pm(\vec{p},t) &=& \sum_{\vec x}
  e^{-i\vec{p}\cdot\vec{x}} 
\, \mbox{tr}\, \left ( \gamma_0 \pm 1 \right ) \, \nonumber \\
&&\qquad \langle\Omega\vert \,
  T\{\, \chi_i(\vec{x},t)\, \bar\chi_j(\vec{0},0)\, \}\, \vert\Omega\rangle
  \, , 
\label{twopt}
\end{eqnarray}
where $\chi_i$ are the hadronic interpolating fields. In the case of
the proton the most commonly used interpolator is given by 
\eqn{\chi_1(x)=\epsilon^{abc}\,
  (\, u^{\transpose a}(x)\, C\gamma_5\, d^b(x)\, )\, u^c(x) \, ,} 
with the corresponding adjoint given by
\eqn{\bar\chi_1(0)=\epsilon^{abc}\,
  (\, \bar{d}^b(0)\, C\gamma_5\, \bar{u}^{\transpose a}(0)\, )\,
  \bar{u}^c(0) \, .}
In order to construct the wave function, the quark fields in the
annihilation operator are each given a spatial dependence, 
\eqn{\chi_1(\vec{x},\vec{y},\vec{z},\vec{w}) = \epsilon^{abc}\, (\,
  u^{\transpose a}(\vec{x}+\vec{y})\, C\gamma_5\, d^b(\vec{x}+\vec{z})\, )\,
  u^c(\vec{x}+\vec{w}) \, , 
 \label{gdepinterp} }
while the creation operator remains local.  This generalizes $G(\vec
p, t)$ to a wave function proportional to $G(\vec p, t; \vec y, \vec
z, \vec w)$. In principle, we could allow each of these coordinates,
$\vec y,\ \vec z,\ \vec w$, to vary across the entire lattice,
however, we can reduce the computational cost by taking advantage of
the hyper-cubic rotational and translational symmetries of the
lattice.  A near-complete description of the probability distribution
of a particular quark within the proton can be formed by separating
two of the quarks along a fixed axis and calculating the third quark's
wave function for every lattice site.  For this study, we will focus
on the probability distribution of the $d$ quark from
Eq.~(\ref{gdepinterp}) with the $u$-quarks being separated along the
$x$ axis through the centre of the distribution, {\it i.e.},
\begin{align}
\chi_1&(\vec{x}, \vec{d_1}, \vec{z}, \vec{d_2}; t) = \label{interpsep}\\ 
&\epsilon^{abc}\, (\,
  u^{\transpose a}(\vec{x}+\vec{d_1},t)\, C\gamma_5\, d^b(\vec{x}+\vec{z},t)\, )\,
  u^c(\vec{x}+\vec{d_2},t) \nonumber\, . 
\end{align}
where $\vec{d_i}=(d_i,0,0)$, $d_1 > 0$, $d_2 = -d_1$ for separations
across an even number of lattice sites and $d_2 = -(d_1-1)$ for an odd
separation. A symmetrised wave function is presented by averaging the
wave functions calculated with the interpolating field in
Eq.~(\ref{interpsep}) combined with the wave functions produced by the
same interpolating field where $d_1\leftrightarrow d_2$.
%  The resulting construct is gauge dependent, and as discussed in the
%  Introduction, we choose to fix the gauge configurations to Landau
%  gauge.

Landau gauge is a smooth gauge that preserves the Lorentz invariance
of the theory.  While the size and shape of the wave function are
gauge dependent, our selection of Landau gauge is supported by our
results.  For example, the ground state wave function of the $d$ quark
in the proton is described accurately by the non-relativistic quark
model using standard values for the constituent quark masses and
string tension of the confining potential \cite{Roberts:2013ipa}.
Therefore this gauge provides a foundation for a more comprehensive
wave function examination.

To isolate energy eigenstates we use the correlation matrix or
variational method~\cite{Michael:1985ne,Luscher:1990ck}.  As we are
interested in the wave functions for states at rest, we select $\vec p
= 0$ in Eq.~(\ref{twopt}).
To ensure that the matrix elements are all $\sim{\cal{O}}(1)$, each
element of $G_{ij}(t)$ is normalized by the
diagonal elements of $G(0)$ as ${G}_{ij}(t) / (
  \sqrt{{G}_{ii}(0)}\, \sqrt{{G}_{jj}(0)} )$ (no sum on $i$ or $j$).
Using an average of $\{U\} + \{U^*\}$ configurations which have equal
weight in the QCD action, our construction of the two-point functions
is real \cite{Leinweber:1990dv,Boinepalli:2006xd}.

A linear superposition of interpolators
$\bar{\phi}^{\alpha}=\sum_{j}{\bar\chi}_{j}u_{j}^{\alpha}$ creating
state $\alpha$ provides the following recurrence relation
\begin{align}
G_{ij}(t_{0}+\triangle t)\, u_{j}^{\alpha} & = e^{-m_{\alpha}\triangle
  t}\, G_{ij}(t_{0})\, u_{j}^{\alpha}  \, ,
 \label{eq:recurrence_relation}
\end{align}  
from which right and left eigenvalue equations are obtained
\begin{align}
[(G(t_{0}))^{-1}\, G(t_{0}+\triangle t)]_{ij}\, u^{\alpha}_{j} & = c^{\alpha}\, u^{\alpha}_{i}, \label{eq:right_evalue_eq}\\
v^{\alpha}_{i}\, [G(t_{0}+\triangle t)\, (G(t_{0}))^{-1}]_{ij} & = c^{\alpha}v^{\alpha}_{j},\label{eq:left_evalue_eq}
\end{align} 
with $c^{\alpha}=e^{-m_{\alpha}\triangle t}$.
The eigenvectors for state $\alpha$, $u_{j}^{\alpha}$ and
$v_{i}^{\alpha}$, provide the eigenstate projected correlation
function
\begin{align}
 G^{\alpha}_{\pm}(t) & \equiv v_{i}^{\alpha}\, G^{\pm}_{ij}(t)\, u_{j}^{\alpha} ,
 \label{projected_cf_final}
\end{align}
with parity $\pm\,$.  The effective mass can then be calculated from
the projected two-point functions by $m(t) = \log \left ( G(t) /
G(t+1) \right )$. While the effective mass is insensitive to a wide
range of 
 parameters \cite{Mahbub:2010rm}, we follow
Ref.~\cite{Mahbub:2010rm} and select $t_0$ to be 2 time slices after
the source with $\Delta t=2$.

Different interpolators exhibit different couplings to the proton
ground and excited states and hence can be used to construct a
variational basis. The limited number of local interpolators restricts
the size of the operator basis \cite{Leinweber:1994nm}. To remedy
this, we exploit the smearing dependence of the coupling of states to
one or more standard interpolating operators in order to construct a
larger variational basis where the $\chi_i$ and $\bar\chi_j$ from
Eq.~(\ref{twopt}) contain a smearing dependence. This method has been
shown to allow access to states associated with resonances such as the
Roper \cite{Mahbub:2010rm,Roberts:2013ipa} and the $\Lambda(1405)$
\cite{Menadue:2011pd}.

The non-local sink operator used to construct the wave function is
unable to be smeared, and hence the standard technique of
Eq.~(\ref{projected_cf_final}) cannot be applied.  However,
Eq.~(\ref{eq:right_evalue_eq}) illustrates it is sufficient to isolate
the state at the source using the right eigenvector.  Thus, the
probability distributions are calculated with each smeared source
operator and the right eigenvectors calculated from the standard
variational analysis are then applied in order to extract the
individual states of interest.  

\begin{figure}
\includegraphics[angle=90,width=0.95\linewidth]{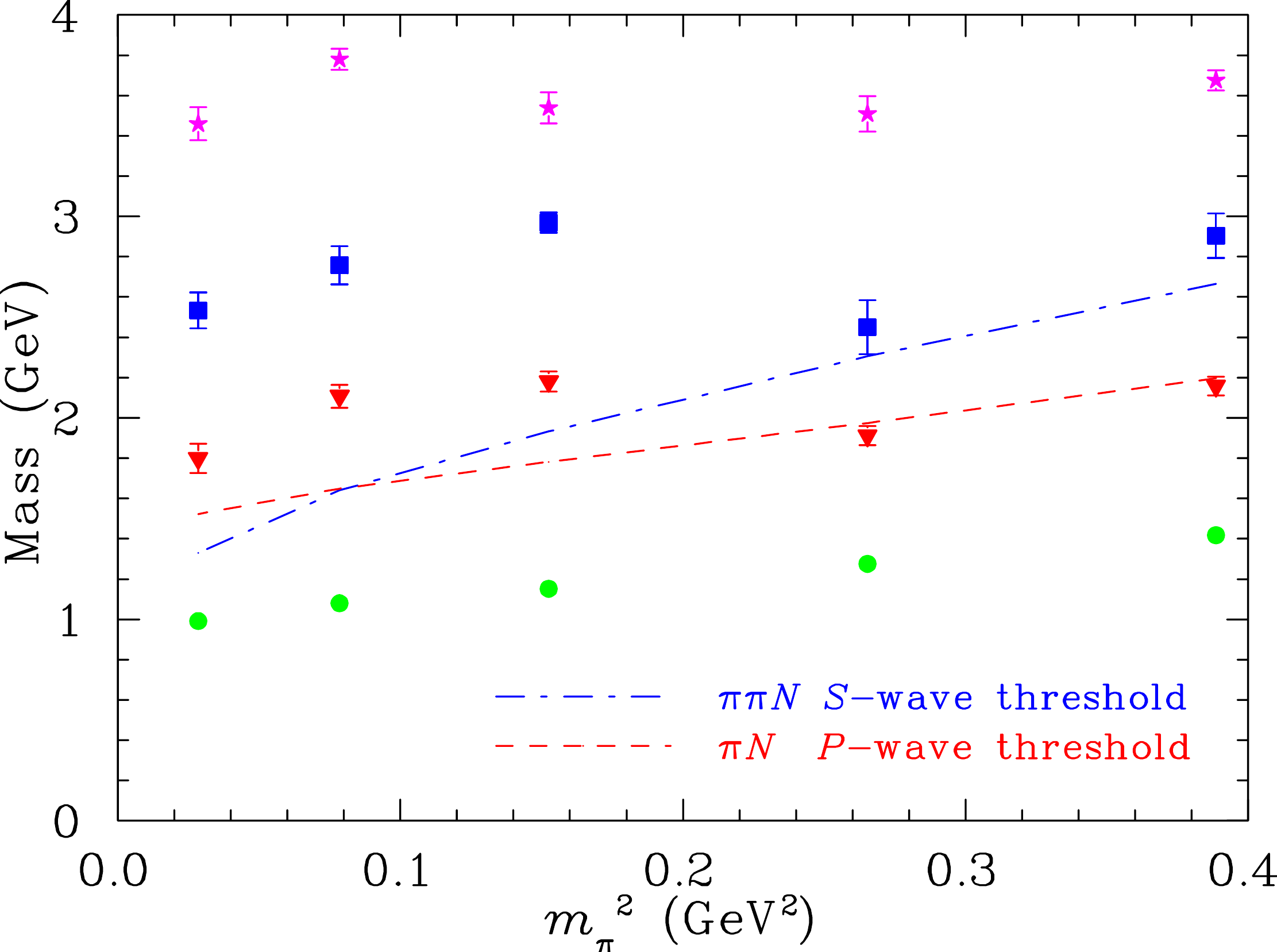}
\caption{The mass dependence of the four lowest-lying even-parity
  eigenstates excited by the $\chi_1$ interpolating field is compared
  with the $S$ and $P$-wave non-interacting multi-particle energy
  thresholds on the finite volume lattice.  Plot symbols track the
  eigenvector associated with each state. }
\label{mass}
\end{figure}

Our focus on $\chi_1$ in this investigation follows from the results
of Ref.~\cite{Mahbub:2013ala}, where the lowest-lying excitation of
the nucleon was shown to be predominantly associated with the $\chi_1$
interpolating field.  The results from their $8 \times 8$ correlation
matrix of $\chi_1$ and $\chi_2 = \epsilon^{abc}\, (\, u^{\transpose
  a}(x)\, C\, d^b(x)\, )\, \gamma_5 \, u^c(x)$ revealed that $\chi_2$
plays a marginal role in exciting the Roper.  The coefficients of the
Roper source eigenvector multiplying $\overline\chi_2$ are near zero.
Further comparison with Ref.~\cite{Mahbub:2013ala}, identifies the
third state extracted herein as the fifth state of the twelve states
identified and the fourth state herein as the tenth state.  Figure
\ref{mass} illustrates the quark mass dependence of these four states
which will be examined in detail herein.

The quark mass flow of these states tracked by their associated
eigenvectors \cite{Mahbub:2013ala} is not smooth and suggests the
presence of avoided level crossings as one transitions from the
heaviest two quark masses to lightest three quark masses.  At the two
heaviest quark masses, it is seems likely these states are dominated
by multi-particle $N \pi$ components, whereas at the lighter three
quark masses, single particle components are more dominant.  We will
search for evidence of this in the wave functions of these states.

In summary, the wave function for the $d$ quark in state $\alpha$
having momentum $\vec p$ observed at Euclidean time $t$ with the $u$
quarks at positions $\vec d_1$ and $\vec d_2$ is
\begin{align}
\psi^\alpha_d&(\vec p, t; \vec{d_1}, \vec{d_2}; \vec{z}) =
\sum_{\vec x}
  e^{-i\vec{p}\cdot\vec{x}}\, \mbox{tr}\, \left ( \gamma_0 + 1 \right
  ) \,  \label{psi_alpha} \\
&\langle\Omega\vert
  \,T\{\, \chi_1(\vec{x}, \vec{d_1}, \vec z, \vec{d_2}; t)\,
  \bar\chi_j(\vec{0}, \vec{0}, \vec{0}, \vec{0}; 0)\,  \}\, 
  \vert\Omega\rangle \,
  u_j^\alpha \, , \nonumber
\end{align}
where $\chi_1(\vec{x}, \vec{d_1}, \vec z, \vec{d_2}; t)$ is given by
Eq.~(\ref{interpsep}).

As discussed above, $\chi_1$ has the spin-flavour construct that is
most relevant to the excitation of the Roper from the QCD vacuum.  As
such, it is an ideal choice for revealing the spatial distribution of
quarks within the Roper.  However, the selection of $\chi_1$ in 
Eq.~(\ref{psi_alpha}) is not unique and other choices are
possible.  For example, the selection of $\chi_2$ would reveal small
contributions to the Roper wave function where vector diquark degrees
of freedom are manifest.  Similarly, $D$-wave contributions could be
resolved through the consideration of a spin-3/2 isospin-1/2
interpolating field at the sink.  Research exploring these aspects of
the wave functions is in progress.

\section{Simulation Results}

\subsection{Lattice Parameters}

We use the $2+1$ flavour $32^3\times 64$ configurations created by the
PACS-CS collaboration \cite{Aoki:2008sm} constructed with the Iwasaki
gauge action \cite{Iwasaki:2011jk} and the $\mathcal{O}(a)$-improved
Wilson action \cite{Sheikholeslami:1985ij} with $\beta=1.90$, giving a
lattice spacing of $0.0907(13)\ \mathrm{fm}$.  
The hopping parameters are 0.13700, 0.13727, 0.13754, 0.13770 and 0.13781
giving pion masses of 702, 570, 411, 296 and 156 MeV respectively.  For
each quark mass we consider 398, 391, 447, 395 and 198 gauge field
configurations respectively, and at the lightest quark mass we
increase statistics through the consideration of four sources per
configuration distributed evenly along the time axis.

To isolate the QCD eigenstates, a $4\times 4$ variational basis
is constructed using the $\chi_1$ operator with four smearing levels;
16, 35, 100 and 200 sweeps \cite{Mahbub:2010rm} of gauge-invariant
Gaussian smearing \cite{Gusken:1989qx}.  These smearing levels
correspond to smearing radii of 2.37, 3.50, 5.92 and 8.55 lattice
units or 0.215, 0.317, 0.537 and 0.775 fm respectively.

The choice of variational parameters $t_{0}=2, \triangle t = 2$ relative 
to the source position is ideal, resulting in the effective mass 
plateaus of the states commencing at $t=t_0=2$ as desired
\cite{Roberts:2013ipa}.  This indicates that the number of states
contributing significantly to the correlation functions of the
correlation matrix at $t_0 = 2$ equals the dimension of the
correlation matrix.  As such we examine the wave functions of all four
states with the caution that the fourth state is most susceptible to
excited state contamination.  In reporting the wave functions we
select the mid point of the correlation matrix analysis at $t=3$.  The
wave functions observed for all our states show an approximate
symmetry over the eight octants surrounding the origin.  To improve
our statistics we average over these eight octants when $d_1=d_2=0$, 
and an average over the four quadrants sharing an axis with the $u$ 
quark separation at all other values of $d_1$ and $d_2$.

We fix to Landau gauge by maximizing the $\mathcal{O}(a^2)$ improved
gauge-fixing functional \cite{Bonnet:1999mj}
\begin{equation}
\mathcal{F}_{Imp} =\sum_{x,\mu}
\mathrm{Re}\,\mathrm{tr}\left(\frac{4}{3} U_\mu(x) -
\frac{1}{12\,u_0}\left(U_\mu(x)U(x+\hat{\mu}) + \mathrm{h.c.}\right)\right) 
\end{equation}
using a Fourier transform accelerated algorithm \cite{Davies:1987vs}.

In carrying out our calculations, we average over the equally weighted
$\{ U \}$ and $\{ U^* \}$ link configurations as an improved unbiased
estimator \cite{Leinweber:1990dv}.  The two-point function is then
perfectly real and the probability density is proportional to the
square of the wave function.  In this analysis, we choose to look at
the zero-momentum probability distributions.

\subsection{Wave Functions and Constituent Quark Model Predictions}

\begin{figure}[t!]
\includegraphics[angle=90,width=0.99\linewidth]{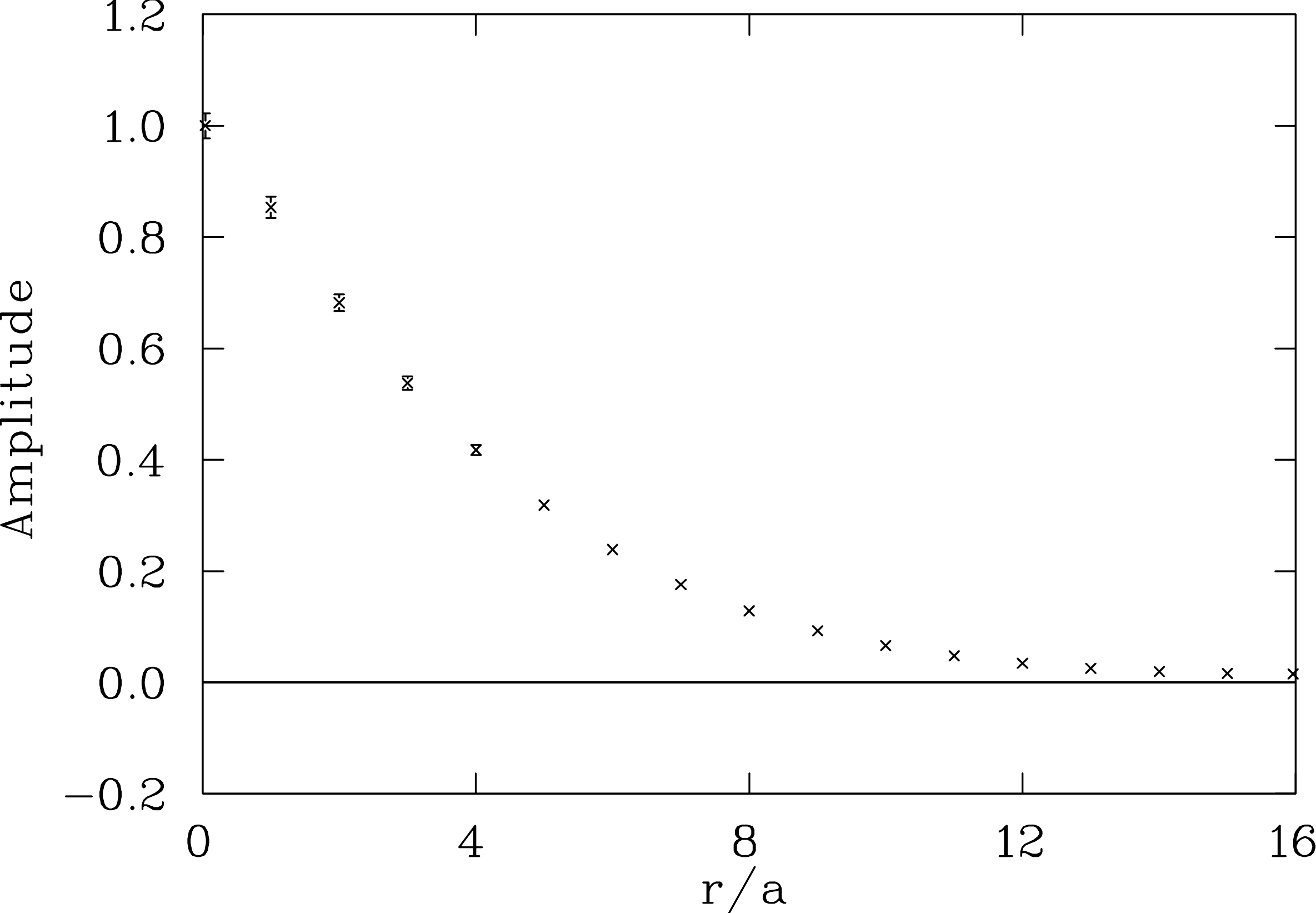}
\includegraphics[angle=90,width=0.99\linewidth]{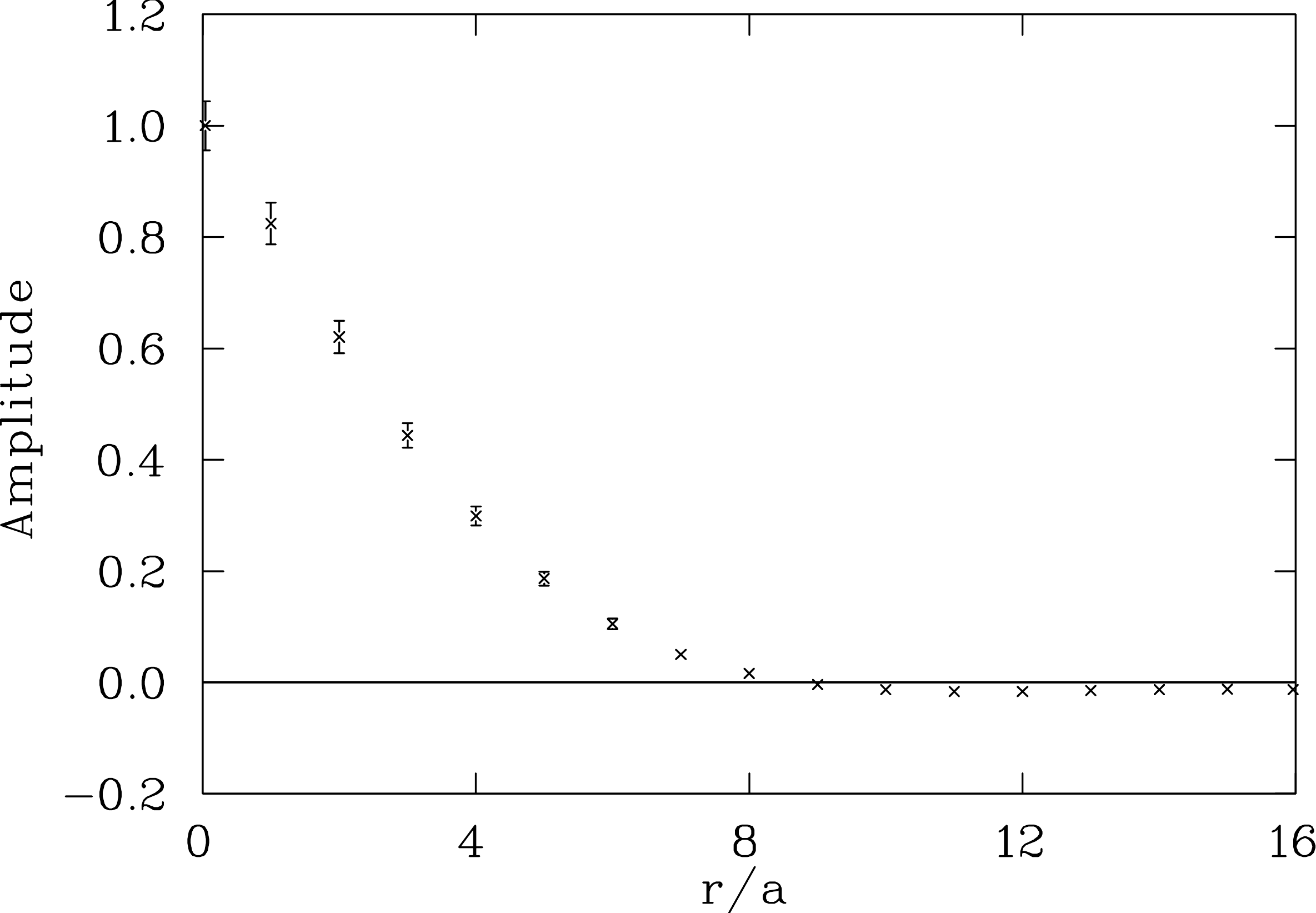}
\includegraphics[angle=90,width=0.99\linewidth]{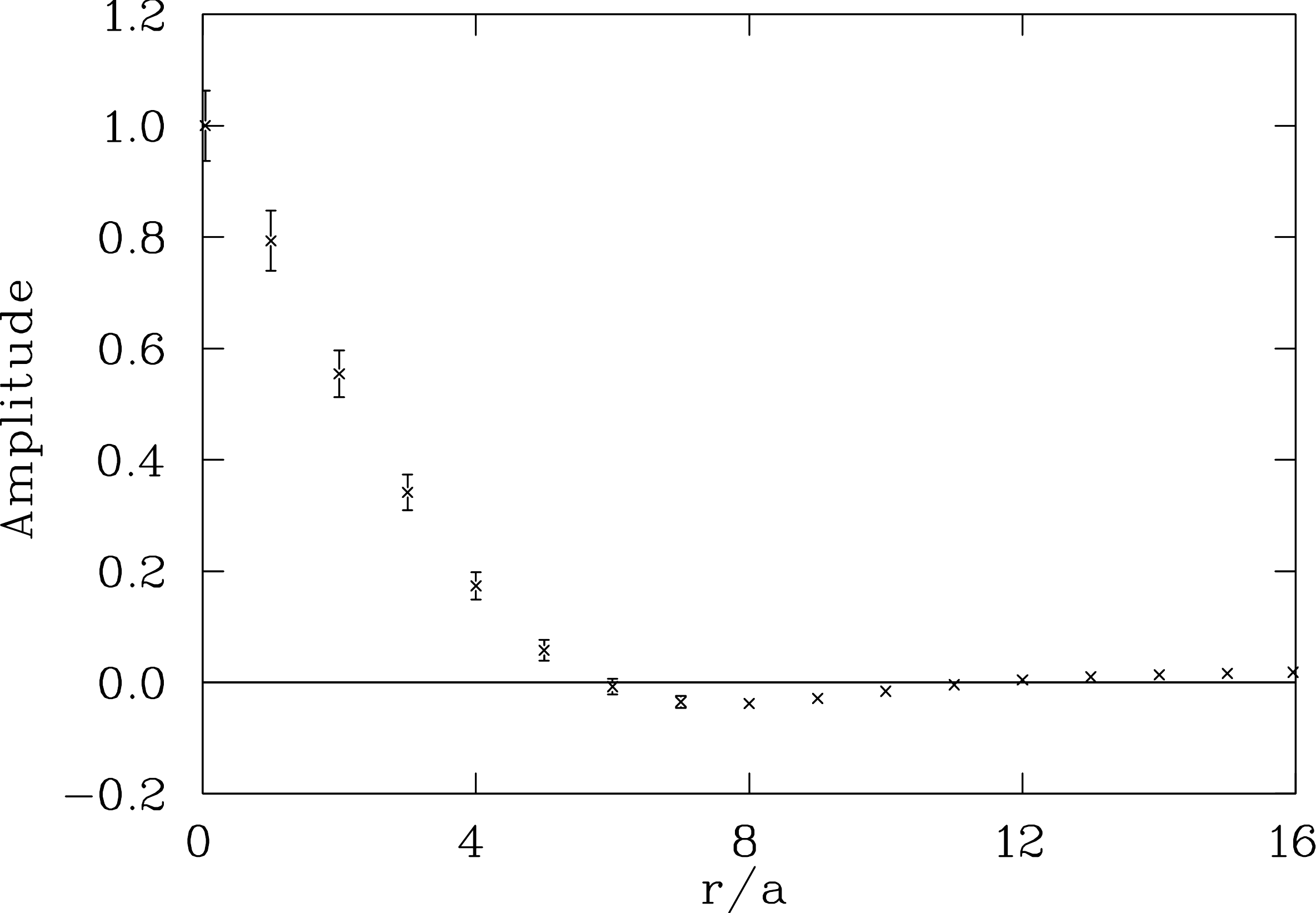}
\caption{The wave function of the $d$ quark in the proton about the
  two $u$ quarks fixed at the origin for the lightest quark mass
  ensemble providing $m_\pi = 156$ MeV.  From top down, the plots
  correspond to the ground, first and second excited states observed
  in our lattice simulation.  The wave function changes sign in the
  excited states and reveals a node structure consistent with $1s$,
  $2S$ and $3S$ states.}
\label{waveFun}
\end{figure}

\begin{figure*}[p]
\includegraphics[angle=90,width=0.3\linewidth]{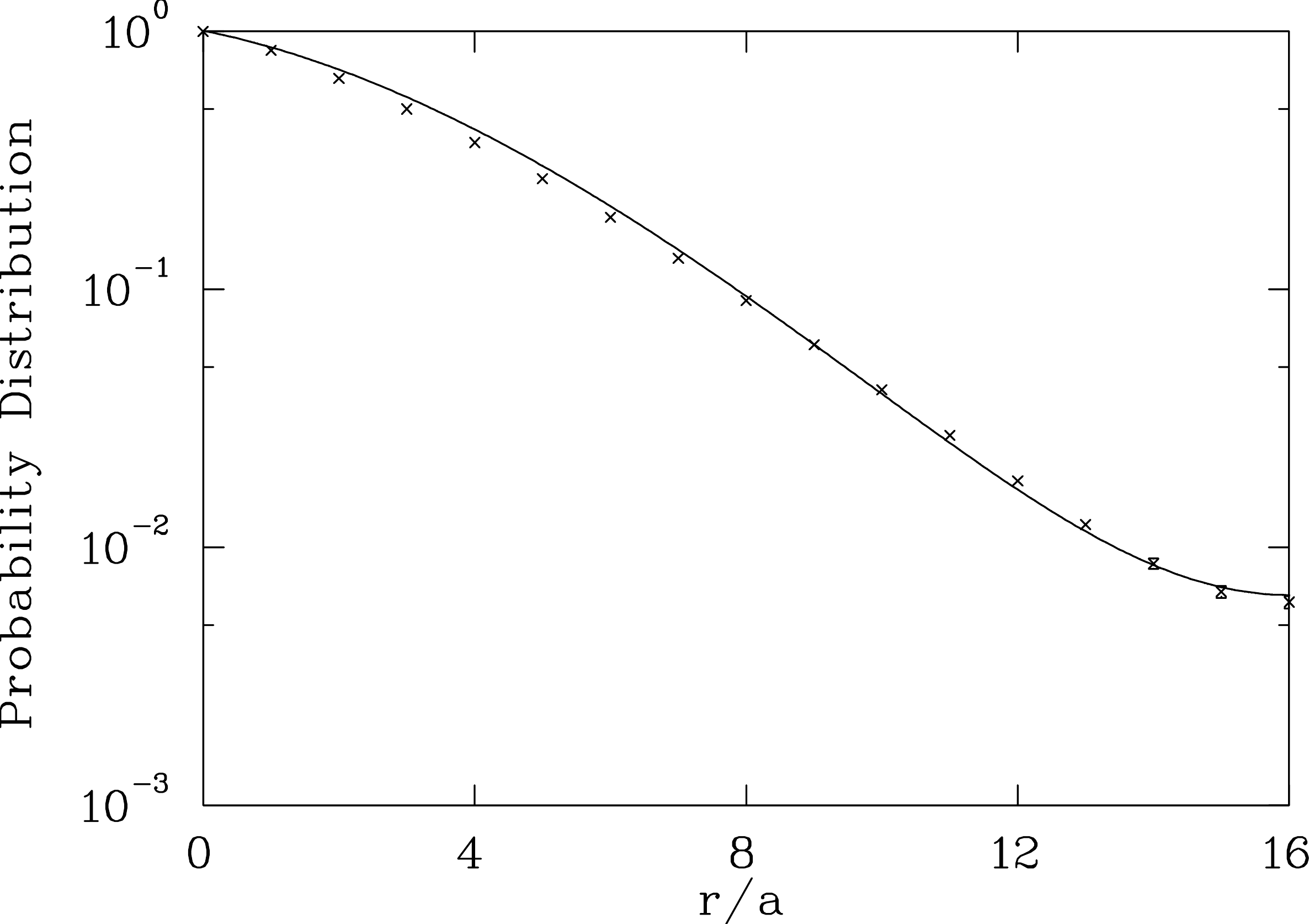}
\includegraphics[angle=90,width=0.3\linewidth]{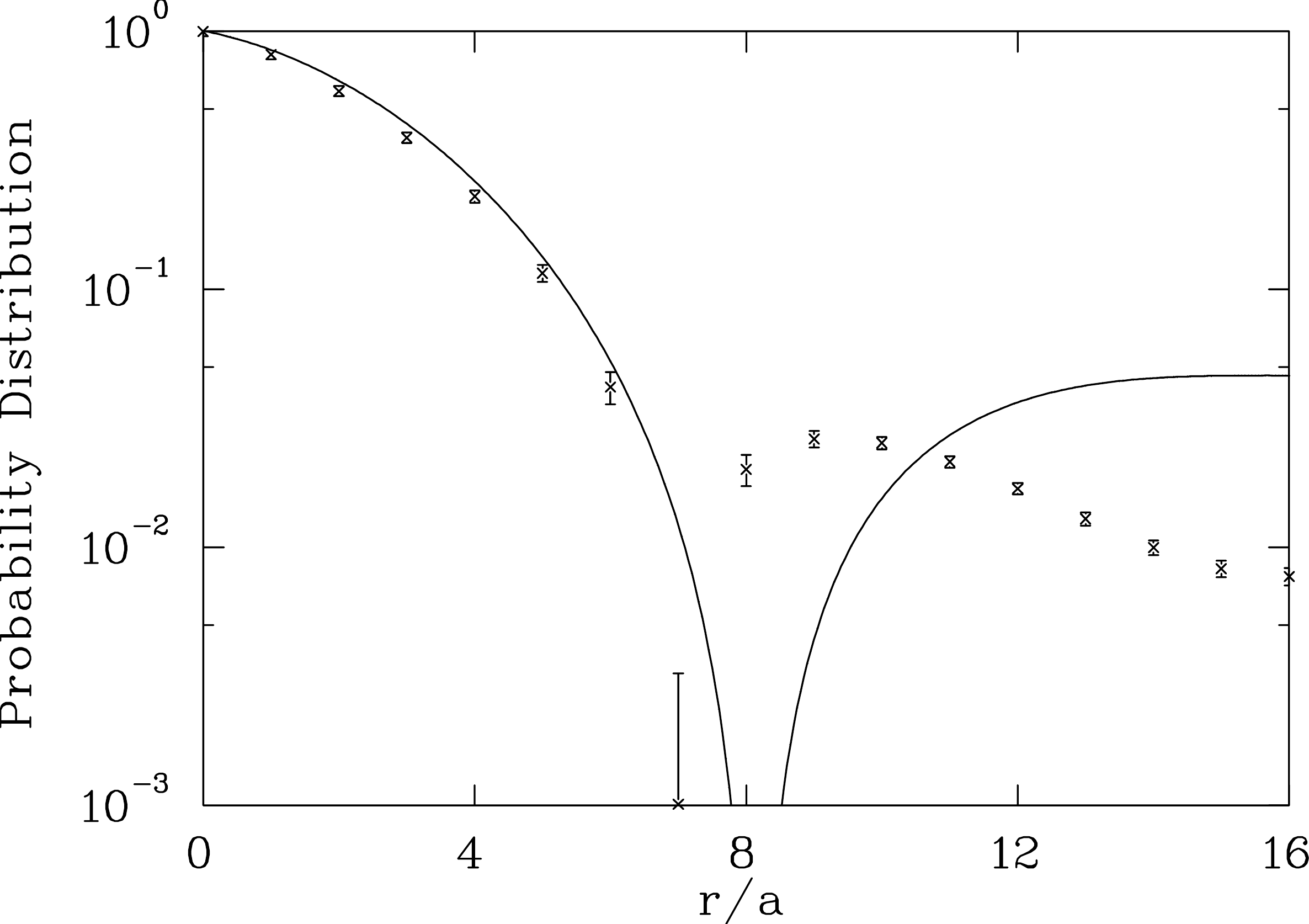}
\includegraphics[angle=90,width=0.3\linewidth]{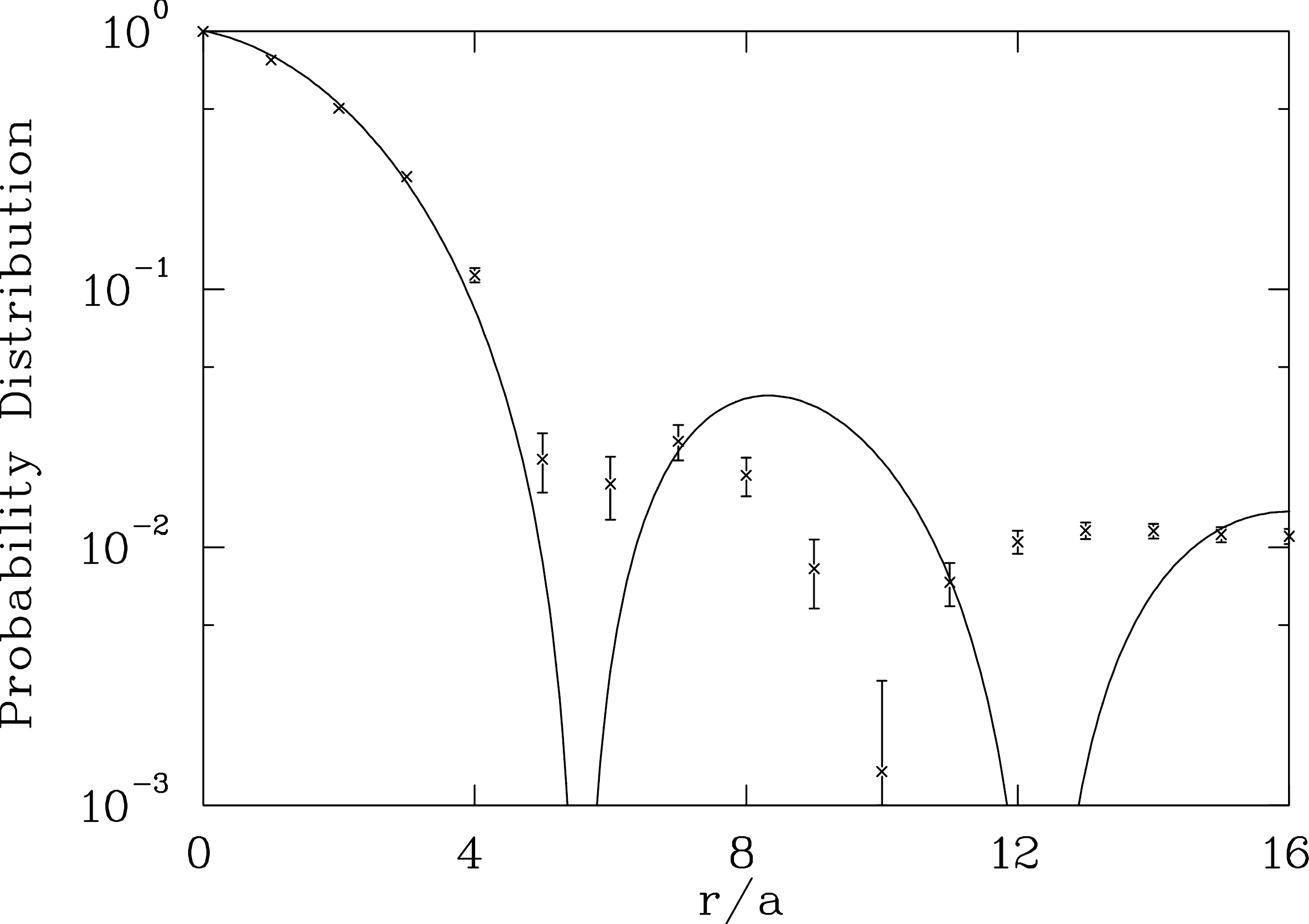}
\includegraphics[angle=90,width=0.3\linewidth]{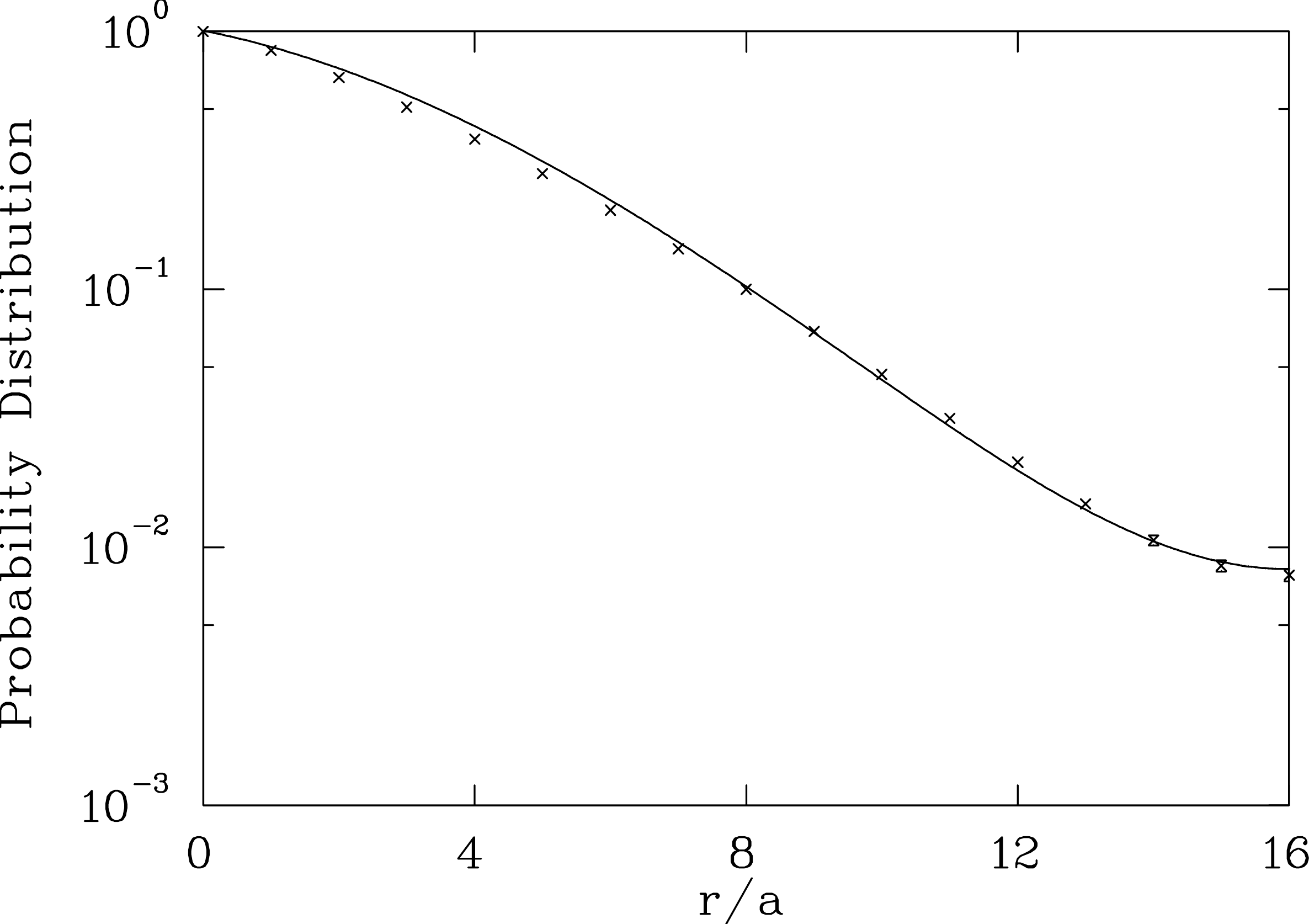}
\includegraphics[angle=90,width=0.3\linewidth]{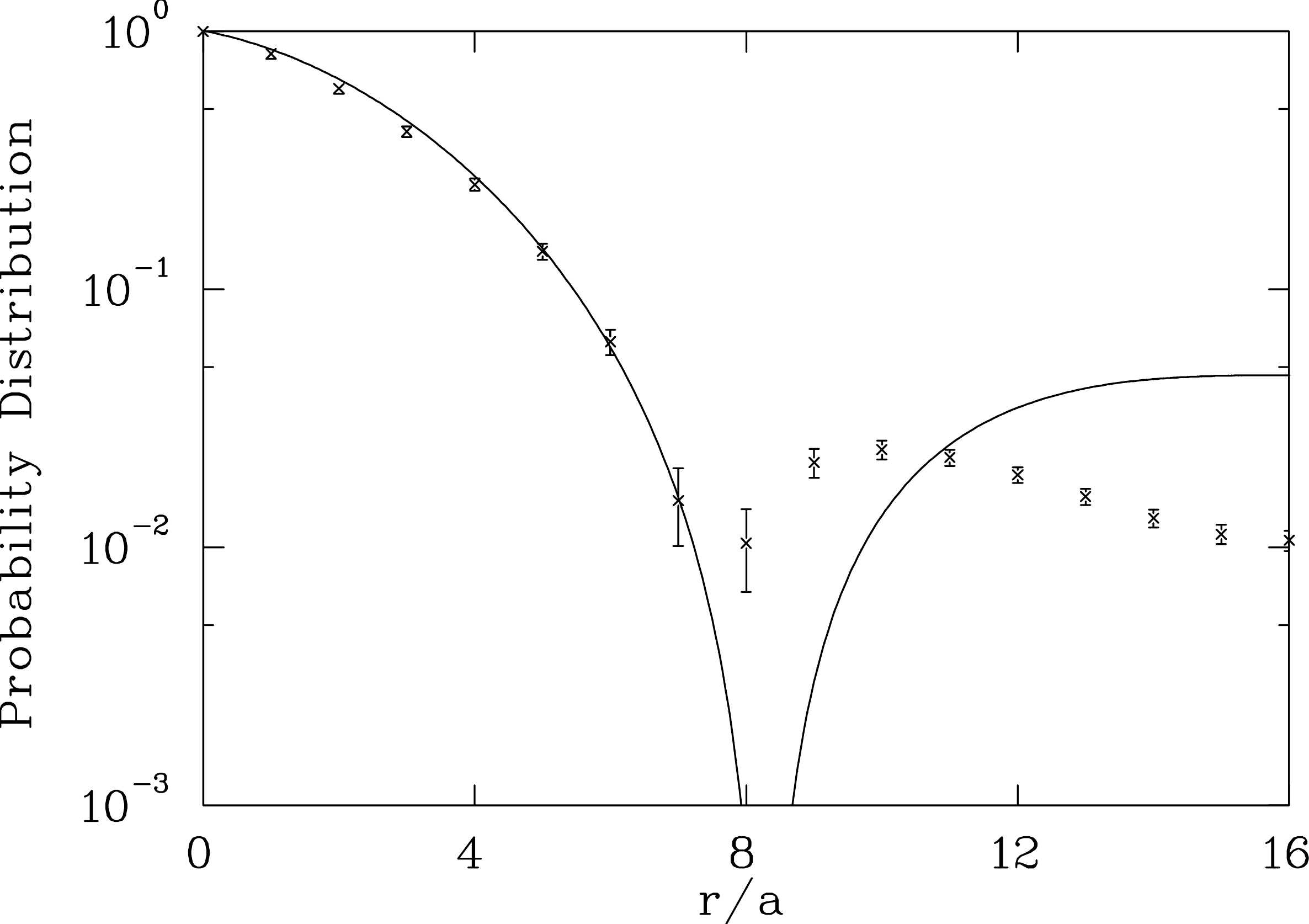}
\includegraphics[angle=90,width=0.3\linewidth]{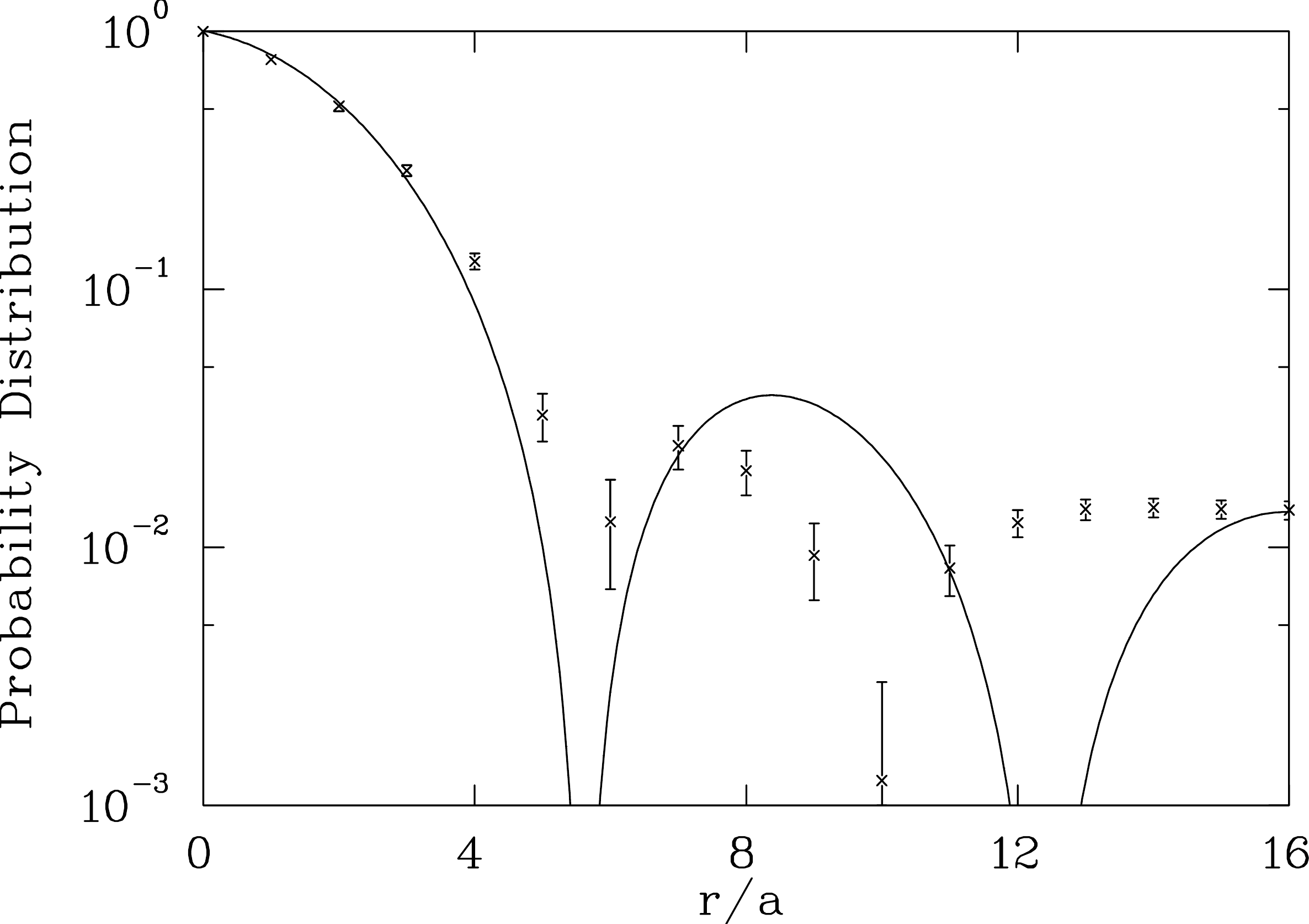}
\includegraphics[angle=90,width=0.3\linewidth]{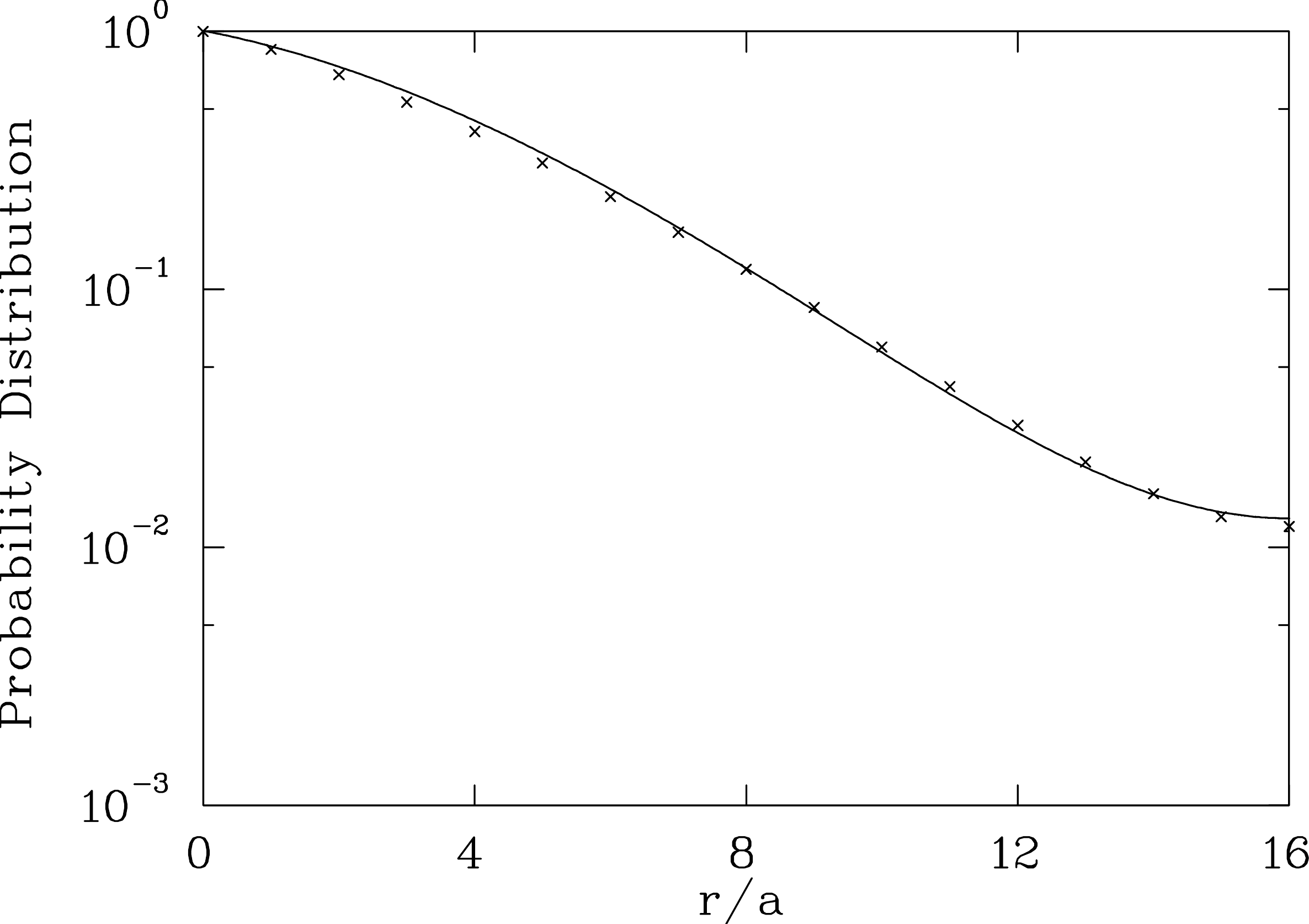}
\includegraphics[angle=90,width=0.3\linewidth]{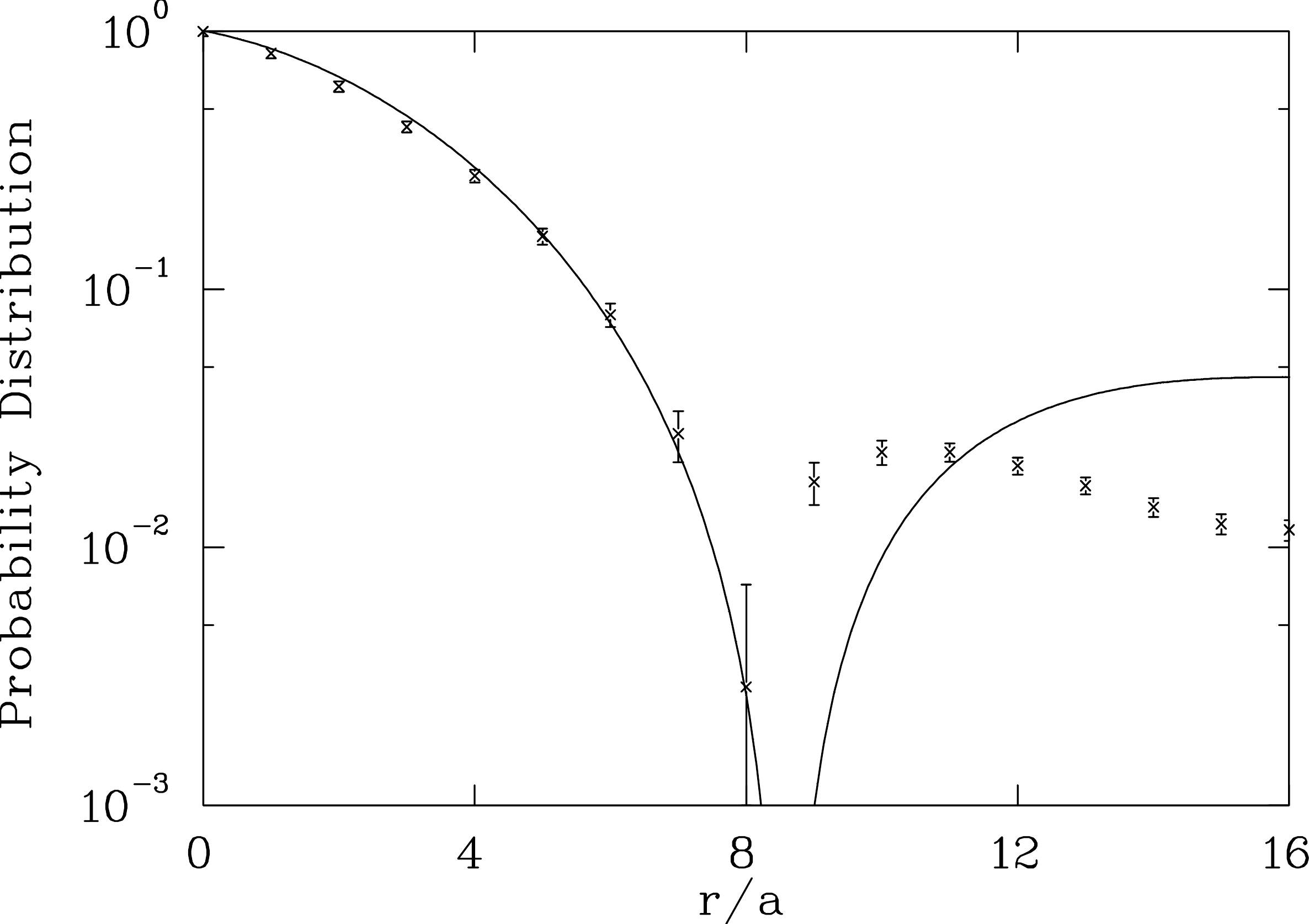}
\includegraphics[angle=90,width=0.3\linewidth]{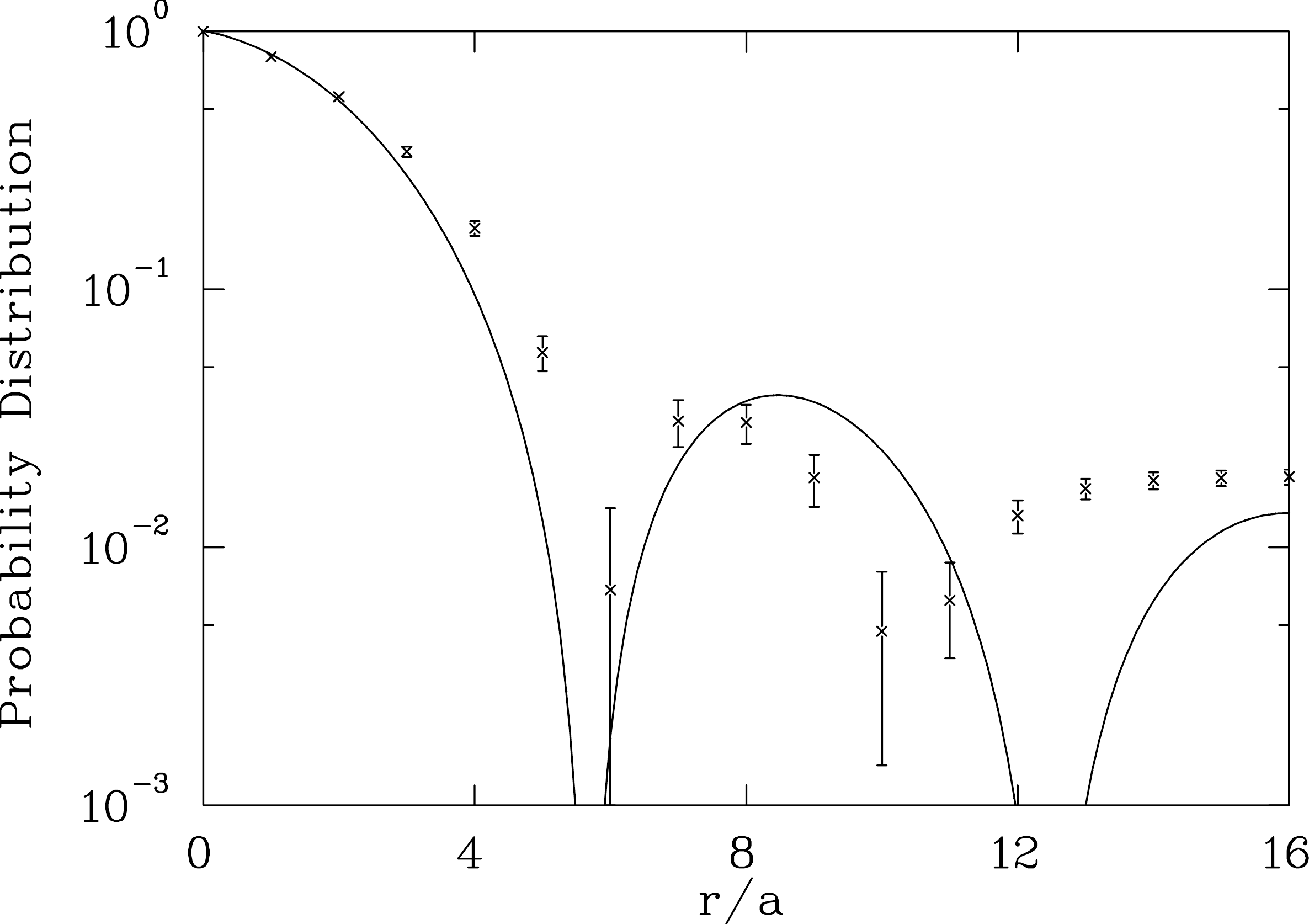}
\includegraphics[angle=90,width=0.3\linewidth]{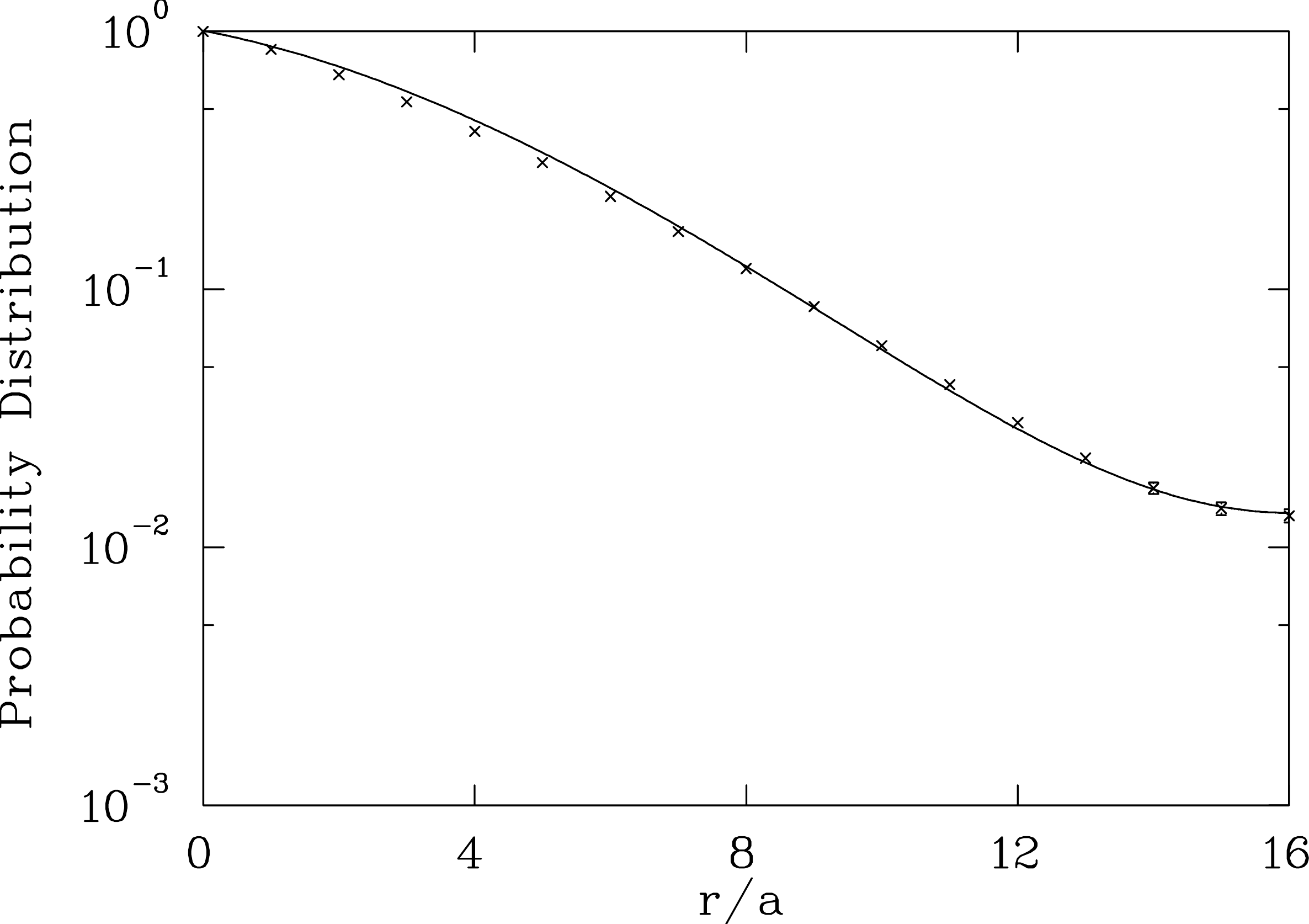}
\includegraphics[angle=90,width=0.3\linewidth]{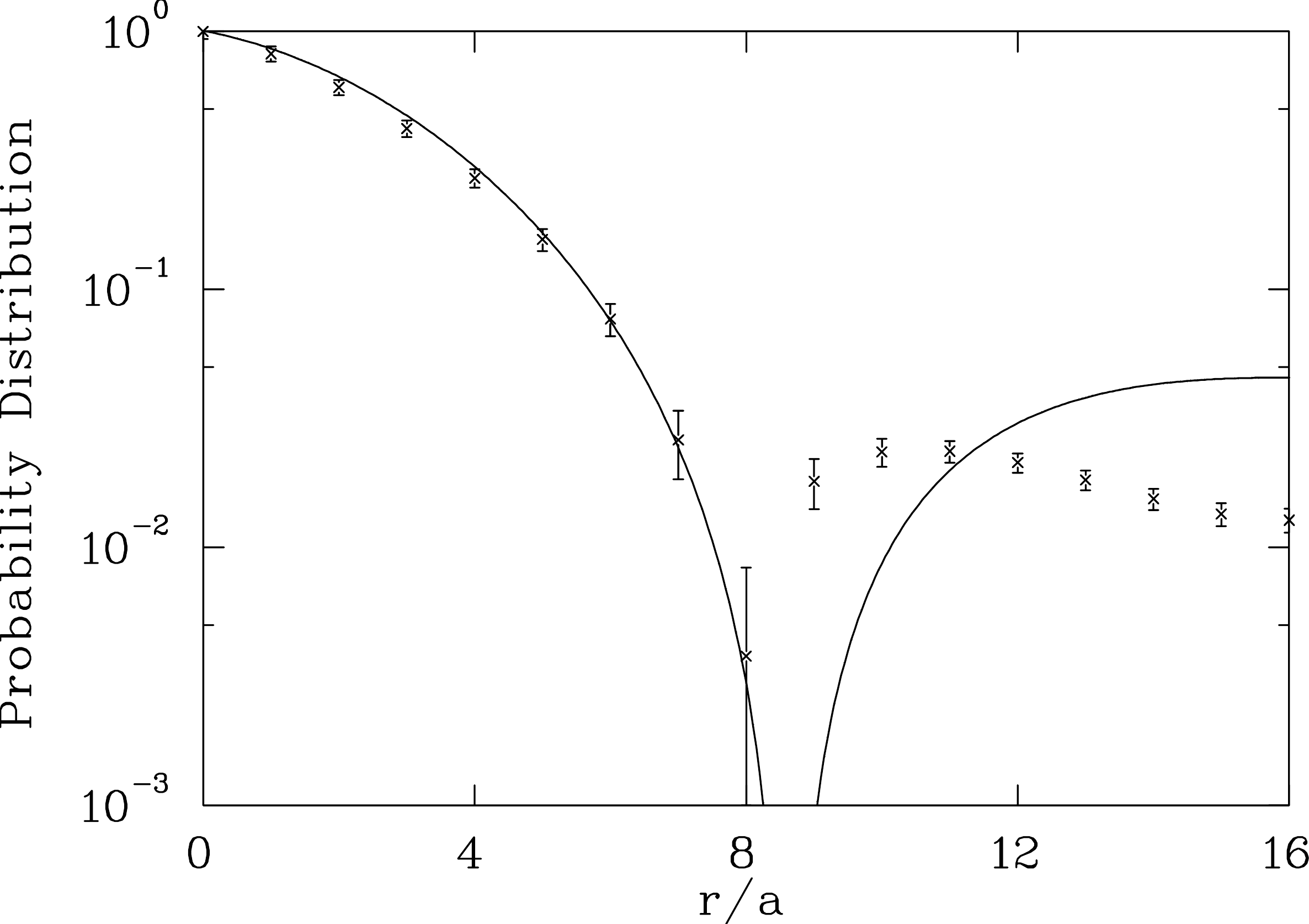}
\includegraphics[angle=90,width=0.3\linewidth]{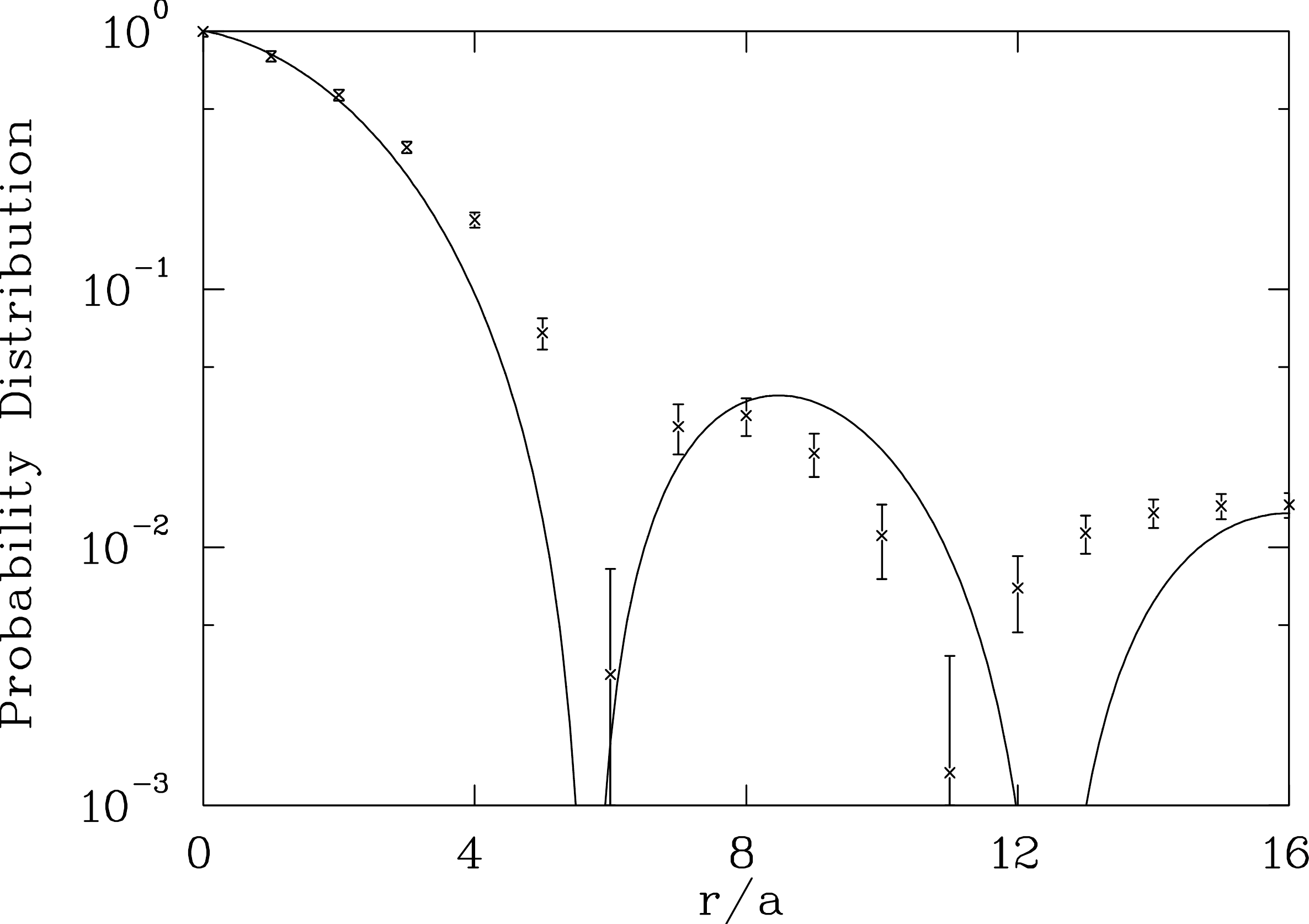}
\includegraphics[angle=90,width=0.3\linewidth]{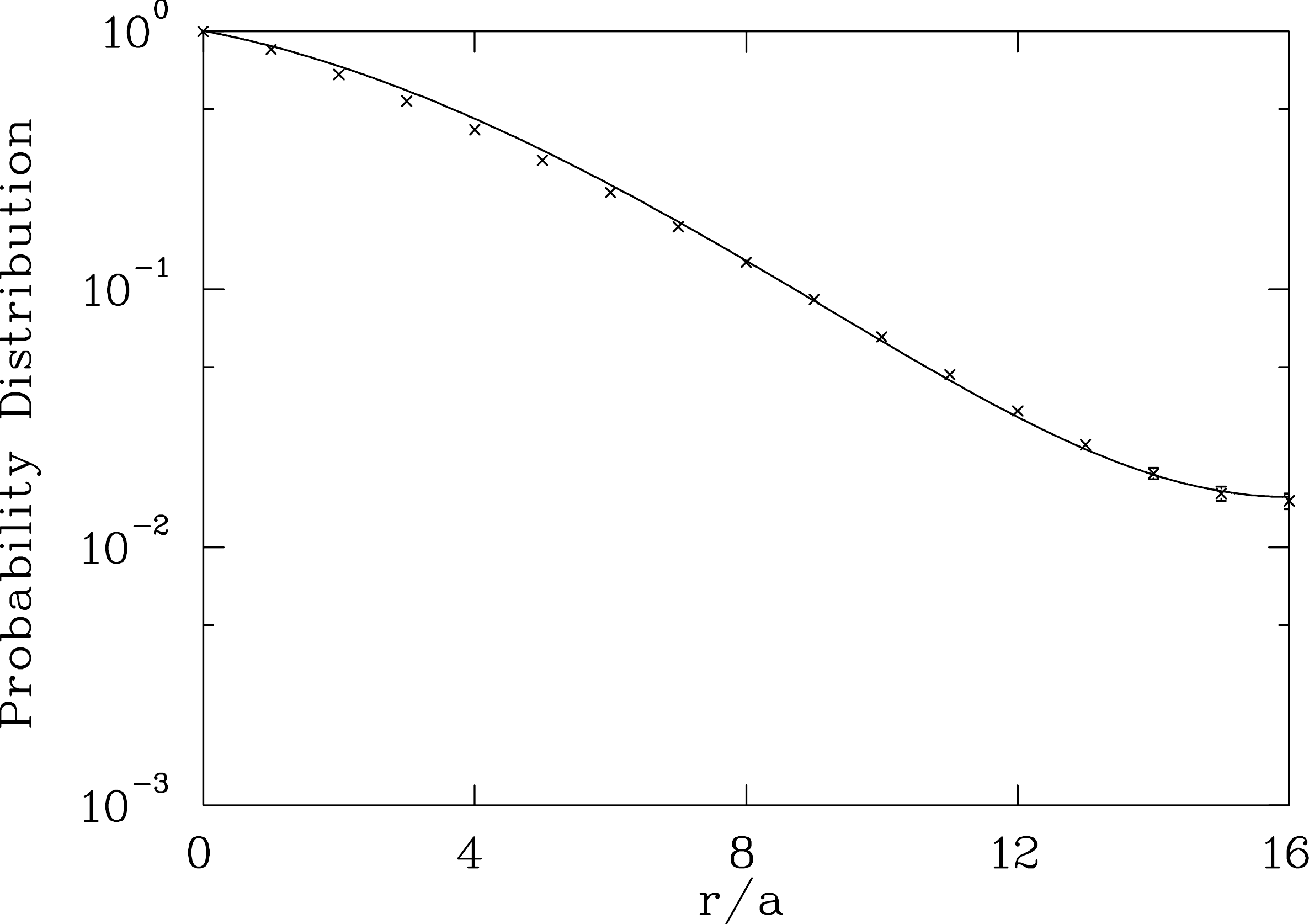}
\includegraphics[angle=90,width=0.3\linewidth]{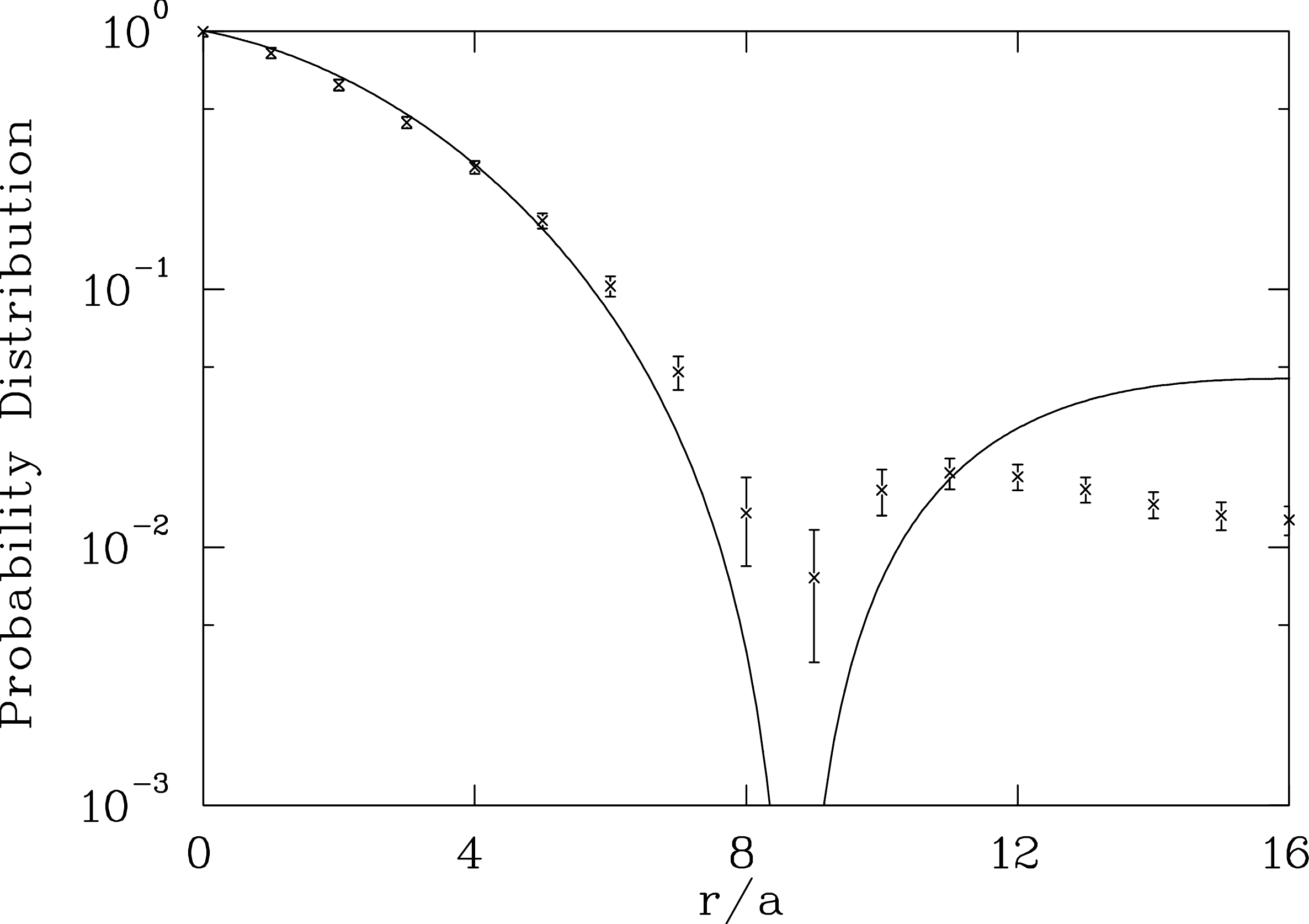}
\includegraphics[angle=90,width=0.3\linewidth]{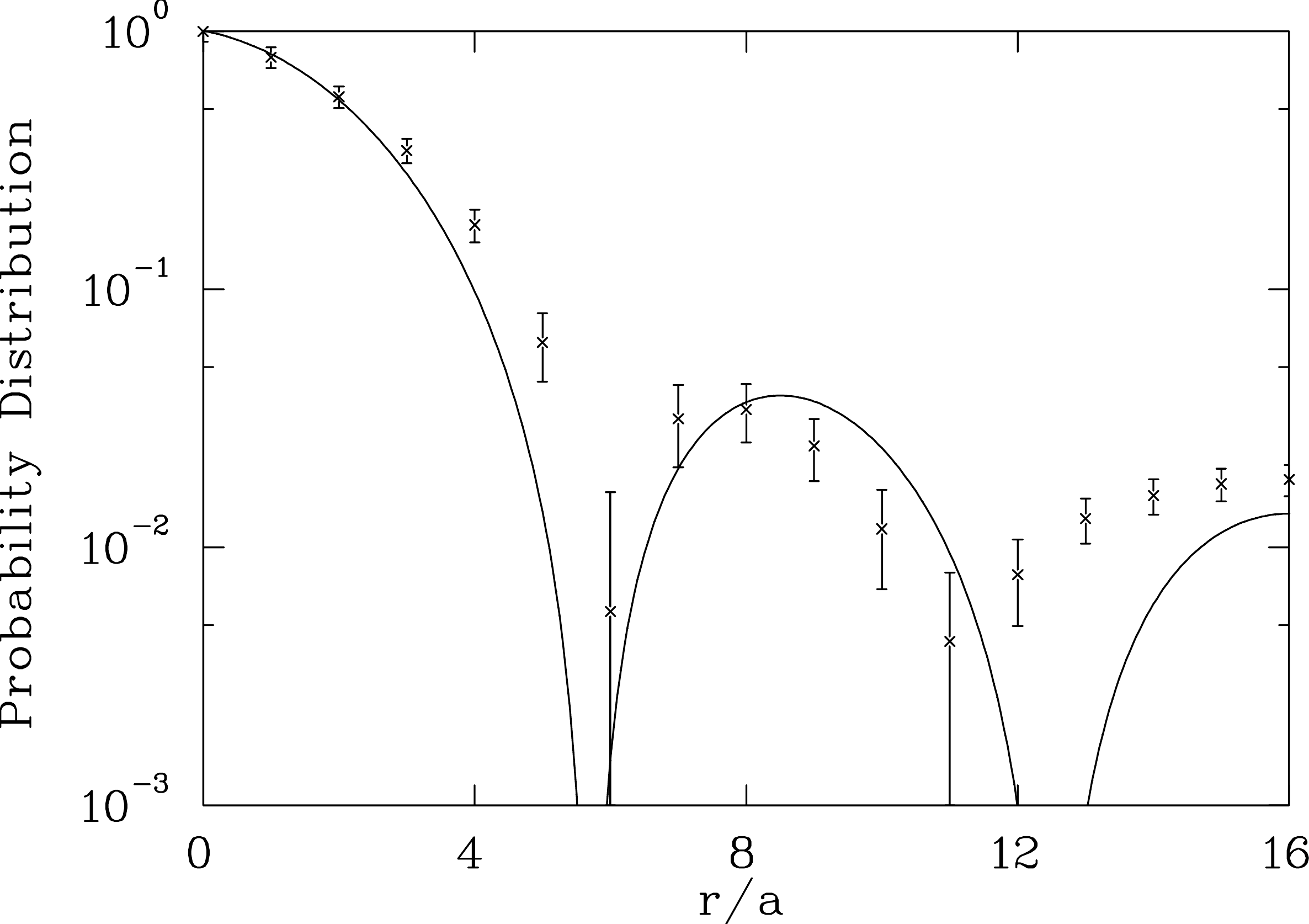}
\caption{The probability distributions for the $d$ quark about two $u$
  quarks fixed at the origin obtained in our lattice QCD calculations
  (crosses) are compared with the quark model prediction (solid curve)
  for the ground (left column), first- (middle column) and second-
  (right column) excited states.  Quark masses range from the heaviest
  (top row) through to the lightest (bottom row).
  The ground state probability distribution of the quark model
  closely resembles the lattice data for all masses considered.
  The first excited states matches the lattice data well at small
  distances, but the node is placed further from the centre of mass in
  the quark model, after which, the lattice data shows a distinct
  second peak, whereas the quark model rises to the boundary.  It is
  interesting that the most significant difference is observed where
  long-distance physics associated with pion-cloud effects not
  included in the quark model are significant.
  For the third state the amplitudes of the shells between the nodes
  of the wave function are predicted well.
}
\label{cqmcomp}
\end{figure*}

Figure \ref{waveFun} presents the wave functions for the first three
states at our lightest quark mass providing $m_\pi = 156$ MeV.  In the
excited states, the wave function changes sign revealing a node
structure consistent with $2S$ and $3S$ excited state wave functions.
To further explore the details of these wave functions, we construct a
probability density from the square of the wave function and plot it
on a logarithmic scale in Fig.~\ref{cqmcomp}.

Our point of comparison with previous models of quark probability
distributions comes from a non-relativistic constituent quark model
with a one-gluon-exchange motivated Coulomb-plus-ramp potential.  The
spin dependence of the model is given in Ref.~\cite{Bhaduri:1980fd}
and the radial Schrodinger equation is solved with boundary conditions
relevant to the lattice data; {\it i.e.} the derivative of the
wave function is set to vanish at a distance $L_x/2$ from the origin.

We consider standard values of the string tension $\sqrt{\sigma} =
440\pm 40$ MeV and optimize the constituent quark mass to minimize the
logarithmic difference between the quark model and lattice QCD
ground-state probability distributions illustrated in the left-hand
column of Fig.~\ref{cqmcomp}.  We find best fit results for
$\sqrt{\sigma} = 400$ MeV and the optimal constituent quark masses
range from 340 to 350 MeV over the range of PACS-CS quark masses
available.  The quark mass dependence is more subtle than expected and
may be associated with the finite volume of the lattice suppressing
changes in the wave function as the quark mass is varied.  At the
lightest quark mass, just above those of Nature, the value of 340 MeV
is in accord with those traditionally used to describe the hadron
spectrum or baryon magnetic moments.

The lattice data for the first three states and all five quark masses
are compared with the constituent quark model in Fig.~\ref{cqmcomp}.
The wave functions are normalized to 1 at the origin.  As the quark
model parameters are determined using only the ground state
probability distribution, the probability densities illustrated for
the excited states are predictions.

An examination of the left-hand column of Fig.~\ref{cqmcomp} reveals
the subtle changes associated with the quark mass.  The probability
distribution of the heaviest ensemble falls off faster and requires a
slightly heavier constituent quark mass to fit the lattice results.
This subtle mass dependence is consistent with early, quenched wave
function studies \cite{Hecht:1992uq}.

Comparing the lattice probability distribution for the $d$ quark in
the first excited state to that predicted by the constituent quark
model in the middle column of Fig.~\ref{cqmcomp}, we see a qualitative
similarity but with important differences.  The quark model predicts
the behavior of the lattice wave function very well within the node
and predicts the position of the node rather well, particularly at the
lightest quark mass.  However, the shape of the wave function tail is
very poorly predicted, suggesting an important role for degrees of
freedom not contained within the quark model.  For example, the long
range pion tail of multi-particle components could alter the
distribution of quarks within the state on the lattice.  The poorest
agreement is for the heaviest ensembles, where the baryon mass is in
close proximity to the $\pi N$ scattering threshold.

Similar comments apply to the second excited state illustrated in the
right-hand column of Fig.~\ref{cqmcomp}.  While the positions of the
nodes are predicted approximately, the amplitudes of the wave function
between the nodes are very accurately predicted by the quark model.
Again the largest discrepancies are for the heaviest states where the
baryon mass is in close proximity to the $\pi \pi N$ scattering
threshold.

\subsection{Quark Mass Dependence of the Probability Distributions}

\begin{figure*}[p]
\includegraphics[clip=true,trim=2.5cm 0.0cm 2.5cm 0.0cm,width=0.24\linewidth]{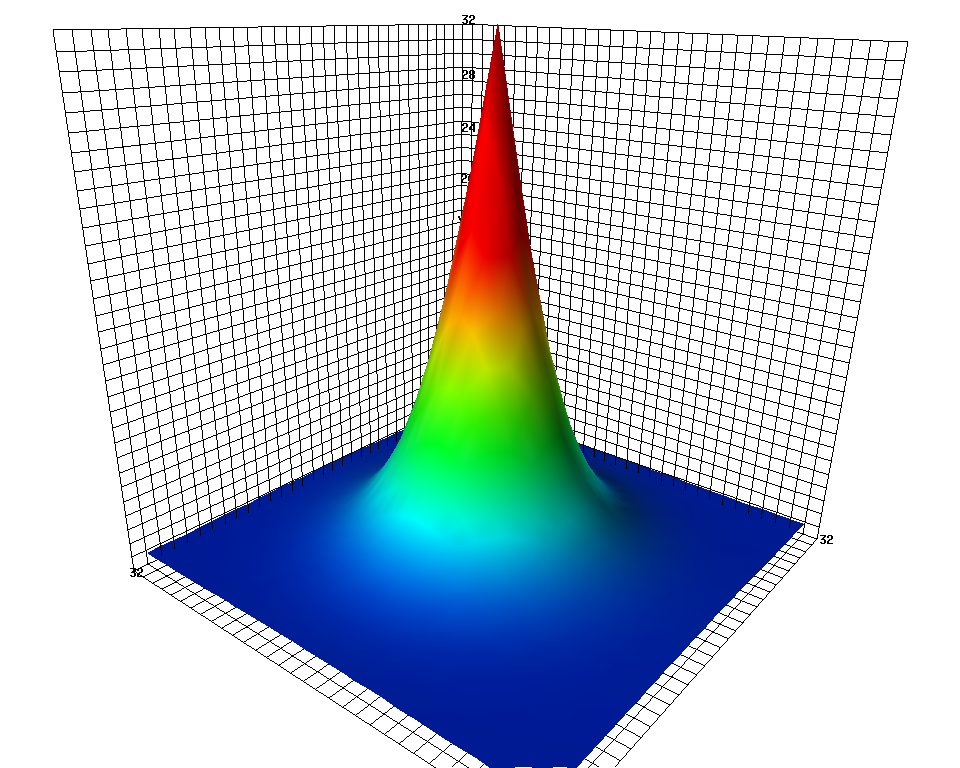}
\includegraphics[clip=true,trim=2.5cm 0.0cm 2.5cm 0.0cm,width=0.24\linewidth]{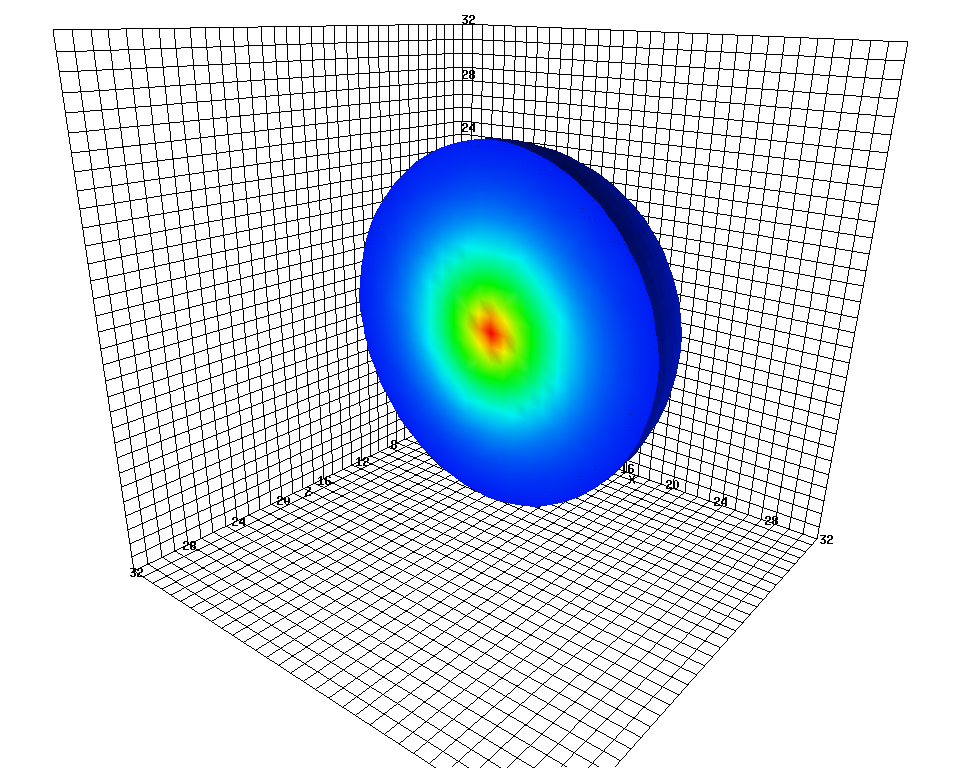}
\includegraphics[clip=true,trim=2.5cm 0.0cm 2.5cm 0.0cm,width=0.24\linewidth]{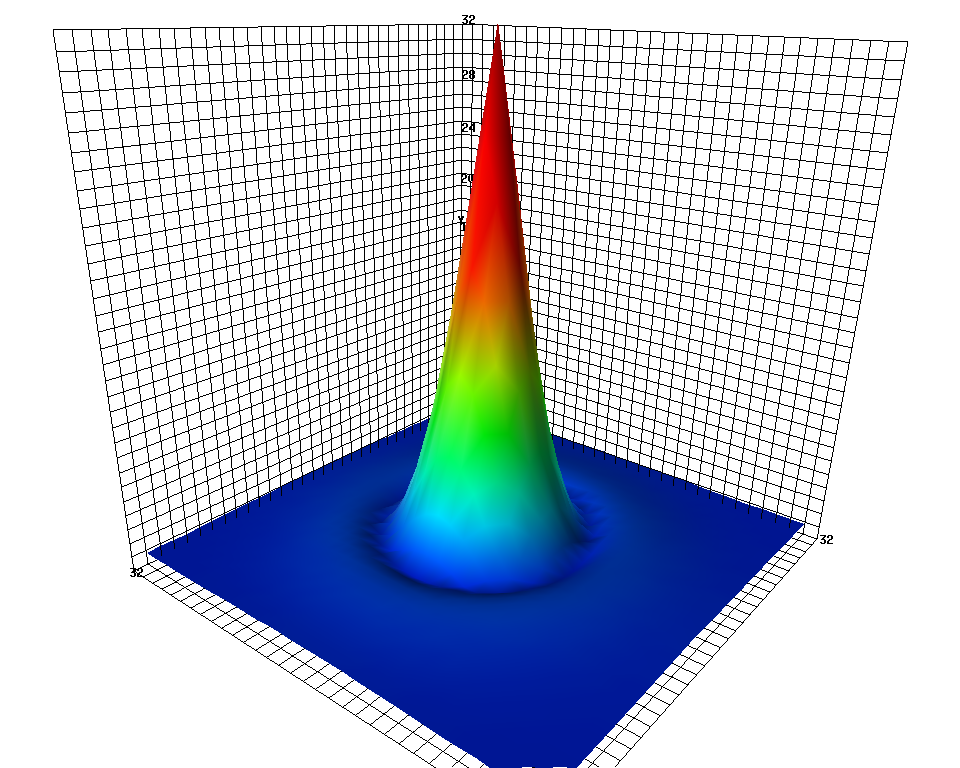}
\includegraphics[clip=true,trim=2.5cm 0.0cm 2.5cm 0.0cm,width=0.24\linewidth]{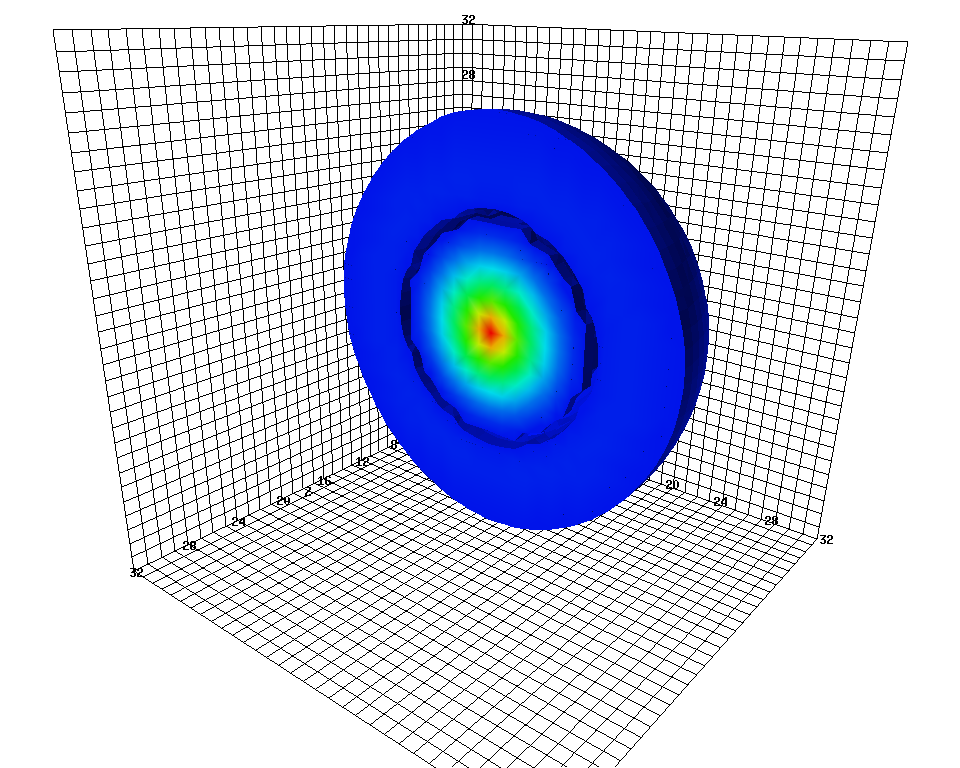}     \\[0.1cm]
\includegraphics[clip=true,trim=2.5cm 0.0cm 2.5cm 0.0cm,width=0.24\linewidth]{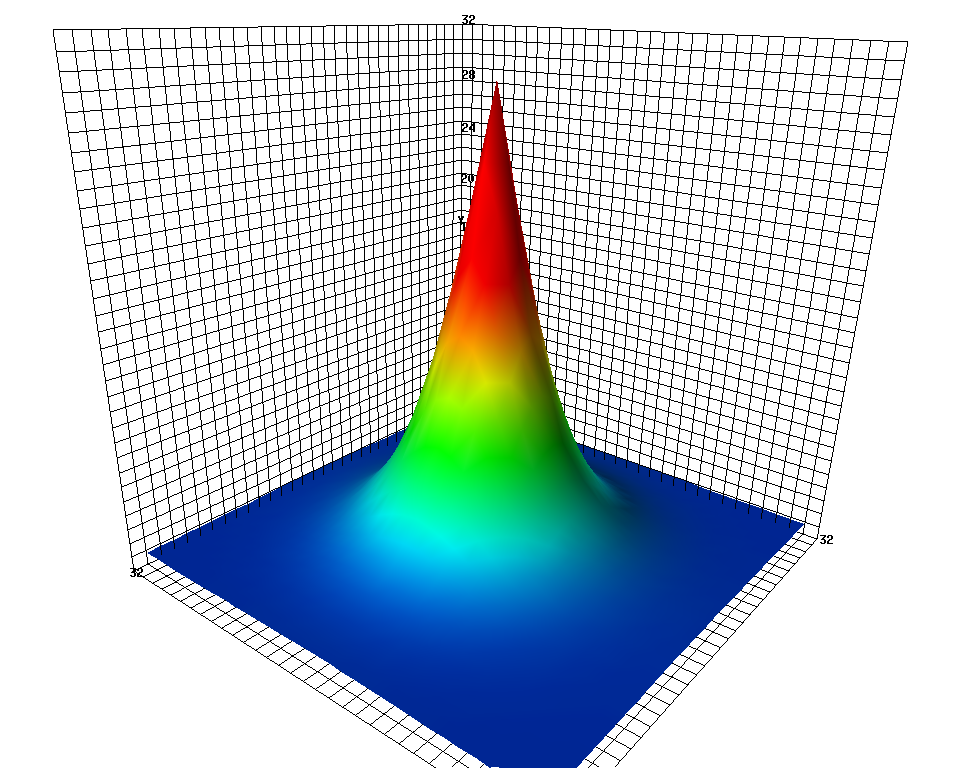}
\includegraphics[clip=true,trim=2.5cm 0.0cm 2.5cm 0.0cm,width=0.24\linewidth]{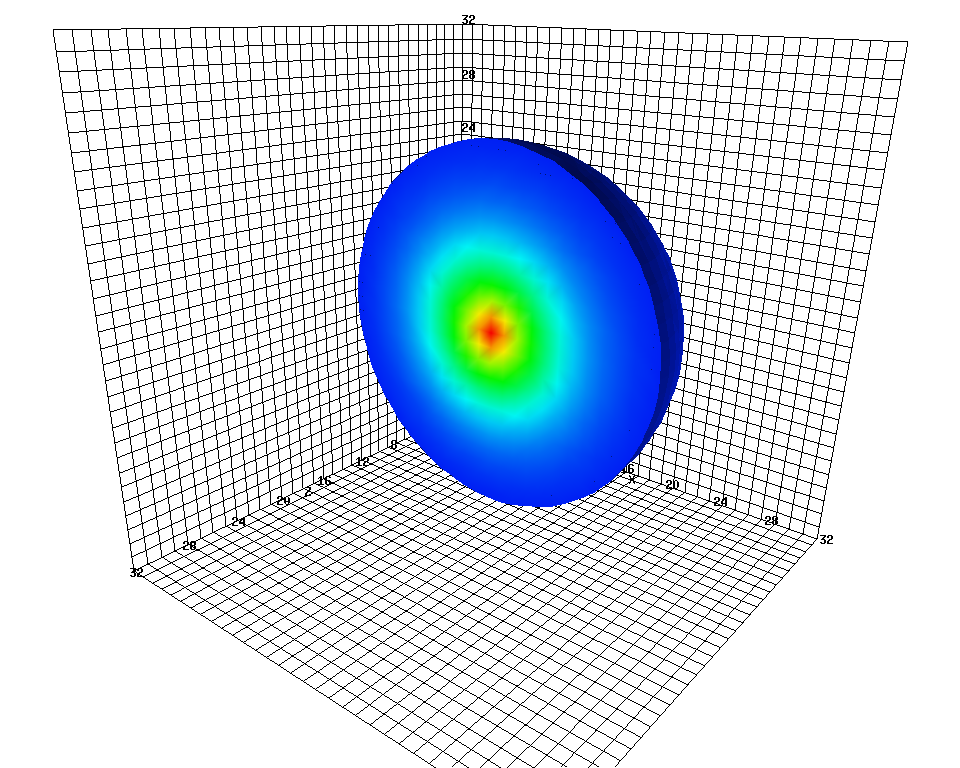}
\includegraphics[clip=true,trim=2.5cm 0.0cm 2.5cm 0.0cm,width=0.24\linewidth]{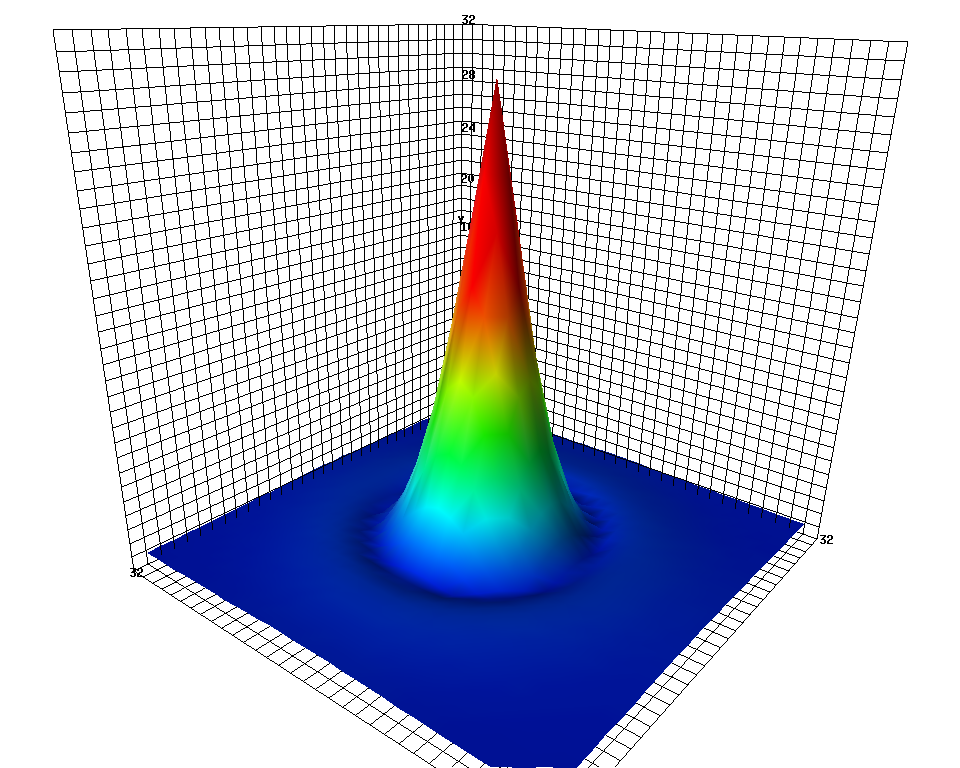}
\includegraphics[clip=true,trim=2.5cm 0.0cm 2.5cm 0.0cm,width=0.24\linewidth]{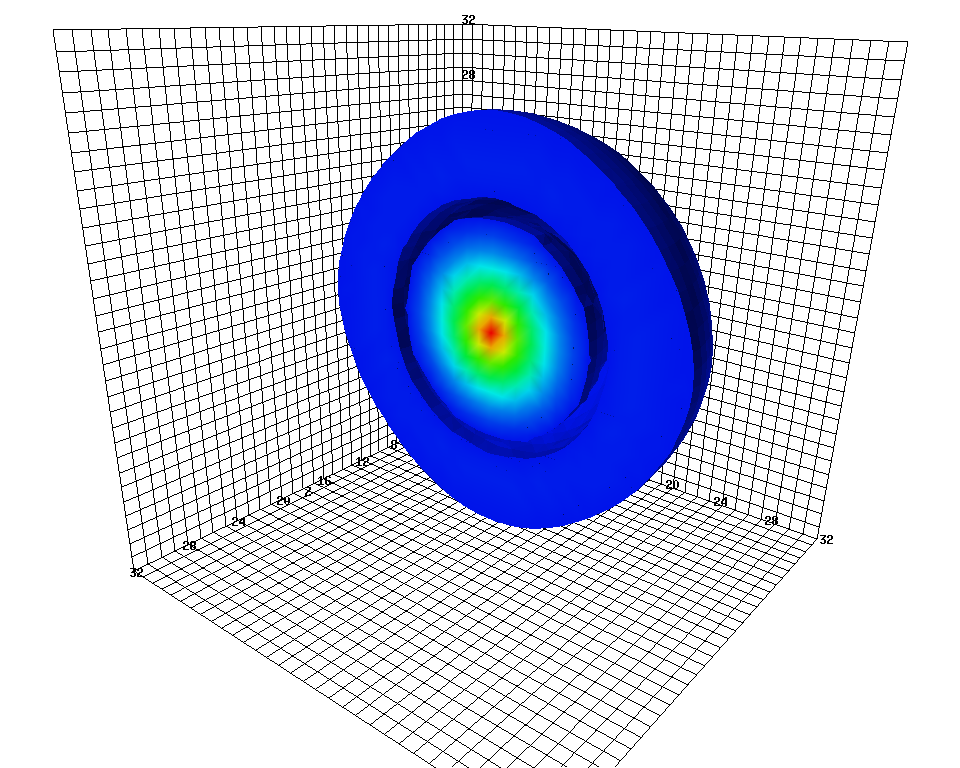}     \\[0.1cm]
\includegraphics[clip=true,trim=2.5cm 0.0cm 2.5cm 0.0cm,width=0.24\linewidth]{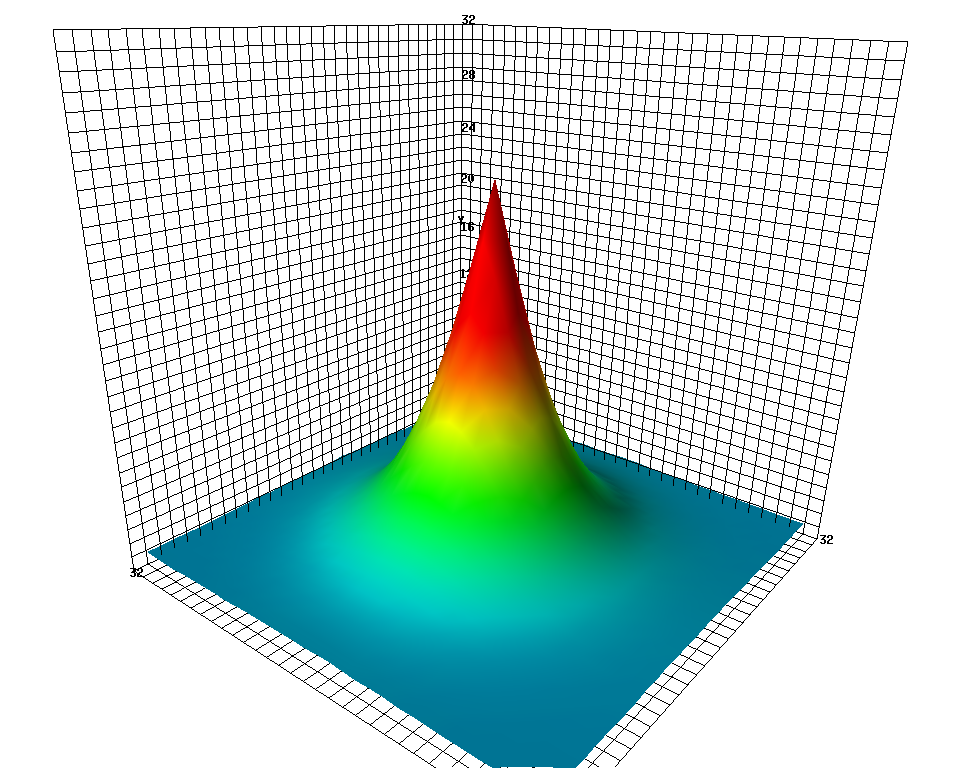}
\includegraphics[clip=true,trim=2.5cm 0.0cm 2.5cm 0.0cm,width=0.24\linewidth]{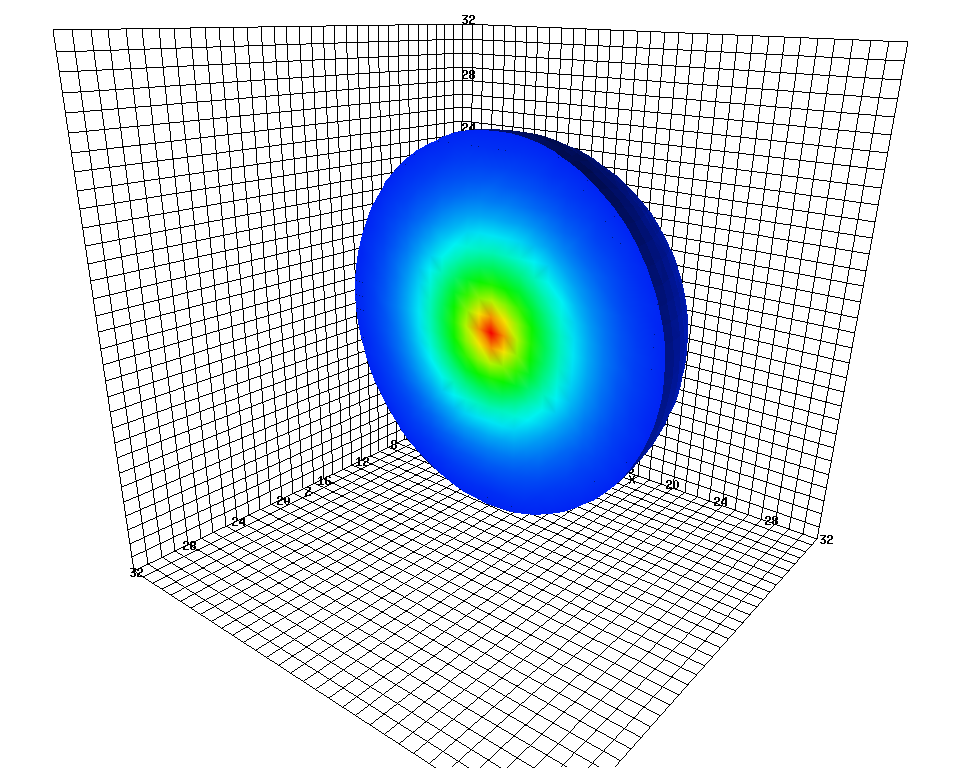}
\includegraphics[clip=true,trim=2.5cm 0.0cm 2.5cm 0.0cm,width=0.24\linewidth]{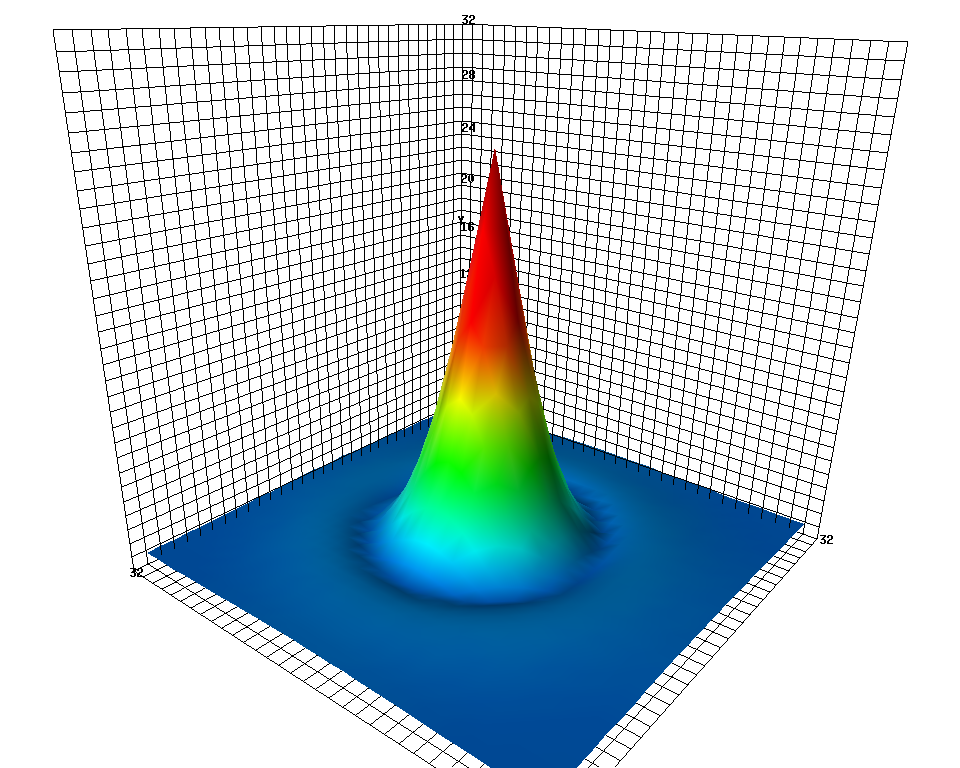}
\includegraphics[clip=true,trim=2.5cm 0.0cm 2.5cm 0.0cm,width=0.24\linewidth]{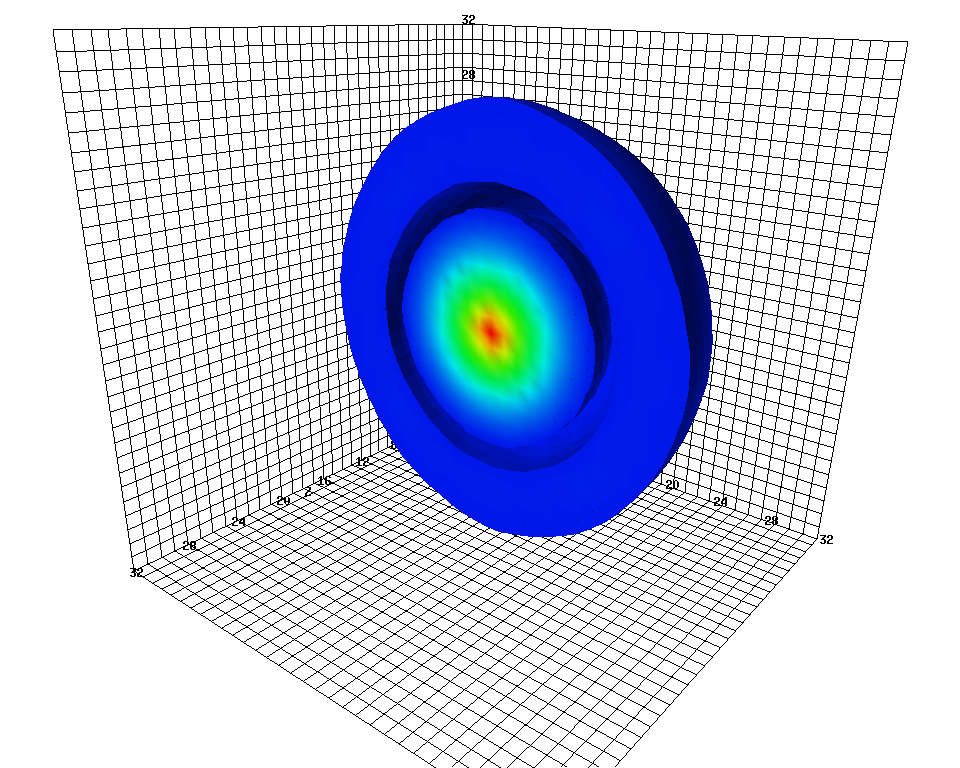}     \\[0.1cm]
\includegraphics[clip=true,trim=2.5cm 0.0cm 2.5cm 0.0cm,width=0.24\linewidth]{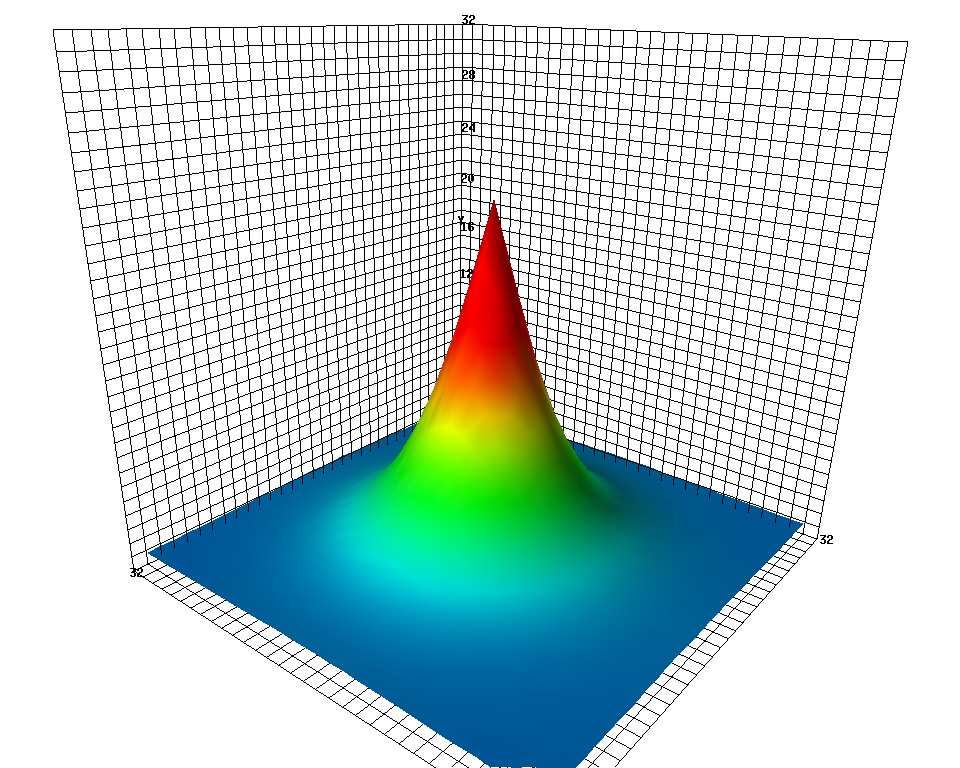}
\includegraphics[clip=true,trim=2.5cm 0.0cm 2.5cm 0.0cm,width=0.24\linewidth]{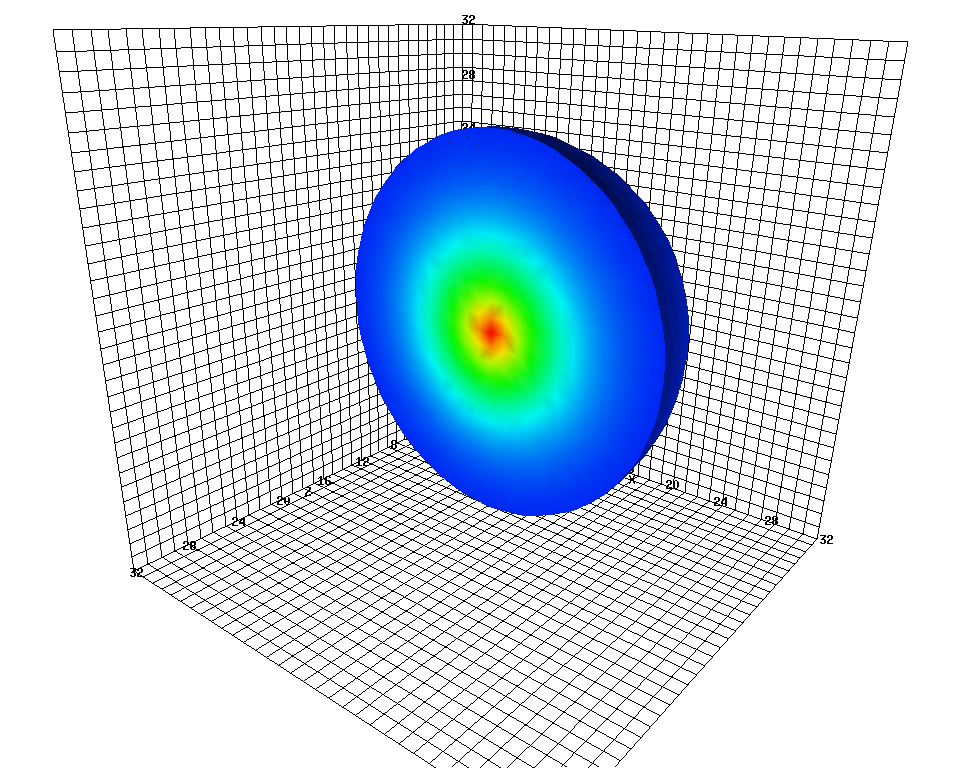}
\includegraphics[clip=true,trim=2.5cm 0.0cm 2.5cm 0.0cm,width=0.24\linewidth]{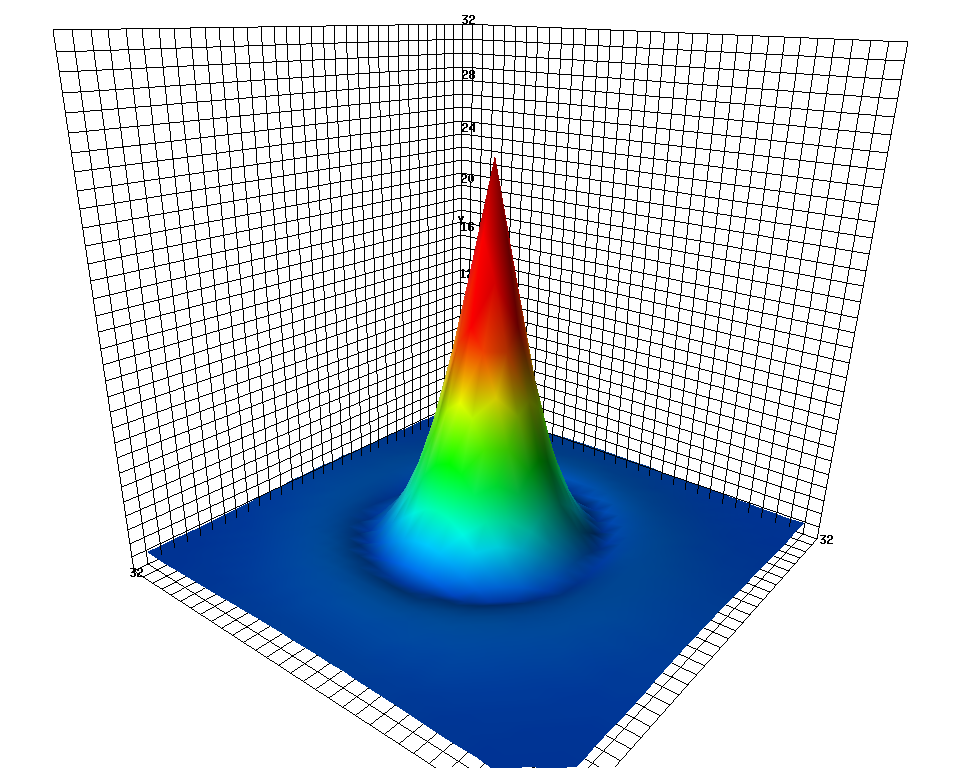}
\includegraphics[clip=true,trim=2.5cm 0.0cm 2.5cm 0.0cm,width=0.24\linewidth]{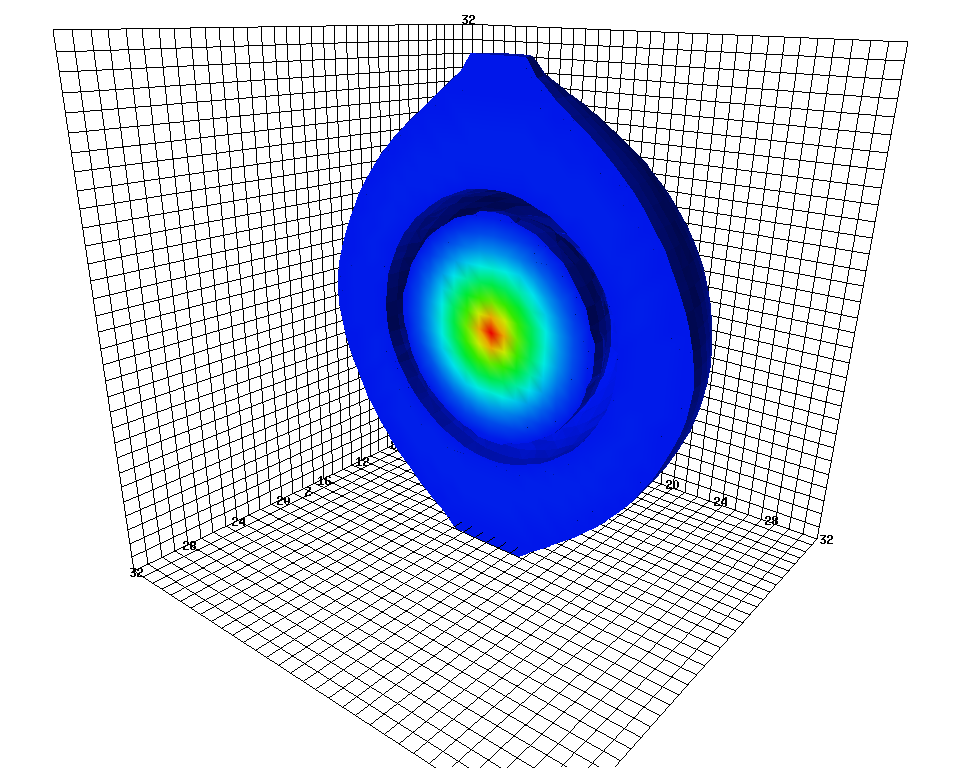}     \\[0.1cm]
\includegraphics[clip=true,trim=2.5cm 0.0cm 2.5cm 0.0cm,width=0.24\linewidth]{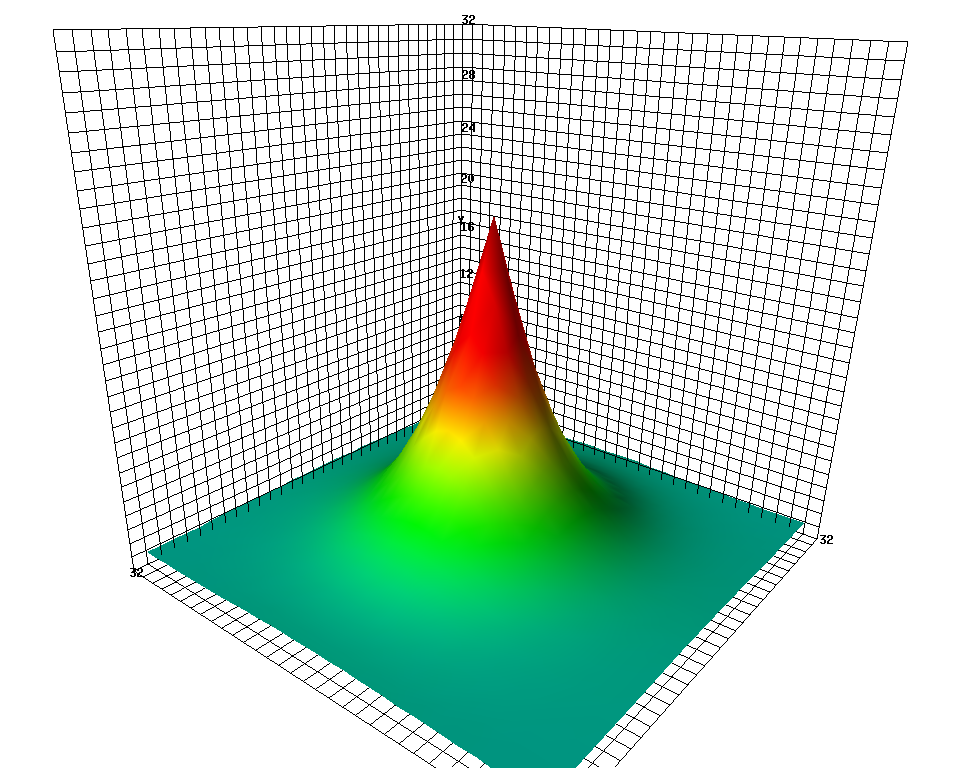}
\includegraphics[clip=true,trim=2.5cm 0.0cm 2.5cm 0.0cm,width=0.24\linewidth]{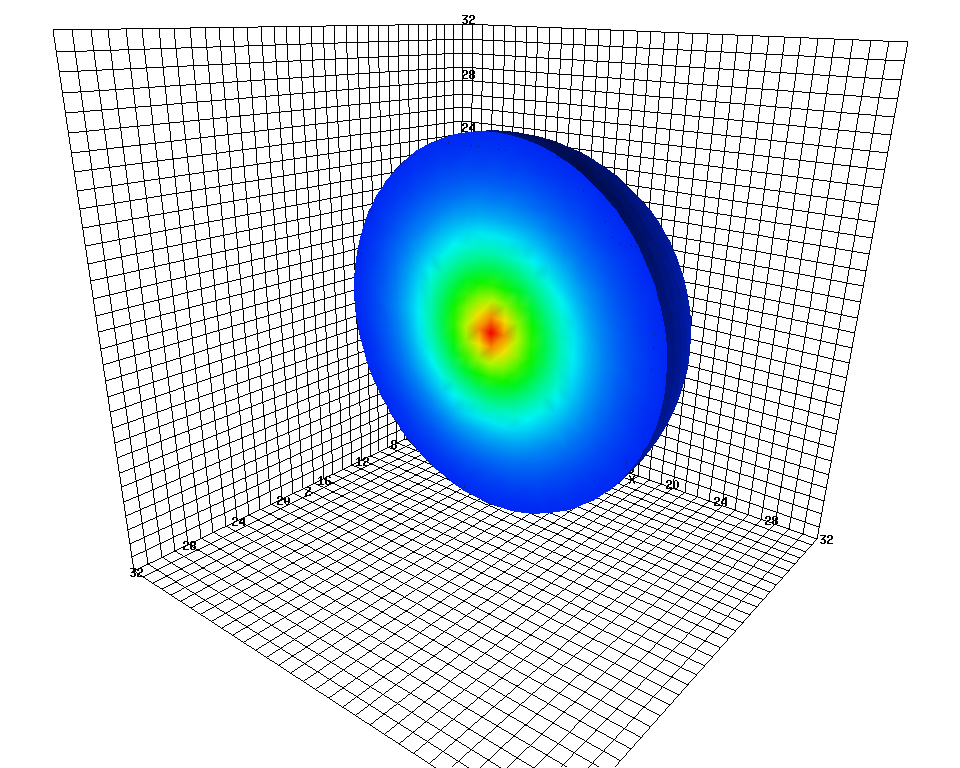}
\includegraphics[clip=true,trim=2.5cm 0.0cm 2.5cm 0.0cm,width=0.24\linewidth]{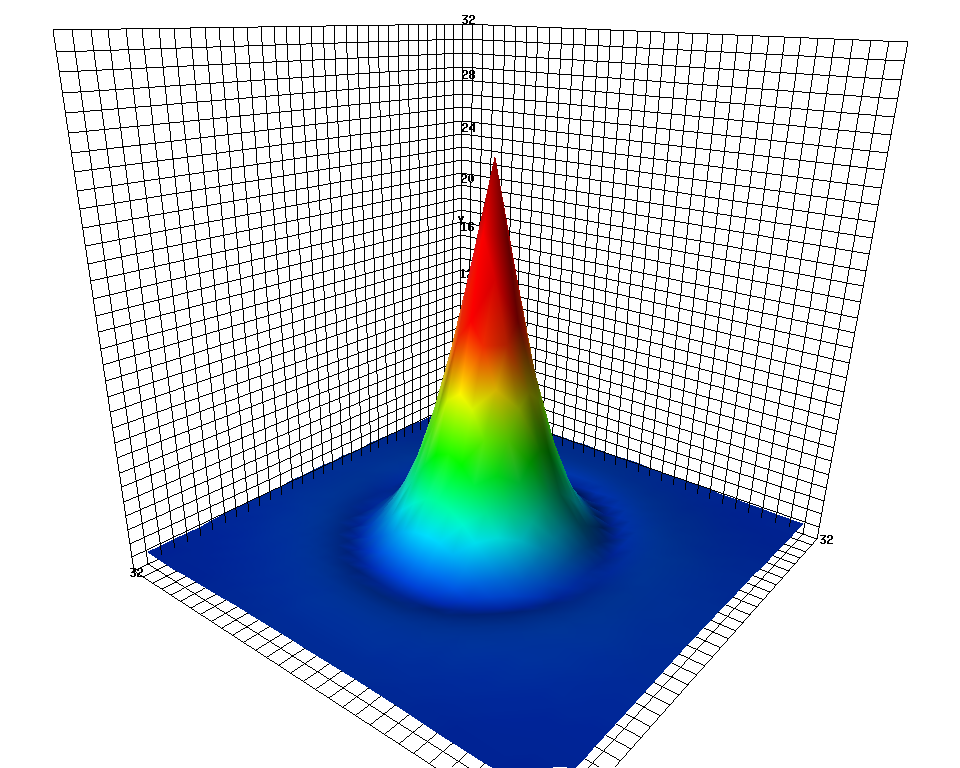}
\includegraphics[clip=true,trim=2.5cm 0.0cm 2.5cm 0.0cm,width=0.24\linewidth]{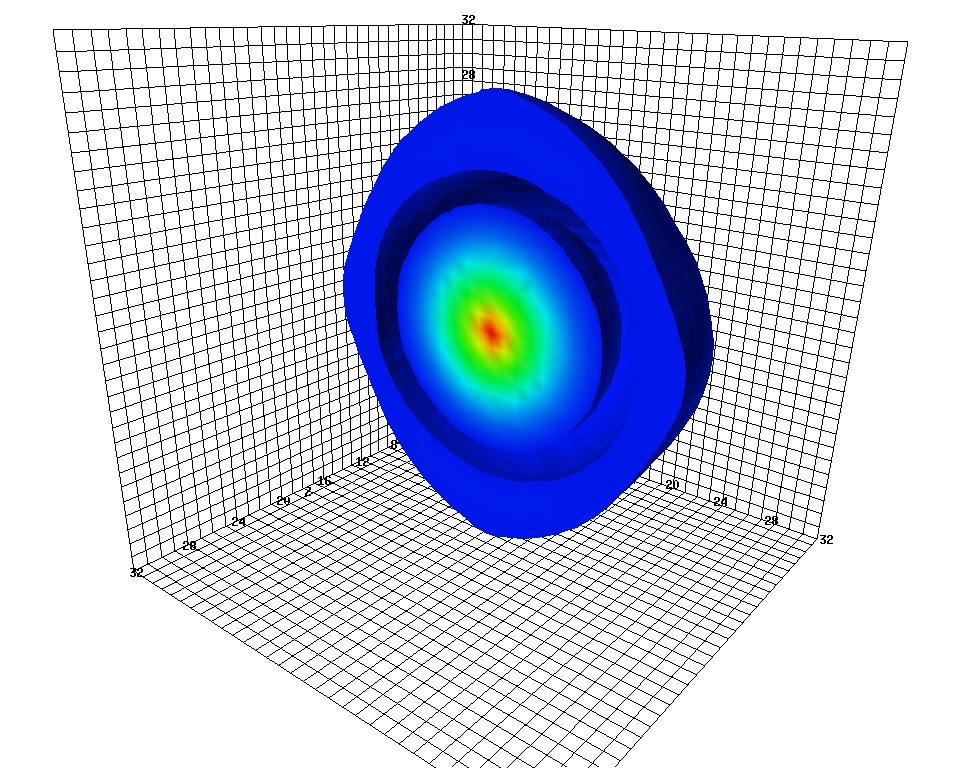}     \\[0.1cm]
\caption{The dependence of the $d$-quark probability distribution on
  the masses of the quarks in the proton (two left-hand columns) and
  its first excited state (two right-hand columns).  The $u$ quarks
  are fixed at the origin at the centre of the plot.  The quark mass
  decreases from heaviest (top row) to lightest (bottom row).  For
  each mass and state, the probability density is normalised to unity
  over the spatial volume of the lattice.  The isovolume threshold for
  rendering the probability distribution in the second and fourth
  columns is $3.0 \times 10^{-5}$. }
\label{MassDepA}
\end{figure*}

\subsubsection{Ground-State Distribution}

The mass dependence of the ground state probability distribution for
the $d$ quark about the two $u$ quarks fixed at the origin is
illustrated in the two left-hand columns of Fig.~\ref{MassDepA}.  The
plots are arranged from heaviest to lightest ensembles with quark mass
decreasing down the page.

Although a Gaussian distribution is used to excite the ground state
from the vacuum, the well-known sharp-peaked shape associated with the
Coulomb potential is reproduced in the probability density for all
quark masses.  This is best observed in the left most column where an
isosurface reports the probability-density values in the plane
containing the two $u$ quarks at the origin.

Because the total probability density is normalised to unity in the
spatial volume, the height of the peak drops as the $d$ quark becomes
light and moves to larger distances from the $u$ quarks.  The
isosurface provides the clearest representation of the mass dependence
of the ground state.

This gentle broadening of the distribution is also reflected in the
isovolume rendering of the projected ground state probability density
in the second column if Fig.~\ref{MassDepA}.  The isovolume has been
cut into the plane containing the $u$ quarks at the origin.
Probability-density values are depicted by a colour map similar to
that used for the isosurface.  The threshold for rendering the
probability distribution is $3.0 \times 10^{-5}$, revealing a smooth
sphere for the surface of the probability distribution.

Finite volume effects do not appear to be significant in the
probability densities of the ground state at any of the quark masses
considered. This is in spite of the fact that the lightest ensemble
has $m_\pi L=2.23$.

\subsubsection{First Excited State}

Lattice results for the $d$-quark probability distribution about the
two $u$ quarks at the origin in the first excited state of the proton
are illustrated in the third and fourth columns of
Fig.~\ref{MassDepA}.  In the light quark-mass regime, this first
excited state is associated with the Roper resonance.  The darkened
ring around the peak of the isosurface indicates a node in the
probability distribution, consistent with a $2S$ radial excitation of
the $d$ quark.  The node is better illustrated in the isovolume
renderings where the probability density drops below the rendering
cutoff of $3.0 \times 10^{-5}$ and leaves a void between the inner and
outer shells of the state.

It is interesting that the narrowest distribution is seen at the
heaviest quark masses, even though these states have energies
coincident with the $\pi N$ scattering threshold.  Enforcing a colour
singlet structure in annihilating the three spatially separated quarks
prevents a direct observation of the two-particle components contained
in the dynamics governing the energy of the state.  In this case the
multi-particle components only modify the three-quark distributions.

The outer edge of the isovolume reveals interesting boundary effects
which may be associated with the necessary finite-volume effects of
multi-particle components mixed in the state.  The deviation from
spherical symmetry in the outer shell will be reflected in the energy
of the excited state observed in the finite-volume lattice simulation.
At the lightest two quark masses, the distortion of the probability
distribution is significant and will correspondingly influence the
eigen-energy.  Even with $m_\pi L= 4.4$ at the second lightest quark
mass, finite volume effects distort the wave function in a significant
manner.  Of course, this interplay between the finite volume and the
energy of the state is key to extracting resonance parameters from
lattice simulation results.

\begin{figure}
\includegraphics[angle=90,width=1.0\linewidth]{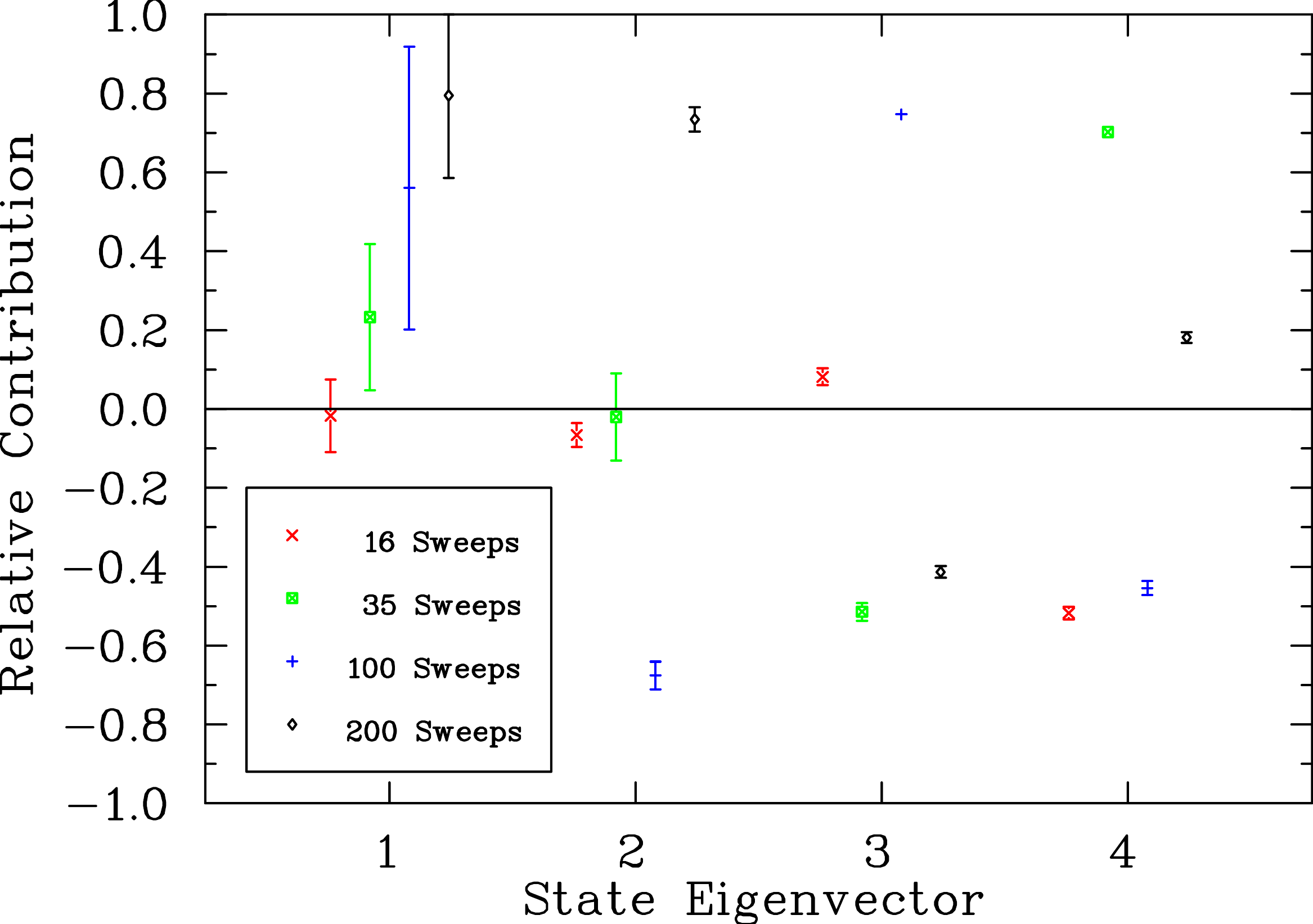}
\caption{The eigenvectors $u_i^\alpha$ describing the contributions of
  each of the source smearing levels to the states $\alpha$ for the
  lightest quark-mass ensemble considered.  Indices $i=1$ to 4
  correspond to 16, 35, 100 and 200 sweeps of gauge-invariant Gaussian
  smearing.  The superposition of positive and negative Gaussian
  smearing levels is consistent with the nodal structure recovered in
  the wave functions.}
\label{EvectPlot}
\end{figure}

The nodal structure of the first excited state also indicates that the
ideal combination of operators to access this state on the lattice is
a superposition of Gaussian distributions of different widths and
opposite signs \cite{Mahbub:2010rm,Mahbub:2013ala}.  Figure
\ref{EvectPlot} presents the eigenvectors $u_i^\alpha$ describing the
contributions of each of the source smearing levels to the states
$\alpha$ for the lightest quark-mass ensemble considered.

For the ground state, all smeared sources contribute positively to the
state.  There is significant interplay between the smeared sources
over the jackknife sub-ensembles giving rise to larger uncertainties
for the preferred operators.  This is not the case for the excited
states where particular superpositions of interpolating fields are
required to isolate the states.
 
For the first excited state, a single large width Gaussian contributes
with a sign opposite to that of a narrower Gaussian, reflecting the
wave function illustrated in the second plot of Fig.~\ref{waveFun}.
The combination of sources creating the second excited state has a
similar pattern, with a narrow Gaussian contributing positively, an
intermediate Gaussian contributing negatively and a wide Gaussian
contributing positively, again reflecting the wave function
illustrated in Fig.~\ref{waveFun} for this state.  This sign
alternating structure is also apparent for the fourth state suggesting
a $3S$ excitation for this state.  We will examine this state further
in the following.

Turning our attention to the mass dependence of the node we note the
movement is somewhat unusual.  While there is a general trend of the
node in the wave function moving outwards as the quark mass decreases,
there is negligible movement in the node between the third and second
lightest quark masses.  We also note how the width of the void in the
probability density increases with as the quarks become lighter.

\begin{figure*}[t!]
\includegraphics[clip=true,trim=2.5cm 0.0cm 2.5cm 0.0cm,width=0.24\linewidth]{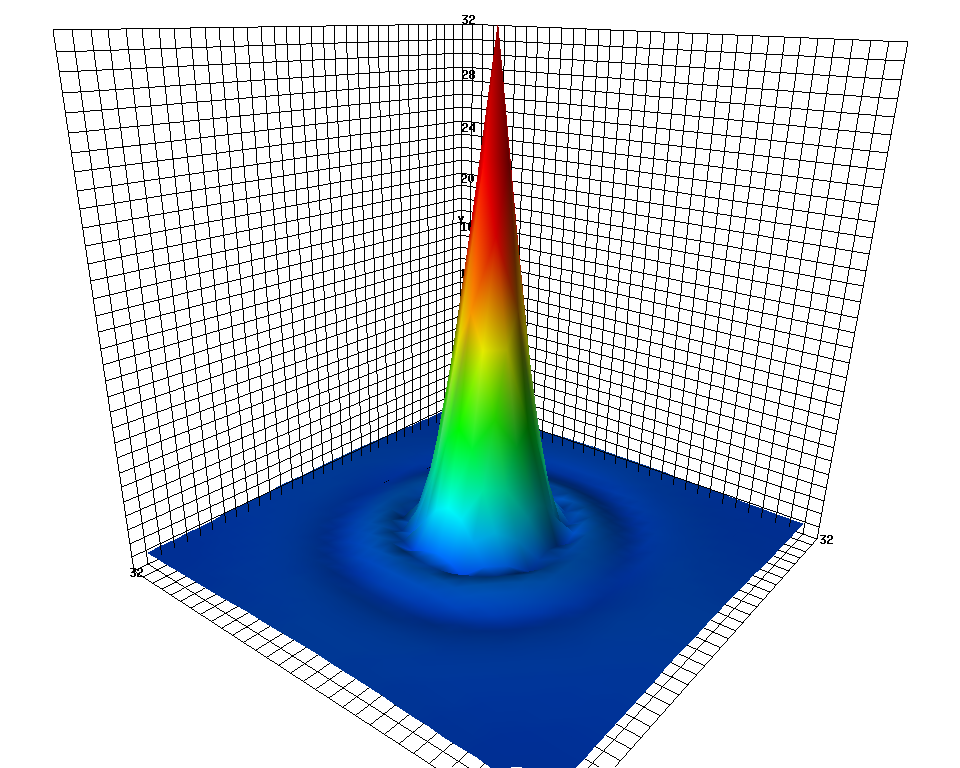}
\includegraphics[clip=true,trim=2.5cm 0.0cm 2.5cm 0.0cm,width=0.24\linewidth]{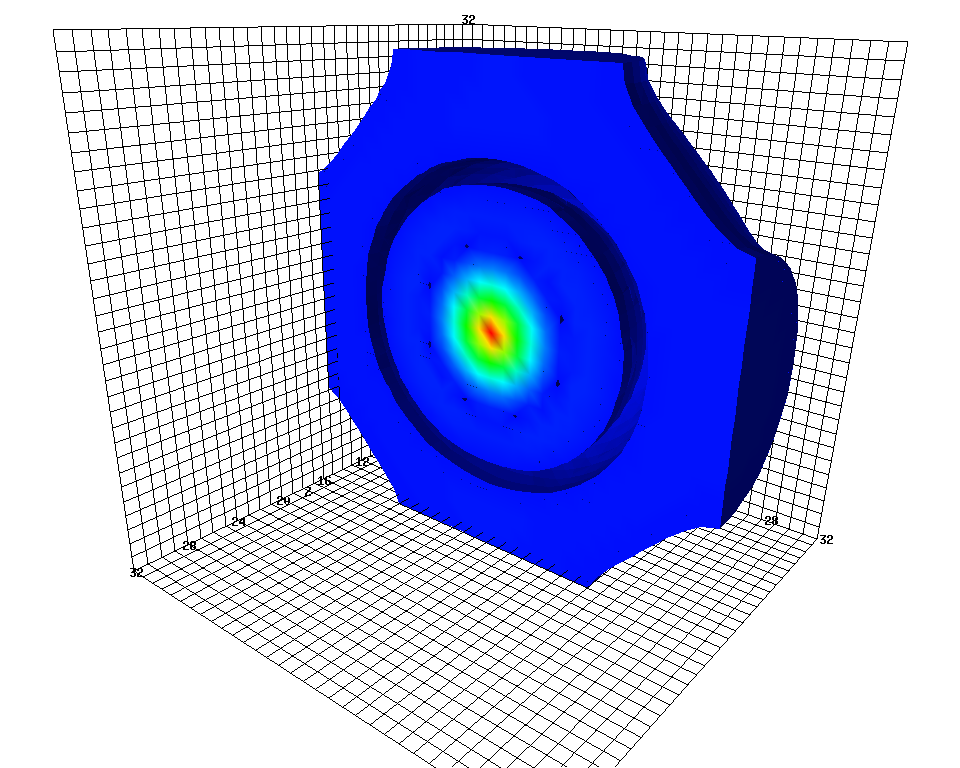}
\includegraphics[clip=true,trim=2.5cm 0.0cm 2.5cm 0.0cm,width=0.24\linewidth]{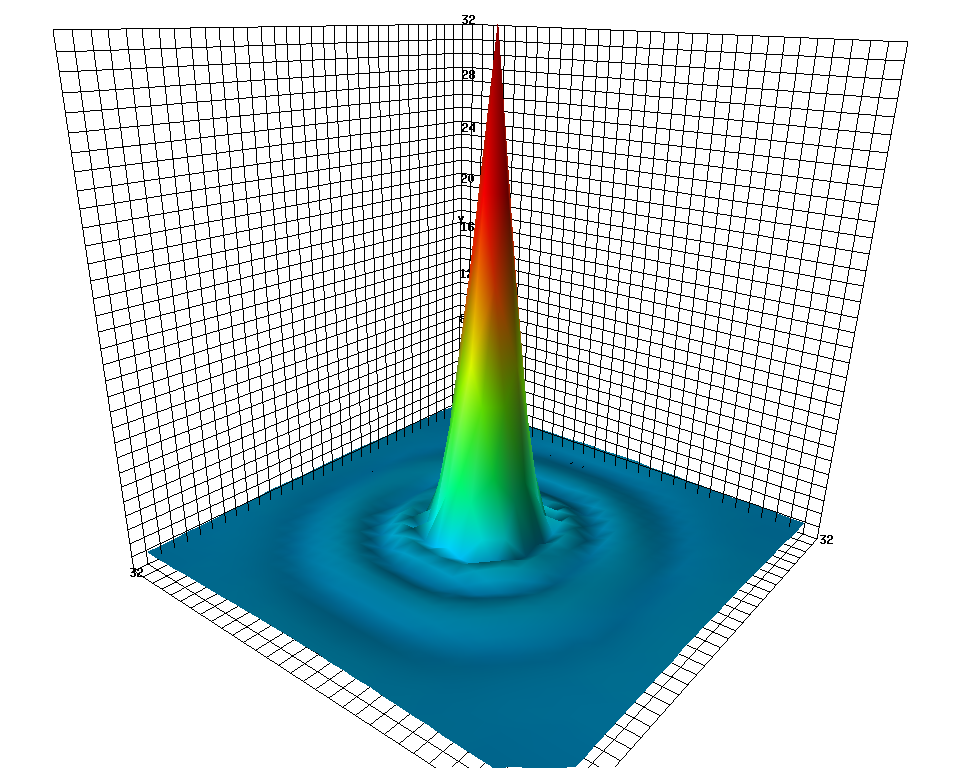}
\includegraphics[clip=true,trim=2.5cm 0.0cm 2.5cm 0.0cm,width=0.24\linewidth]{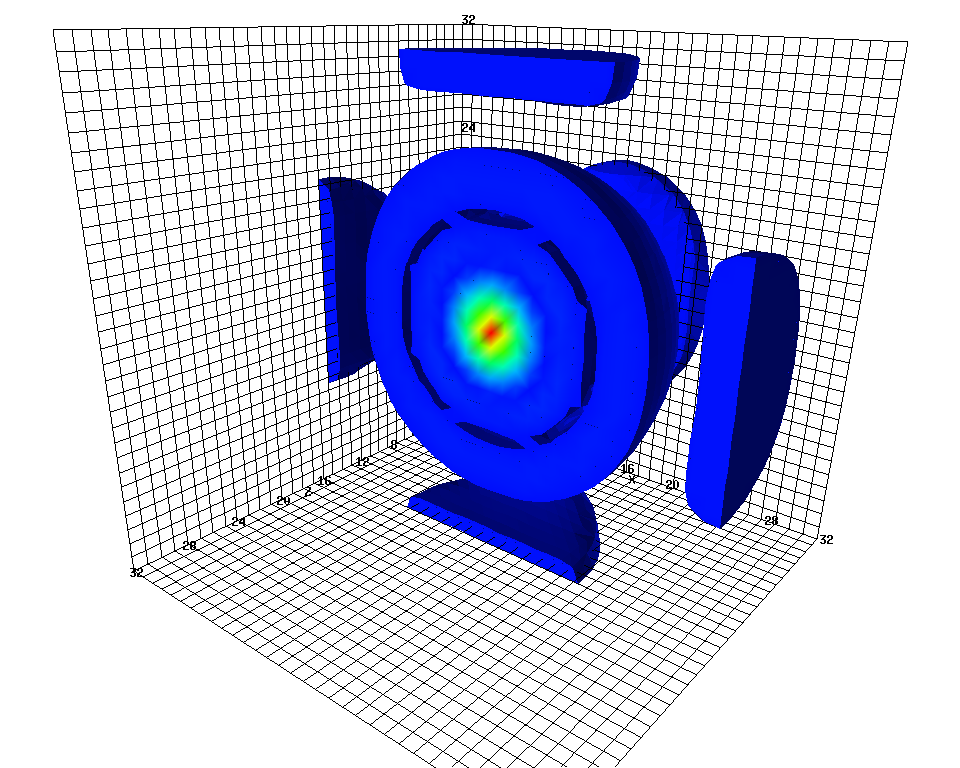}     \\[0.2cm]
\includegraphics[clip=true,trim=2.5cm 0.0cm 2.5cm 0.0cm,width=0.24\linewidth]{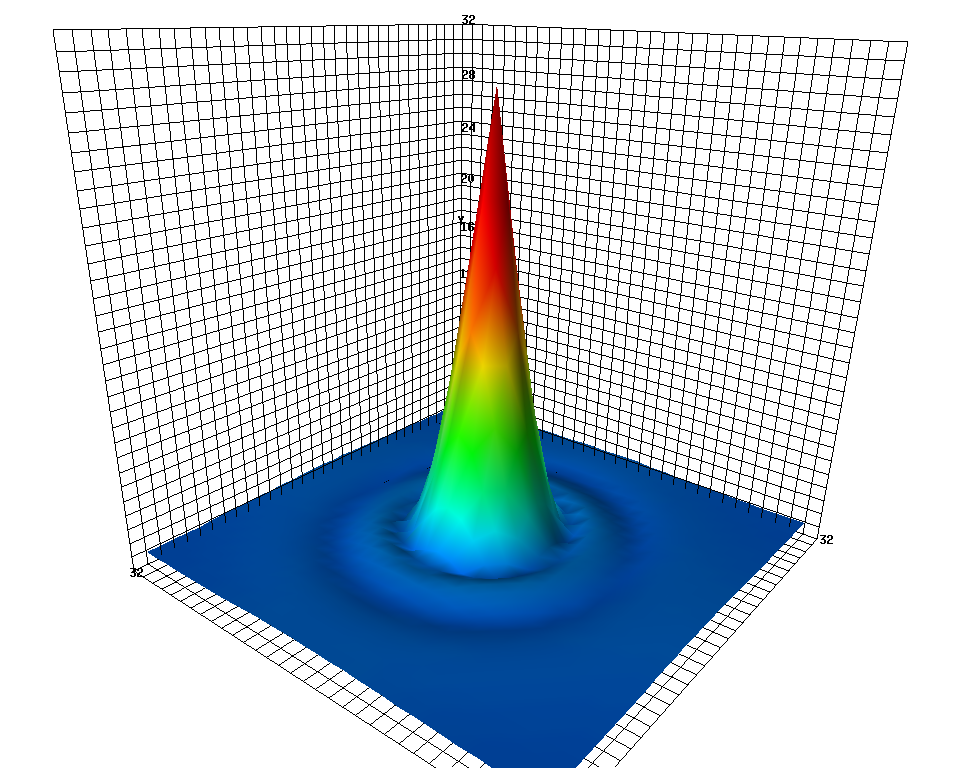}
\includegraphics[clip=true,trim=2.5cm 0.0cm 2.5cm 0.0cm,width=0.24\linewidth]{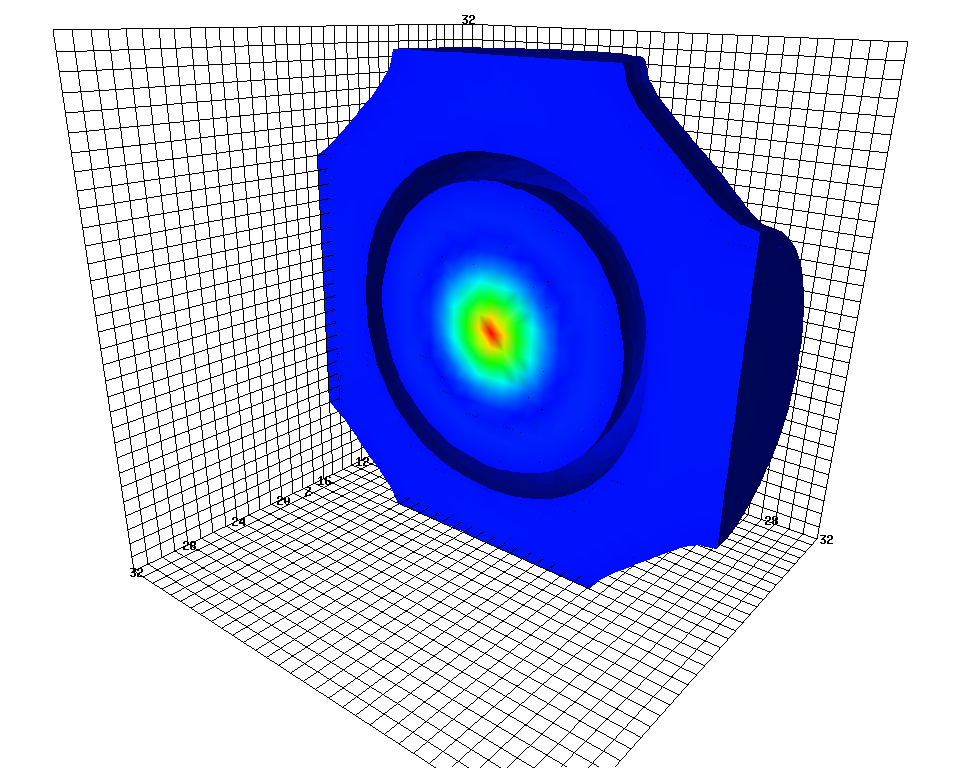}
\includegraphics[clip=true,trim=2.5cm 0.0cm 2.5cm 0.0cm,width=0.24\linewidth]{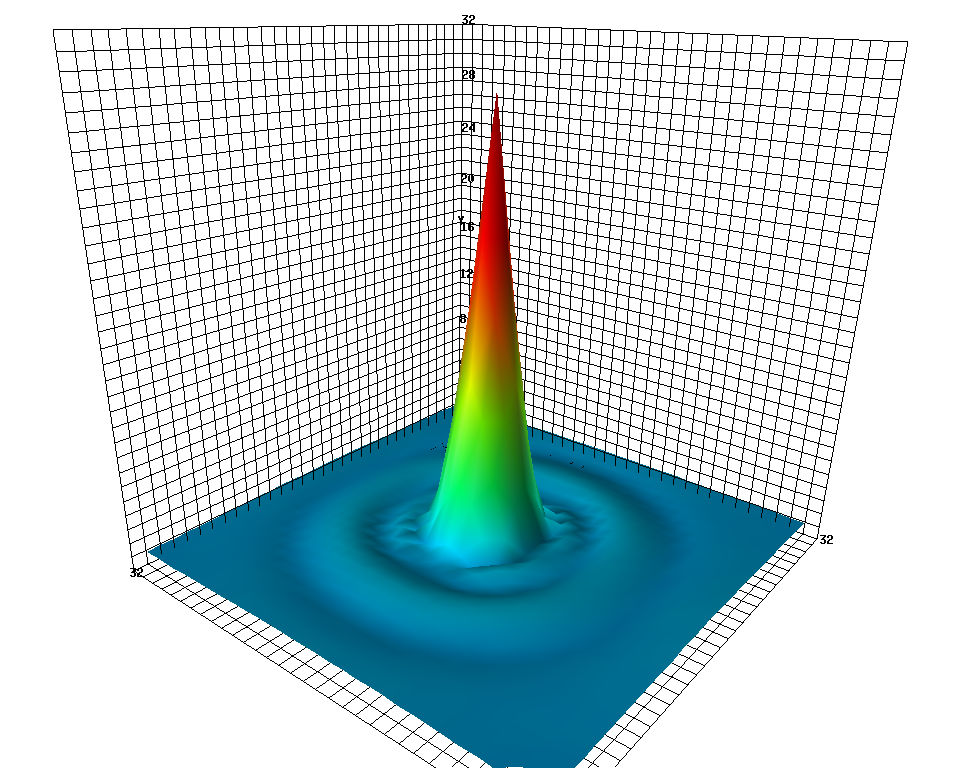}
\includegraphics[clip=true,trim=2.5cm 0.0cm 2.5cm 0.0cm,width=0.24\linewidth]{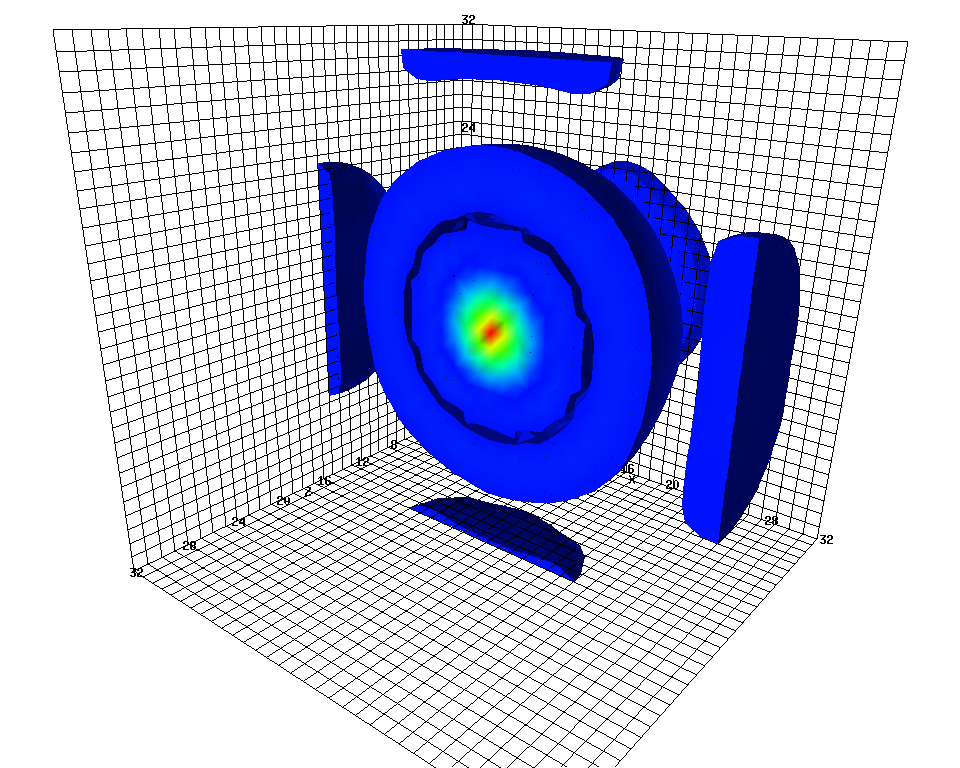}     \\[0.2cm]
\includegraphics[clip=true,trim=2.5cm 0.0cm 2.5cm 0.0cm,width=0.24\linewidth]{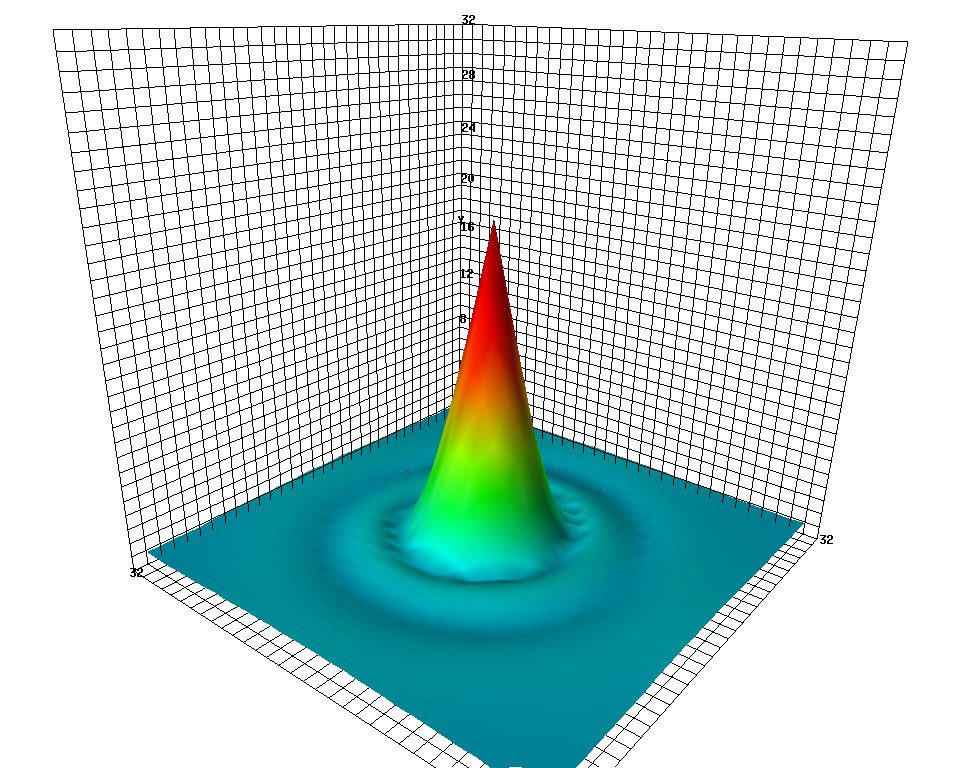}
\includegraphics[clip=true,trim=2.5cm 0.0cm 2.5cm 0.0cm,width=0.24\linewidth]{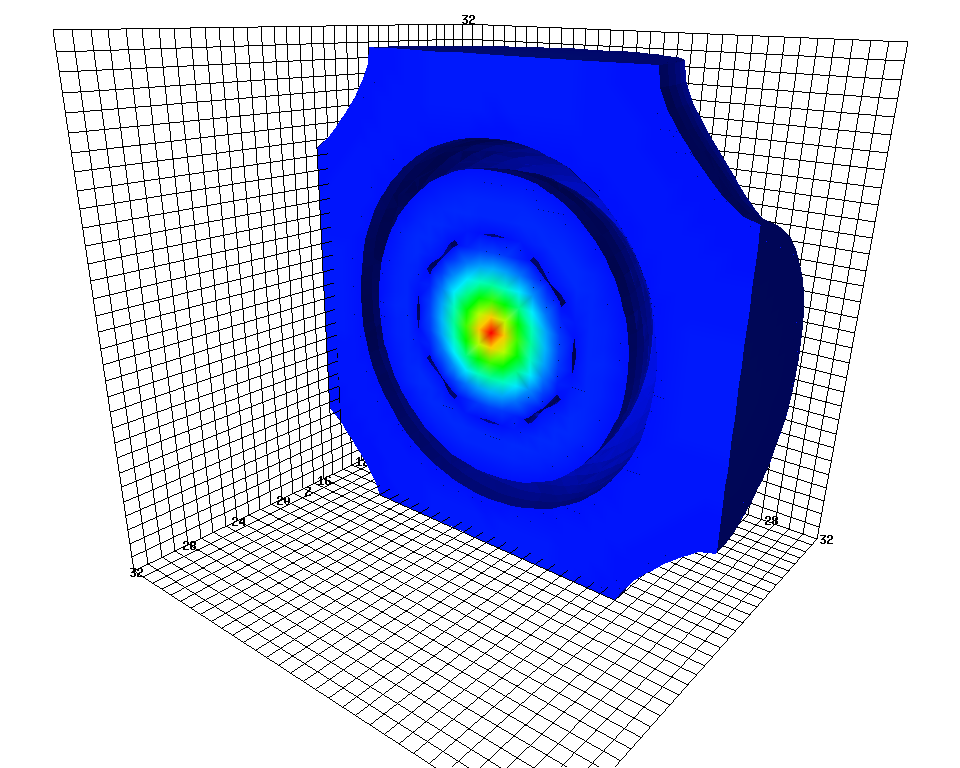}
\includegraphics[clip=true,trim=2.5cm 0.0cm 2.5cm 0.0cm,width=0.24\linewidth]{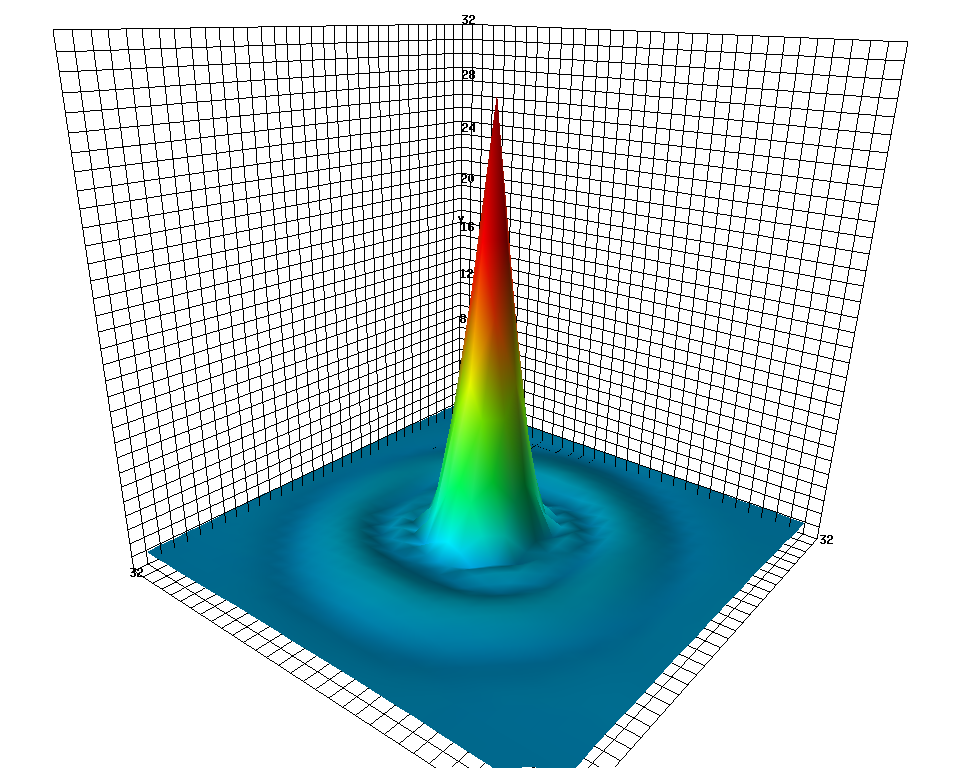}
\includegraphics[clip=true,trim=2.5cm 0.0cm 2.5cm 0.0cm,width=0.24\linewidth]{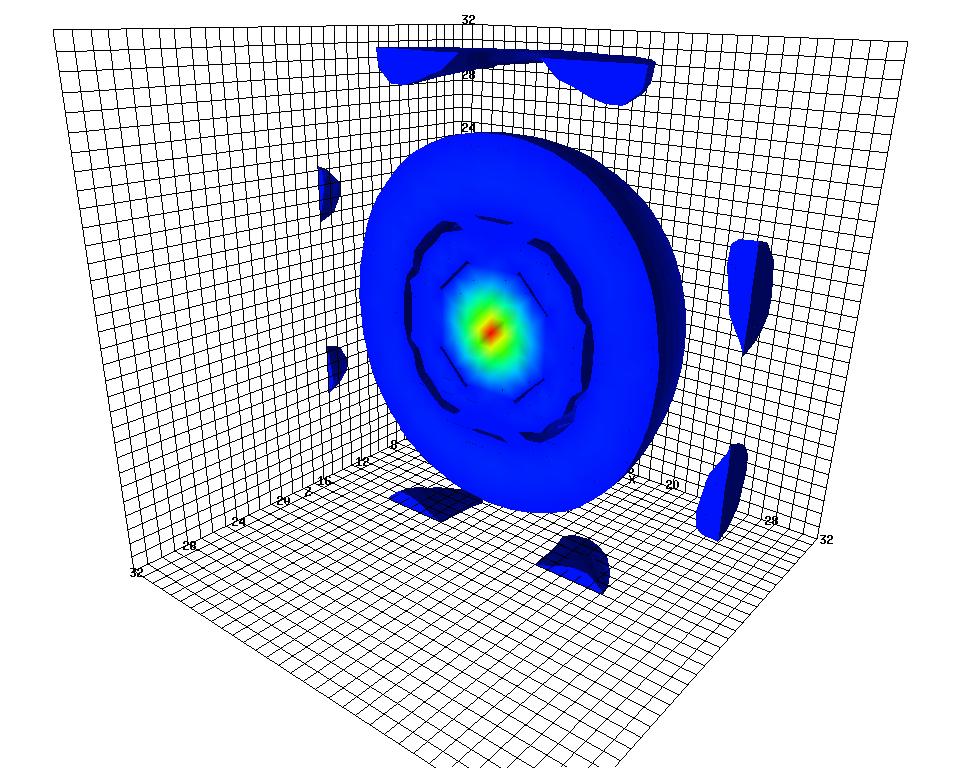}     \\[0.2cm]
\includegraphics[clip=true,trim=2.5cm 0.0cm 2.5cm 0.0cm,width=0.24\linewidth]{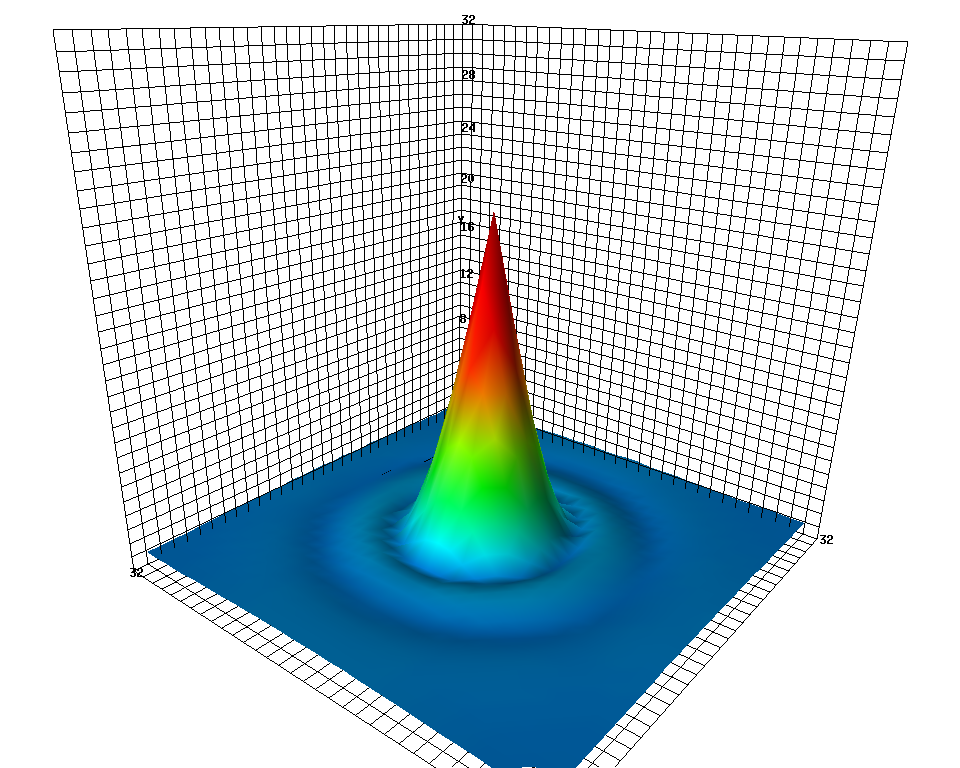}
\includegraphics[clip=true,trim=2.5cm 0.0cm 2.5cm 0.0cm,width=0.24\linewidth]{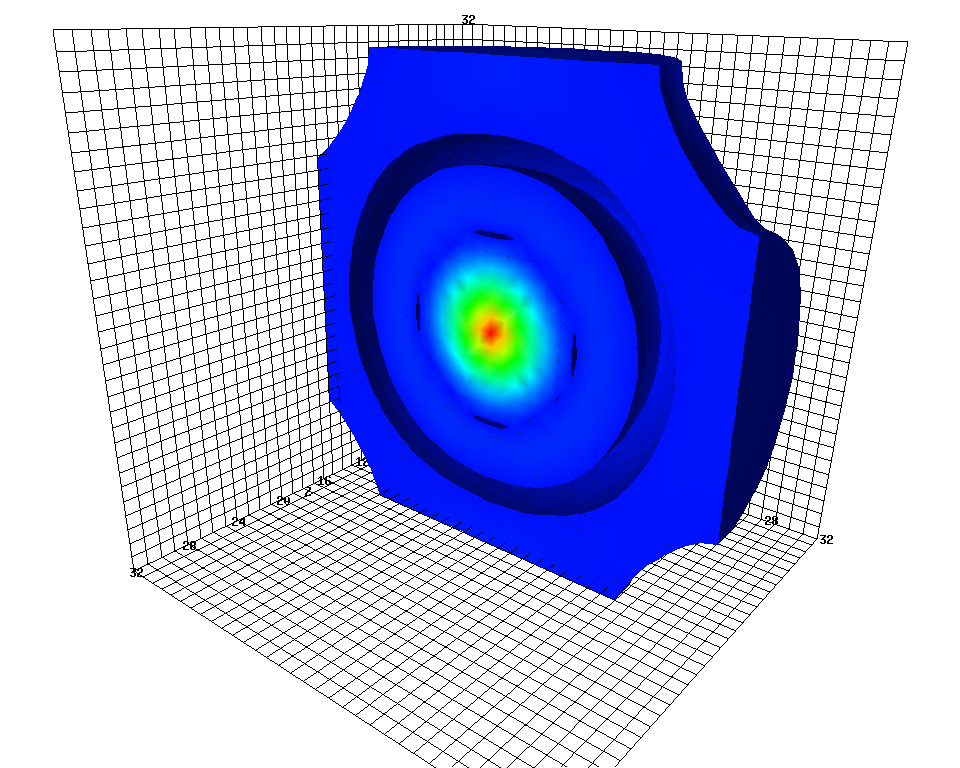}
\includegraphics[clip=true,trim=2.5cm 0.0cm 2.5cm 0.0cm,width=0.24\linewidth]{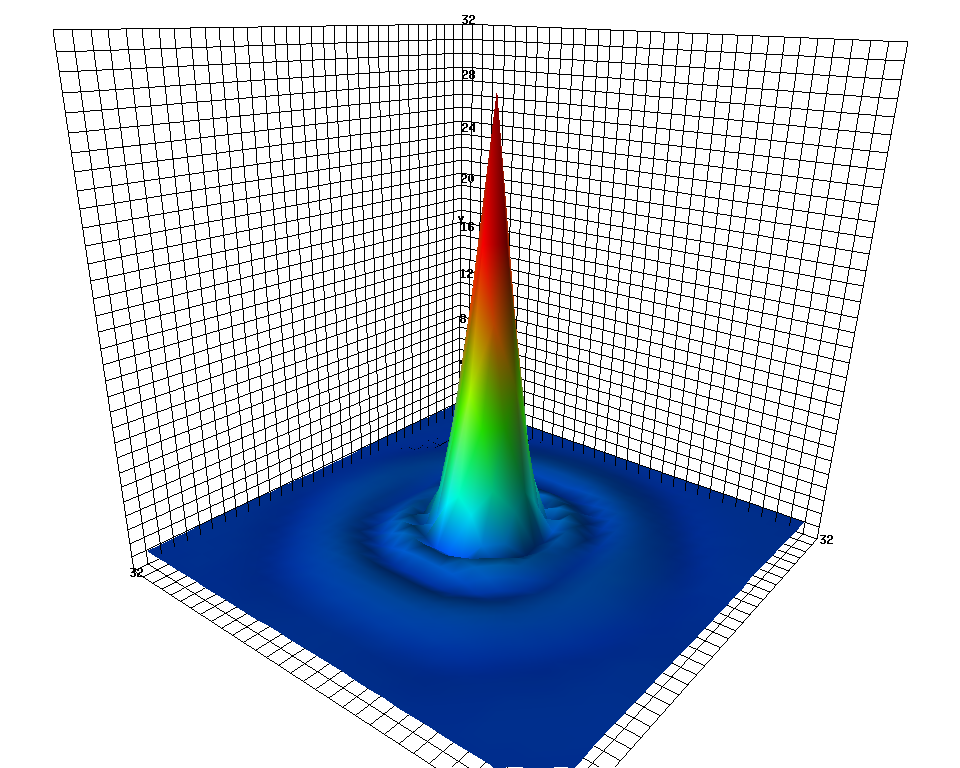}
\includegraphics[clip=true,trim=2.5cm 0.0cm 2.5cm 0.0cm,width=0.24\linewidth]{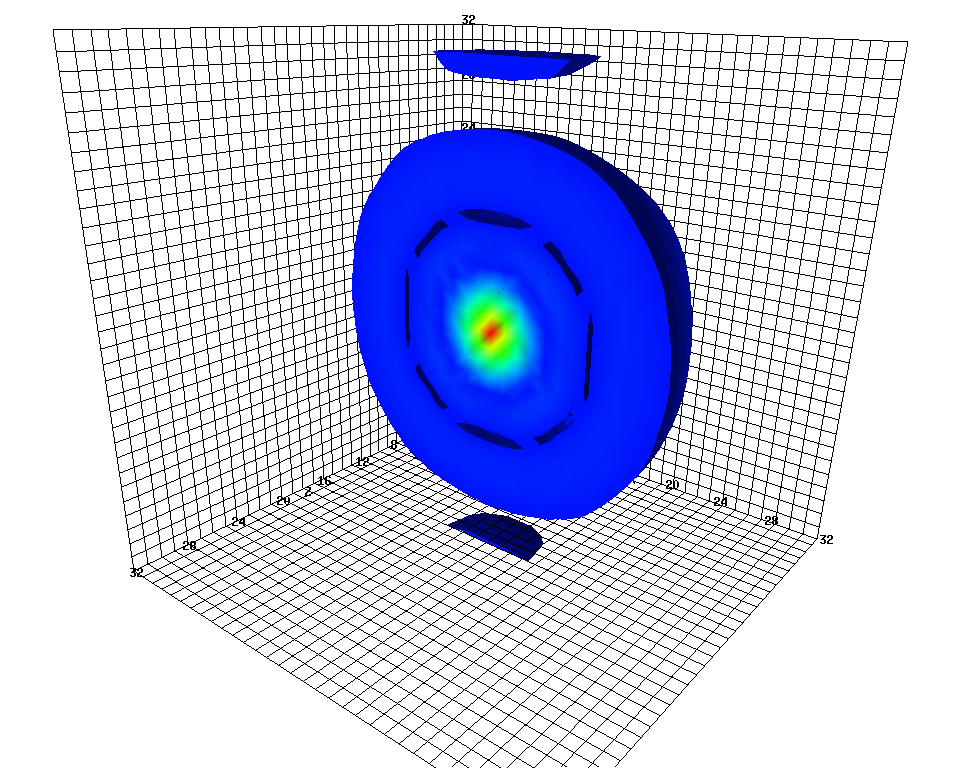}     \\[0.2cm]
\includegraphics[clip=true,trim=2.5cm 0.0cm 2.5cm 0.0cm,width=0.24\linewidth]{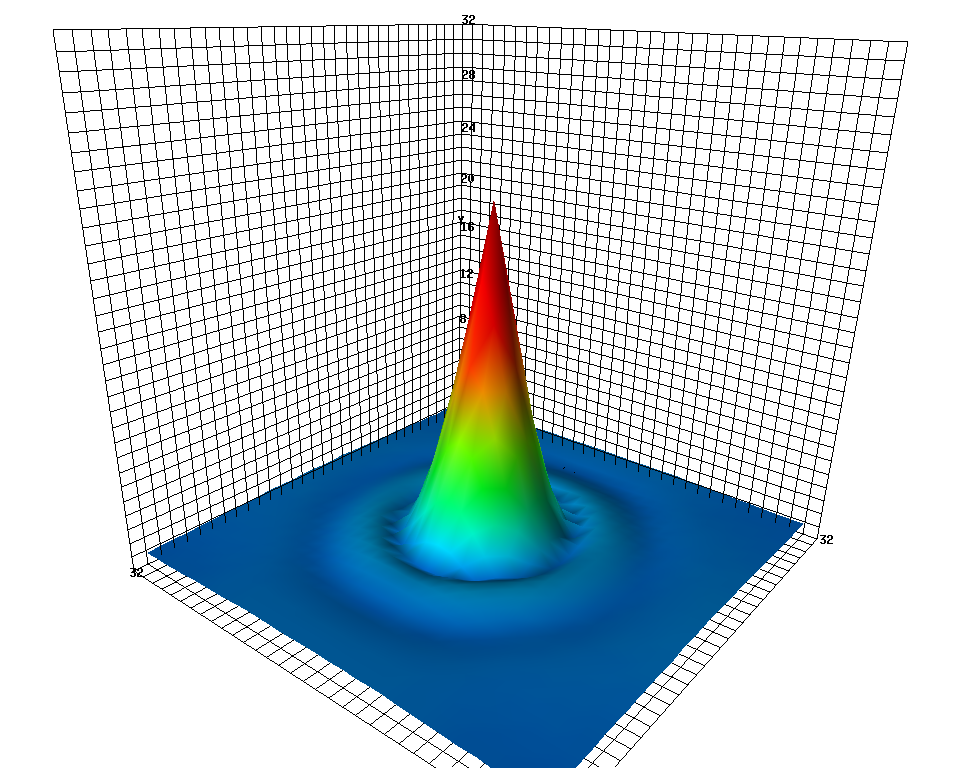}
\includegraphics[clip=true,trim=2.5cm 0.0cm 2.5cm 0.0cm,width=0.24\linewidth]{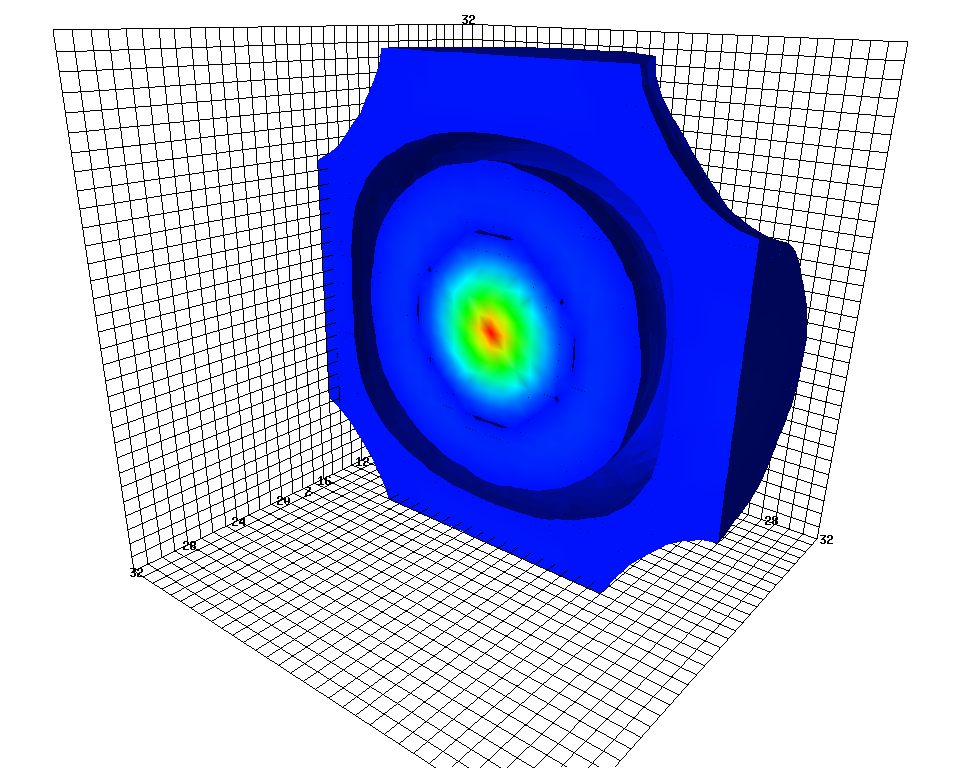}
\includegraphics[clip=true,trim=2.5cm 0.0cm 2.5cm 0.0cm,width=0.24\linewidth]{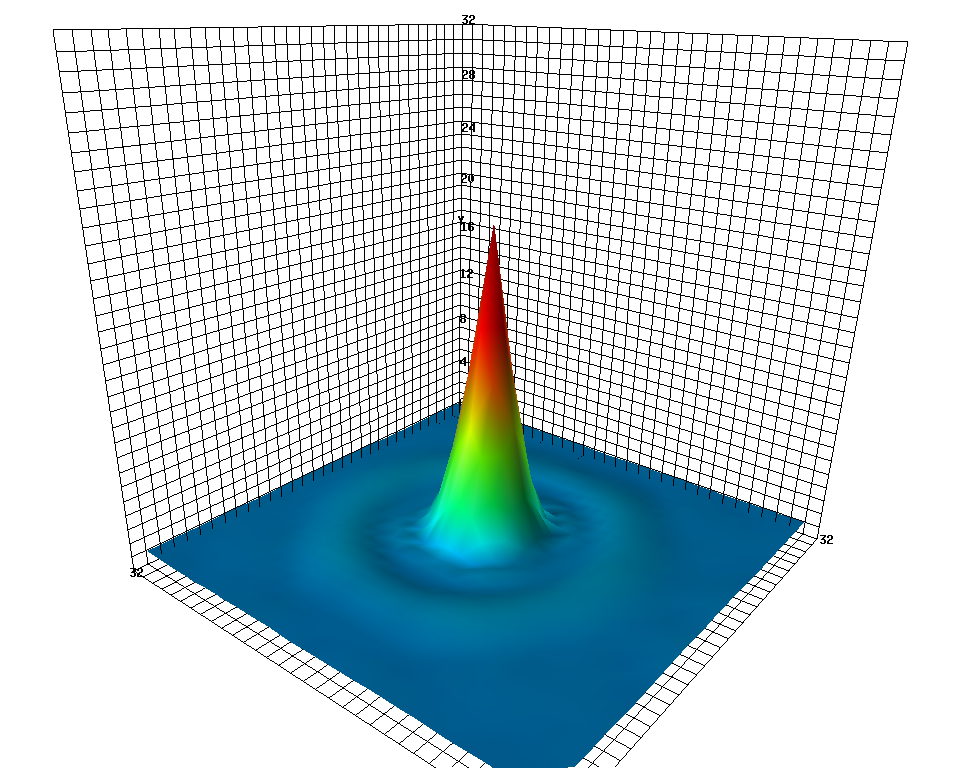}
\includegraphics[clip=true,trim=2.5cm 0.0cm 2.5cm 0.0cm,width=0.24\linewidth]{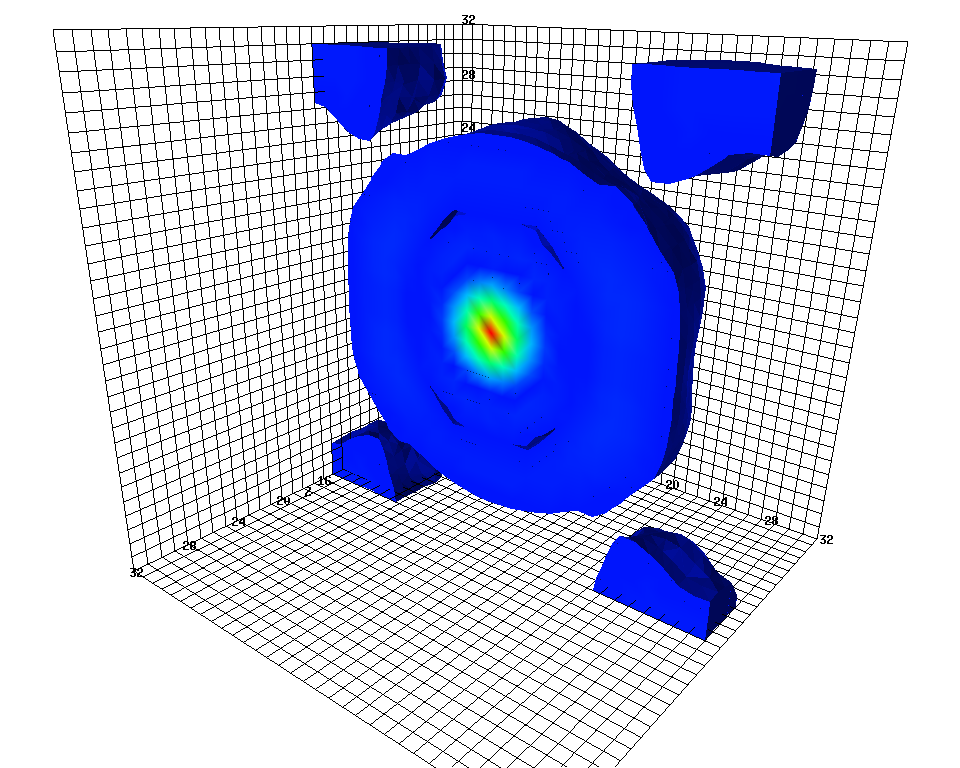}     \\[0.2cm]
\caption{The dependence of the $d$-quark probability distribution on
  the masses of the quarks in the proton for the second (two left-hand
  columns) and third (two right-hand columns) $S$-wave excited states
  of the proton observed herein.  The $u$ quarks are fixed at the
  origin at the centre of the plot.  The quark mass decreases from
  heaviest (top row) to lightest (bottom row).  For each mass and
  state, the probability density is normalised to unity over the
  spatial volume of the lattice.  The isovolume threshold for
  rendering the probability distribution in the second and fourth
  columns is $2.0 \times 10^{-5}$ and $3.0 \times 10^{-5}$
  respectively.  While the former renders the outer shell coherently,
  the latter better reveals the node structure of the $3S$
  distribution.}
\label{MassDepB}
\end{figure*}

\subsubsection{Second Excitation}

The probability distributions for the second excitation of the proton
observed in this study are illustrated in the two left-hand columns of
Fig.~\ref{MassDepB}.  Two nodes are evident at all quark masses,
consistent with a $3S$ radial excitation for the $d$ quark.  The first
inner node is thin at the heavier masses and difficult to see in the
isovolume renderings.  Finite volume effects are readily observed in
the outer-most shell.

For the heaviest mass, finite volume effects at the nodes are minimal.
The nodes are spherical in shape and are largely unaffected by the
boundary.  Again there is a trend of the nodes moving further from the
centre as the quarks become light.  

However, it is the middle quark mass considered that has the broadest
distribution.  The quark mass flow of the eigenstate energies suggests
avoided level crossings are important between the third and fourth
heaviest quark masses.  It may be a strong mixing with multi-particle
states that is giving rise to the broad distribution of quarks at the
middle quark mass.

For the lightest two quark masses the outer node has taken on a
squared-off shape, having been distorted by the boundary of the
lattice.  Again, this is an indication that, even though the ground
state wave function presents as spherical for this quark mass, this
excited state is showing clear finite volume effects.  Even at
relatively modest quark masses, the wave functions of states above the
decay thresholds show an important relationship with the finite volume
of the lattice.

\subsubsection{Third Excitation}

The two right-hand columns of Fig.~\ref{MassDepB} illustrate the mass
dependence of the $d$-quark probability distribution for the the
highest excitation of the proton observed in our analysis.  The
presence of three nodes in the wave function is best observed at the
heaviest and second lightest quarks masses.  

The inner-most node is easily observed in the surface plots.  However
it is very sharp and does not render in an obvious manner in the
isovolume illustrations.  The second node is easily rendered and the
third node is very broad.  To illustrate this node structure the
outer-most shell has become fragmented in the isovolume plots.
The fragments reveal the strong finite volume effects on this state.

What is interesting is the manner in which the finite volume effects
on the outer shell change as a function of quark mass.  At the
heaviest quark mass, the outer shell is strongest along the sides of
the lattice.  By the time one encounters the lightest ensemble, the
outer shell has moved to the corners as if there is no longer room for
the outer shell along the sides of the 2.9 fm lattice.

\section{Quark Separation}
\setcounter{subsubsection}{0}

\begin{figure*}[tbh!]
\includegraphics[clip=true,trim=2.5cm 0.0cm 2.5cm 0.0cm,width=0.24\linewidth]{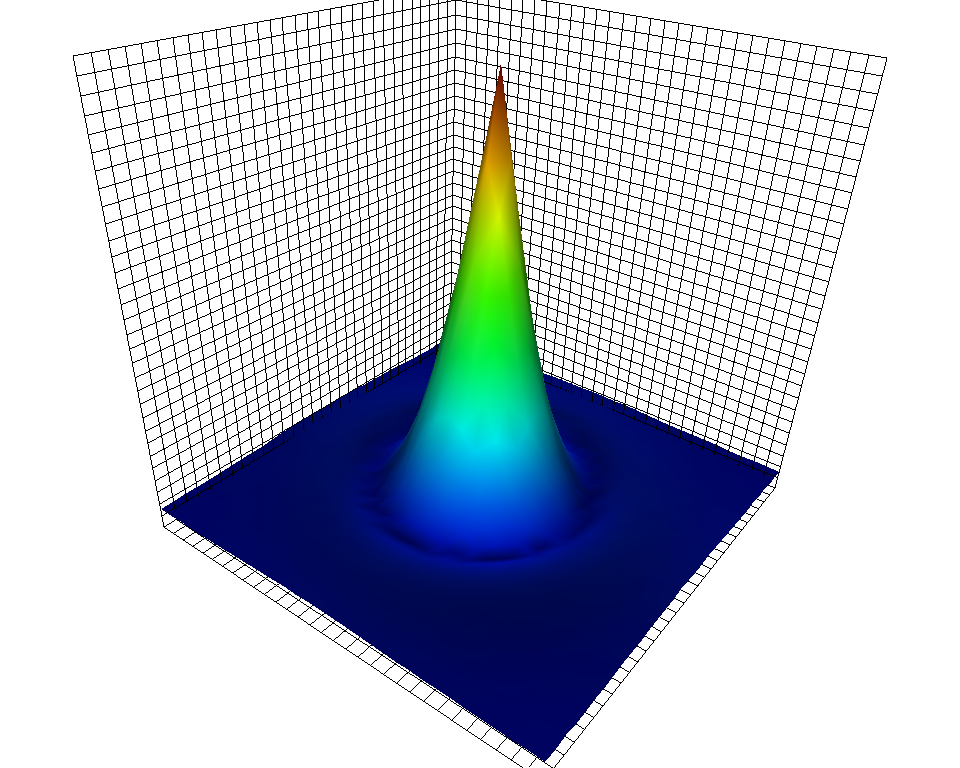}
\includegraphics[clip=true,trim=2.5cm 0.0cm 2.5cm 0.0cm,width=0.24\linewidth]{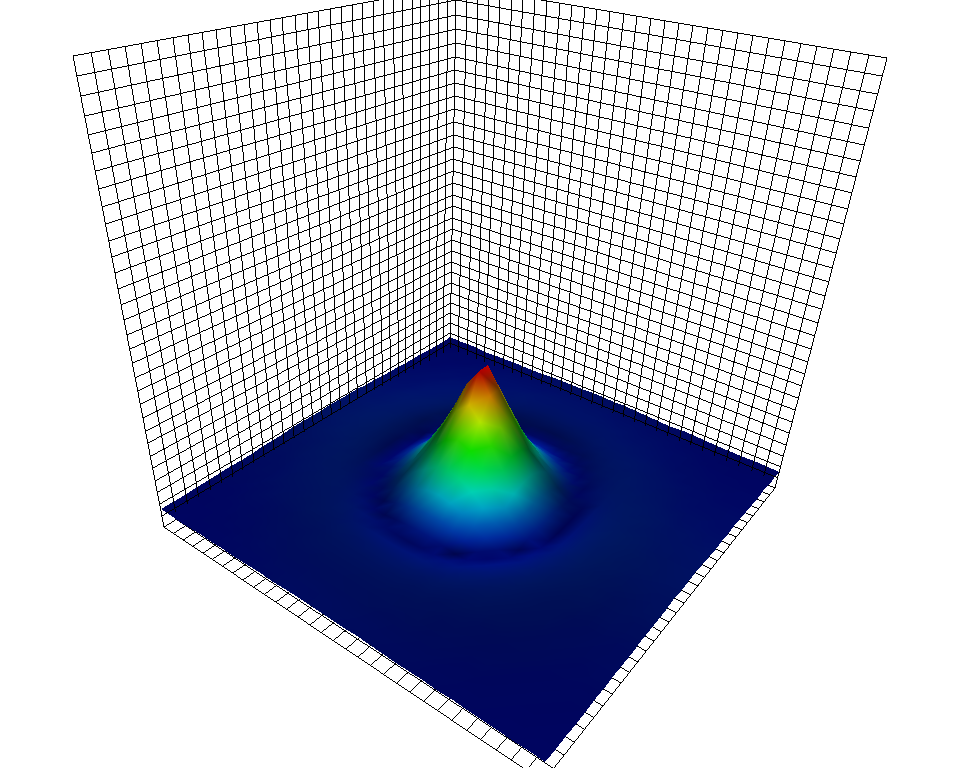}
\includegraphics[clip=true,trim=2.5cm 0.0cm 2.5cm 0.0cm,width=0.24\linewidth]{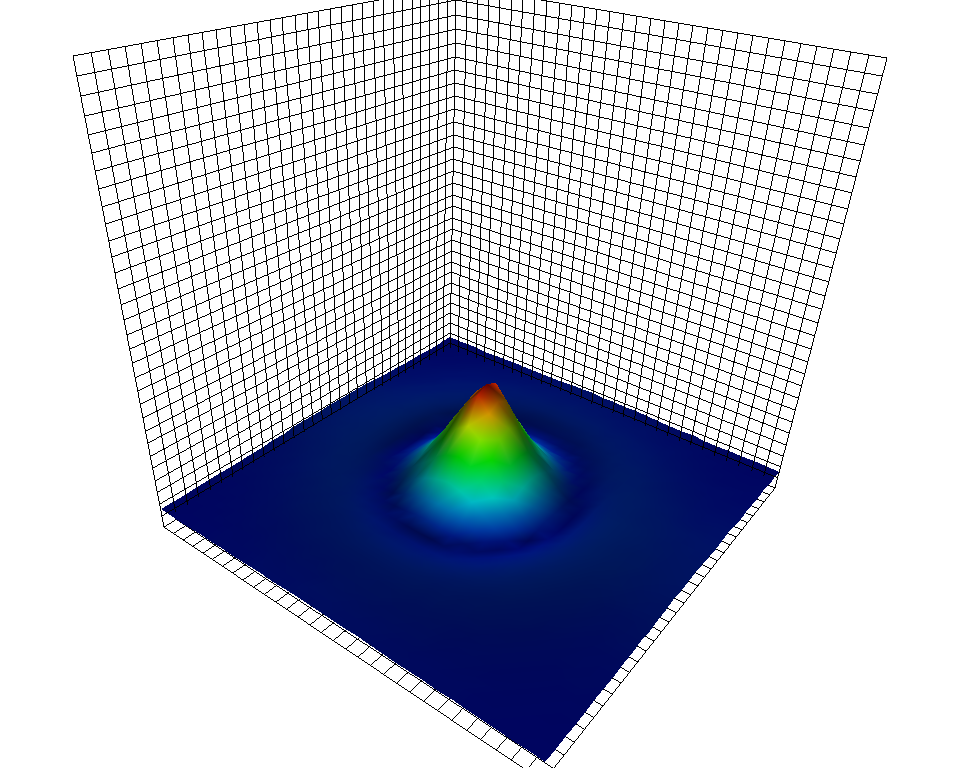}
\includegraphics[clip=true,trim=2.5cm 0.0cm 2.5cm 0.0cm,width=0.24\linewidth]{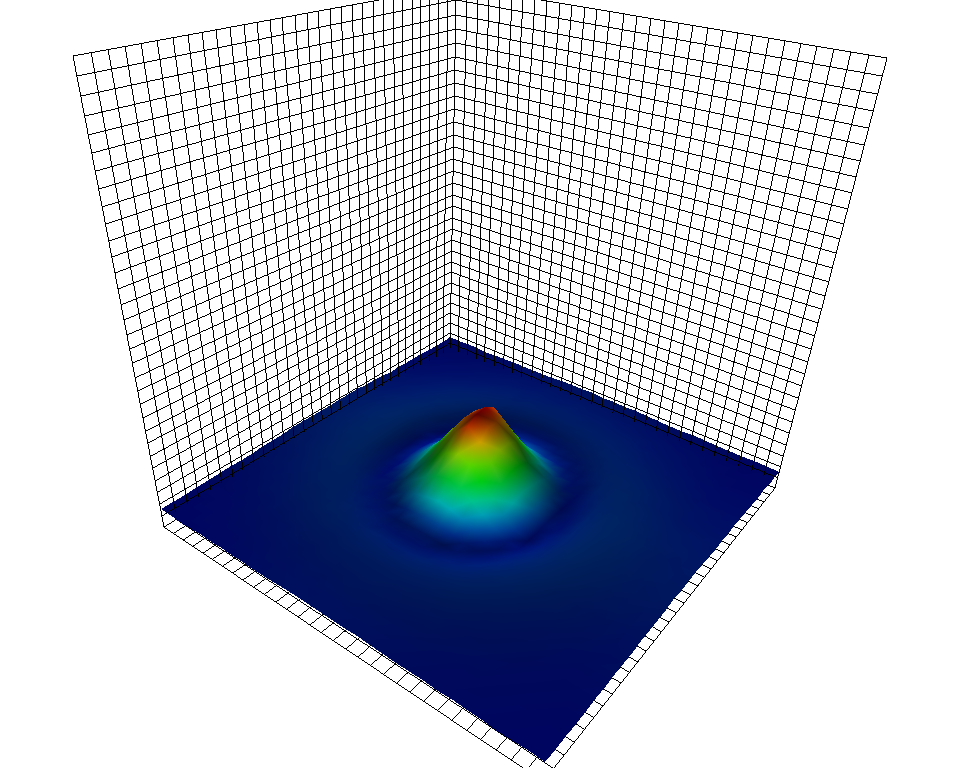}
\caption{The dependence of the $d$-quark probability distribution on
  the positions of the $u$ quarks in the first even-parity excited
  state of the proton at the second lightest quark mass considered.
  The $u$ quarks are fixed on the $x$-axis running from back right
  through front left through the centre of the plot.  The $u$ quarks
  are fixed a distance of $d_1$ and $d_2$ from the origin located at
  the centre.  From left to right, the distance $d = d_1 - d_2$
  increases, taking values 0, 1, 2 and 3 times the lattice spacing
  $a=0.0907$ fm.  }
\label{QuarkSep0}
\end{figure*}

In order to investigate a more complete picture of the wave functions
of the states isolated herein, we choose to focus on the
second-lightest quark mass ensemble providing
$m_\pi=293\,\mathrm{MeV}$ and examine the dependence of the $d$ quark
probability distribution on the positions of the two $u$ quarks
composing the states.  This mass provides the best compromise between
finite-volume effects, quark mass and the ensemble size governing the
signal quality and associated uncertainties.

In continuing to investigate the wave function of the $d$-quark, we
consider the separation of the $u$ quarks along the $x$ axis as
described in Eq.~(\ref{interpsep}).  All integer separations, $d = 
d_1 - d_2$, between zero and half the lattice extent in the $x$
direction ({\it i.e.\ }16 lattice units) are considered.

Figure~\ref{QuarkSep0} illustrates the probability distribution of the $d$
quark for $u$ quarks separated by 0, 1, 2 and 3 lattice steps in the
first excited state associated with the Roper resonance.  The most
notable feature is the rapid reduction in the overlap of the
interpolator with the state as the two $u$ quarks are moved away from
the origin.  While some broadening of the distribution peak is
apparent, it is clear that using a normalization suitable for zero $u$
quark separation is not effective for illustrating the probability
distribution at large $u$ quark separations.

To better illustrate the underlying shape of the wave functions, the
probability distributions are normalised to keep the maximum value of
the probability density constant.  For small $u$ quark separations,
the centre peak height of the distribution is held constant, but for
larger separations the maximum value can be elsewhere in the
distribution.

\begin{figure*}[p]
\includegraphics[clip=true,trim=2.5cm 0.0cm 2.5cm 0.0cm,width=0.24\linewidth]{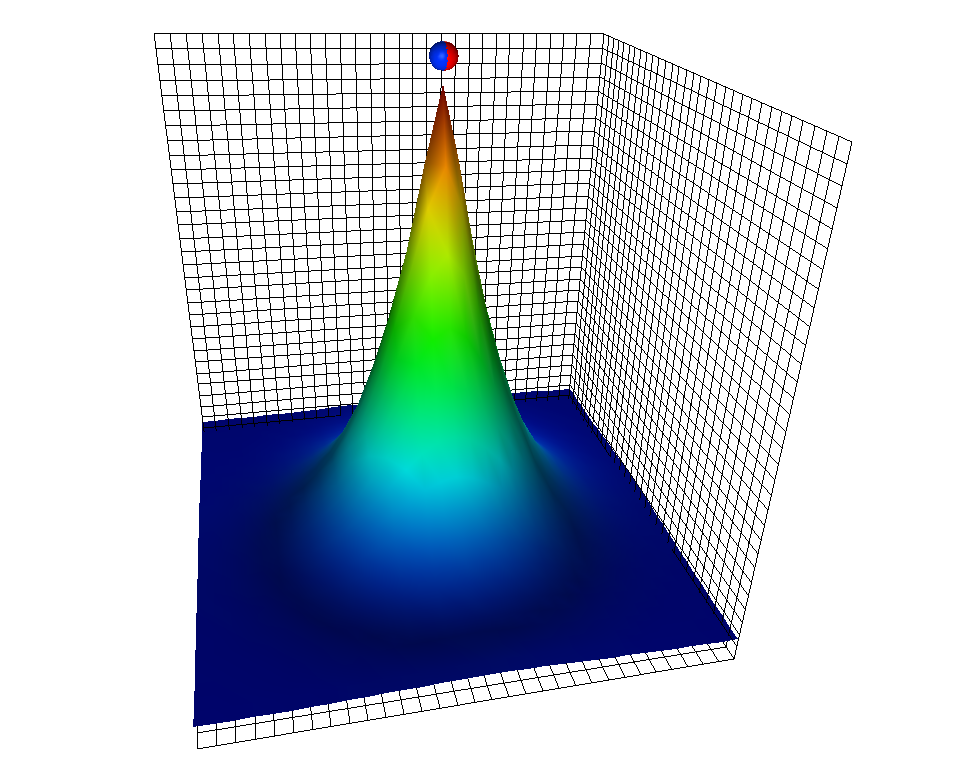}
\includegraphics[clip=true,trim=2.5cm 0.0cm 2.5cm 0.0cm,width=0.24\linewidth]{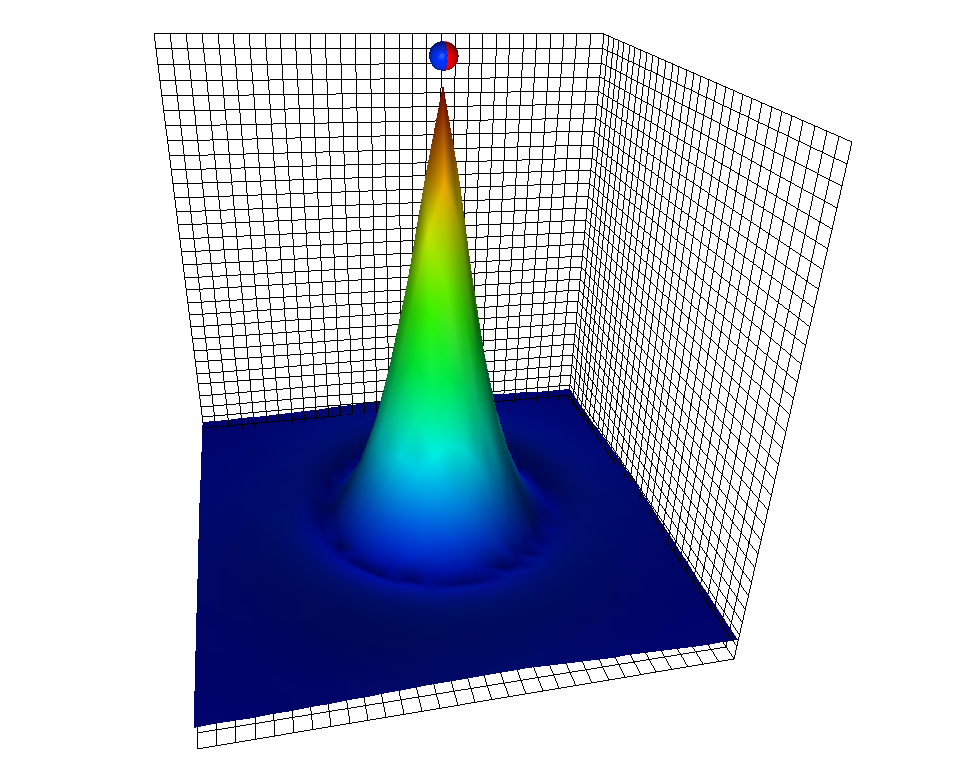}
\includegraphics[clip=true,trim=2.5cm 0.0cm 2.5cm 0.0cm,width=0.24\linewidth]{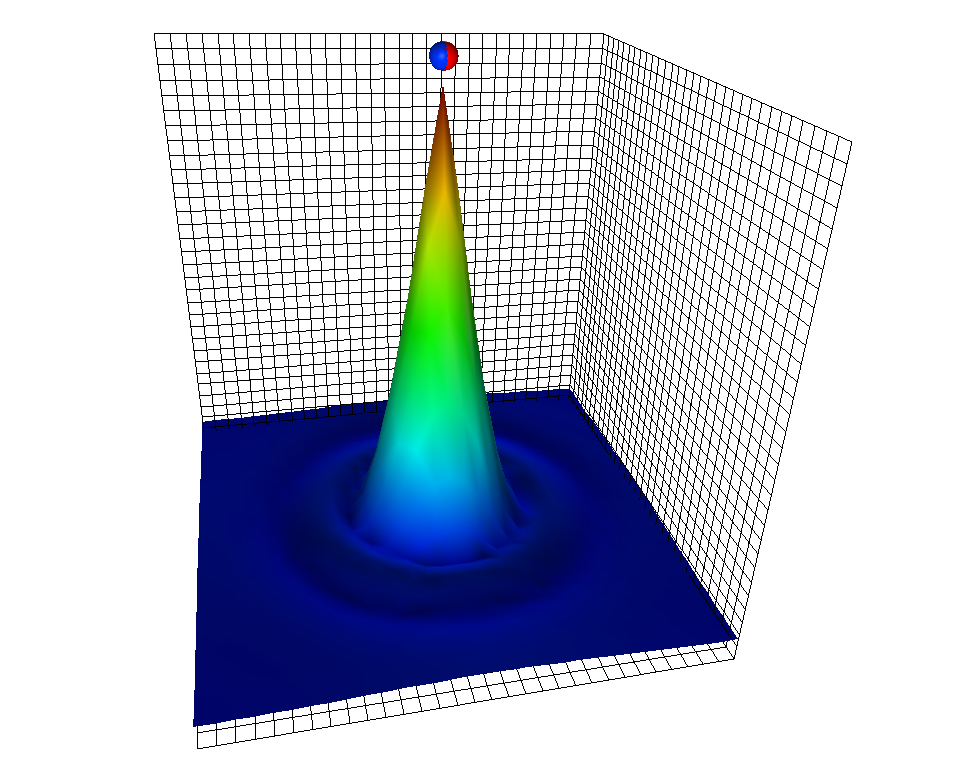}
\includegraphics[clip=true,trim=2.5cm 0.0cm 2.5cm 0.0cm,width=0.24\linewidth]{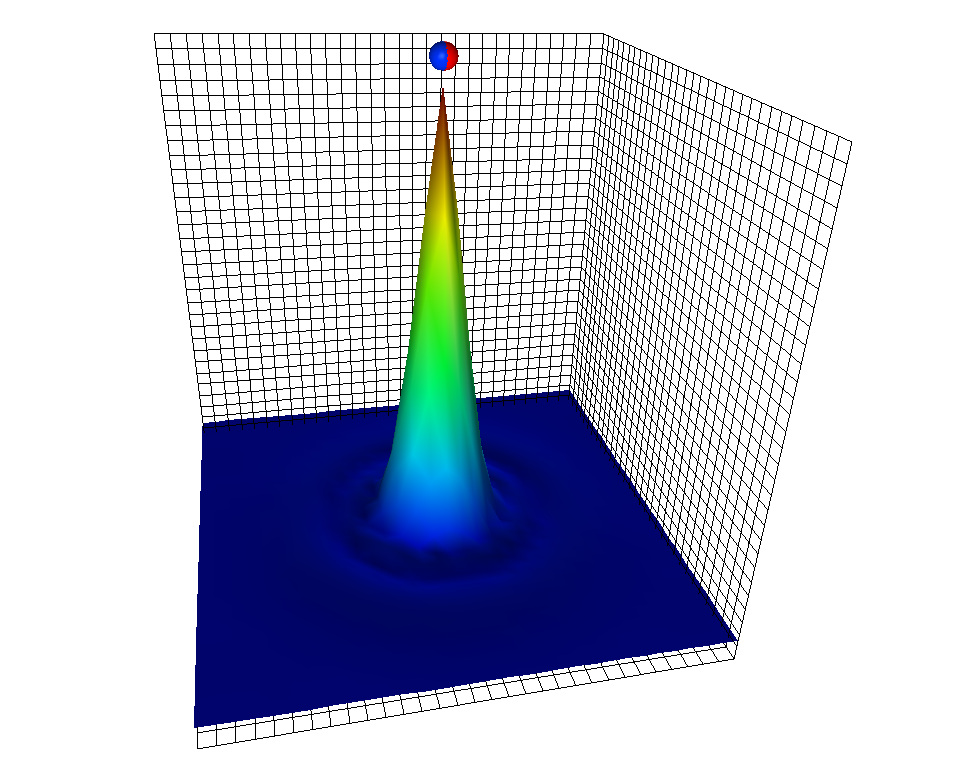} \\[0.1cm]
\includegraphics[clip=true,trim=2.5cm 0.0cm 2.5cm 0.0cm,width=0.24\linewidth]{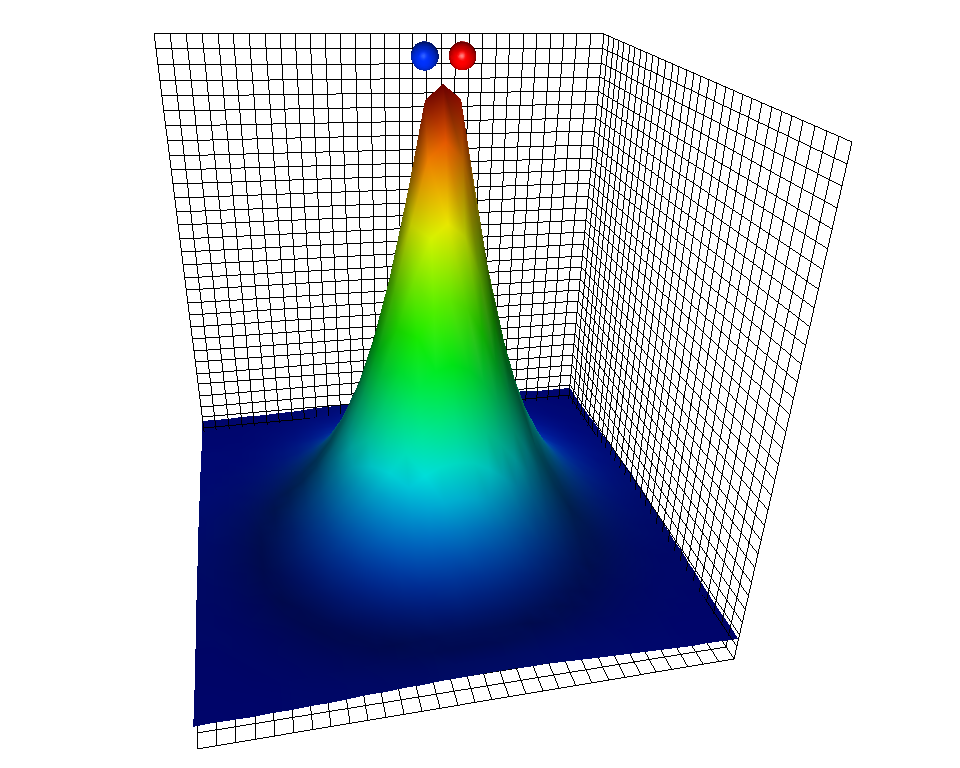}
\includegraphics[clip=true,trim=2.5cm 0.0cm 2.5cm 0.0cm,width=0.24\linewidth]{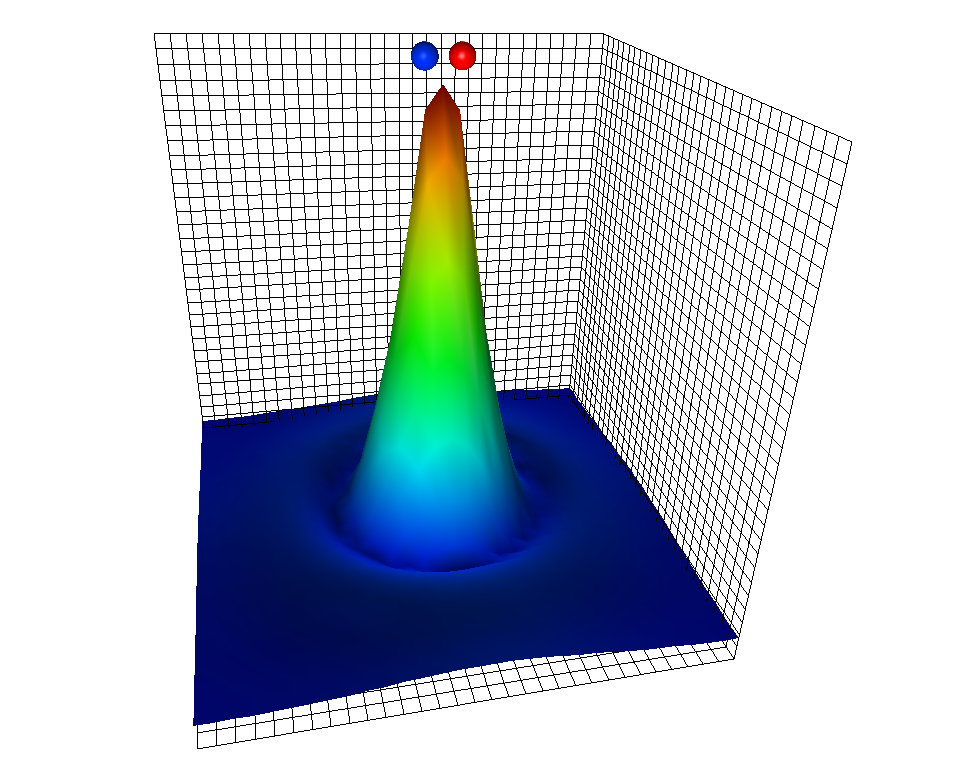}
\includegraphics[clip=true,trim=2.5cm 0.0cm 2.5cm 0.0cm,width=0.24\linewidth]{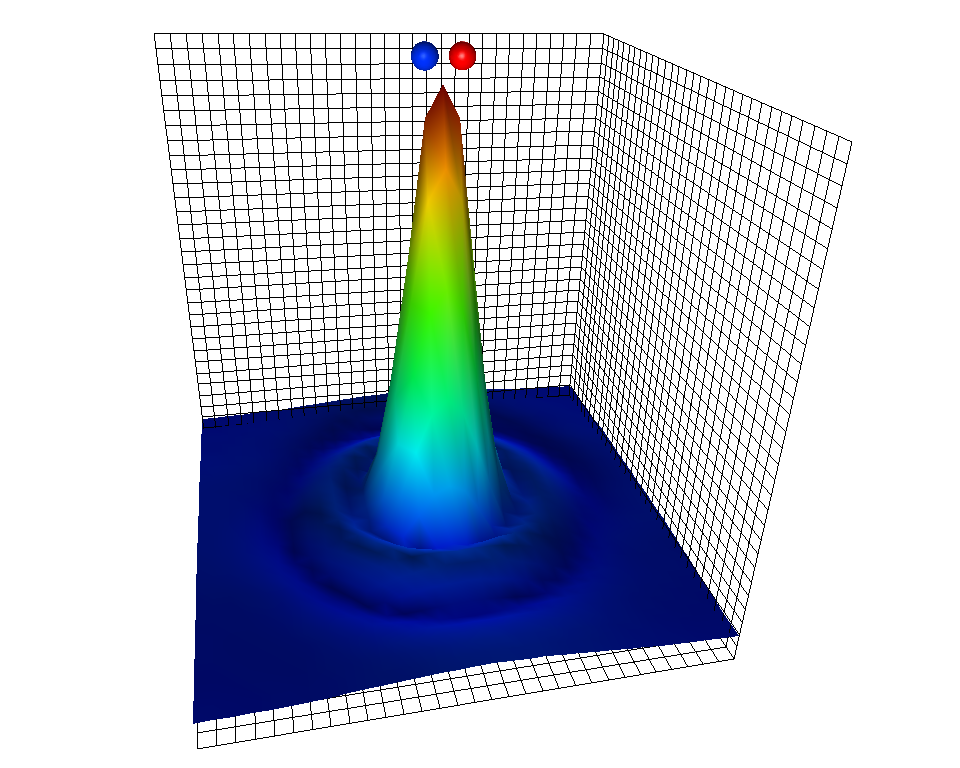}
\includegraphics[clip=true,trim=2.5cm 0.0cm 2.5cm 0.0cm,width=0.24\linewidth]{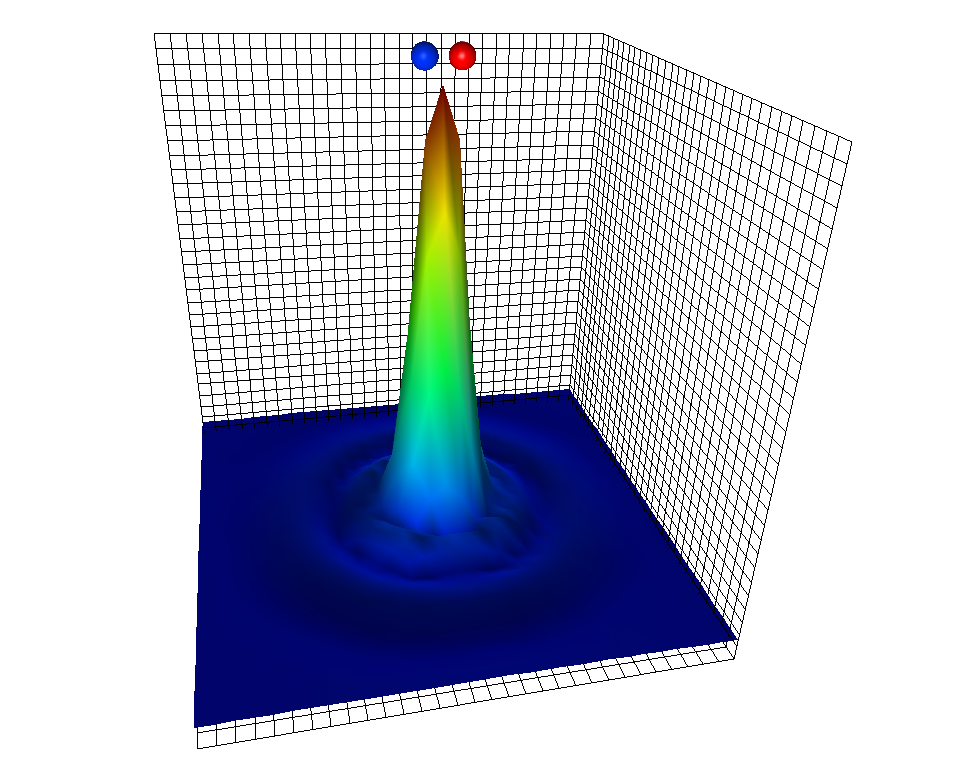} \\[0.1cm]
\includegraphics[clip=true,trim=2.5cm 0.0cm 2.5cm 0.0cm,width=0.24\linewidth]{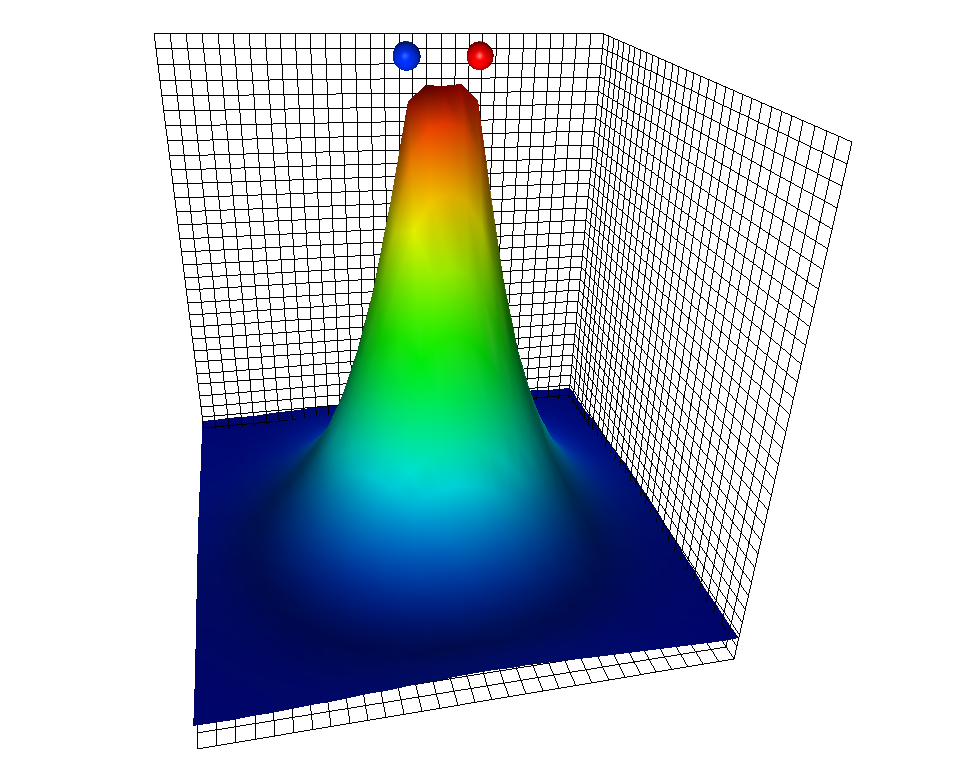}
\includegraphics[clip=true,trim=2.5cm 0.0cm 2.5cm 0.0cm,width=0.24\linewidth]{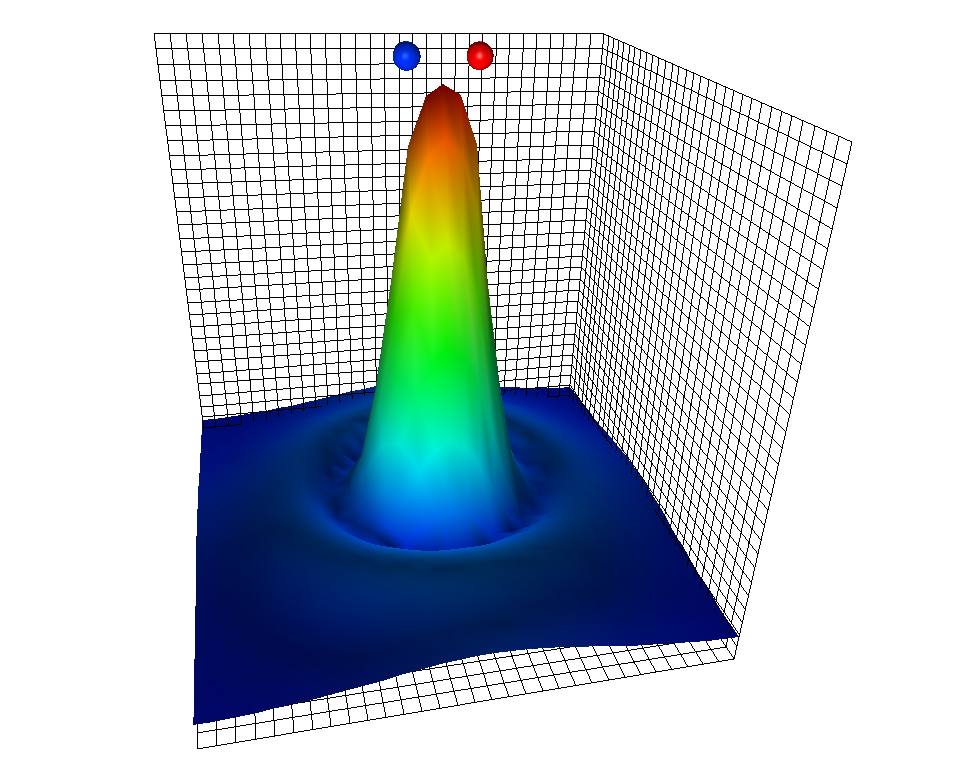}
\includegraphics[clip=true,trim=2.5cm 0.0cm 2.5cm 0.0cm,width=0.24\linewidth]{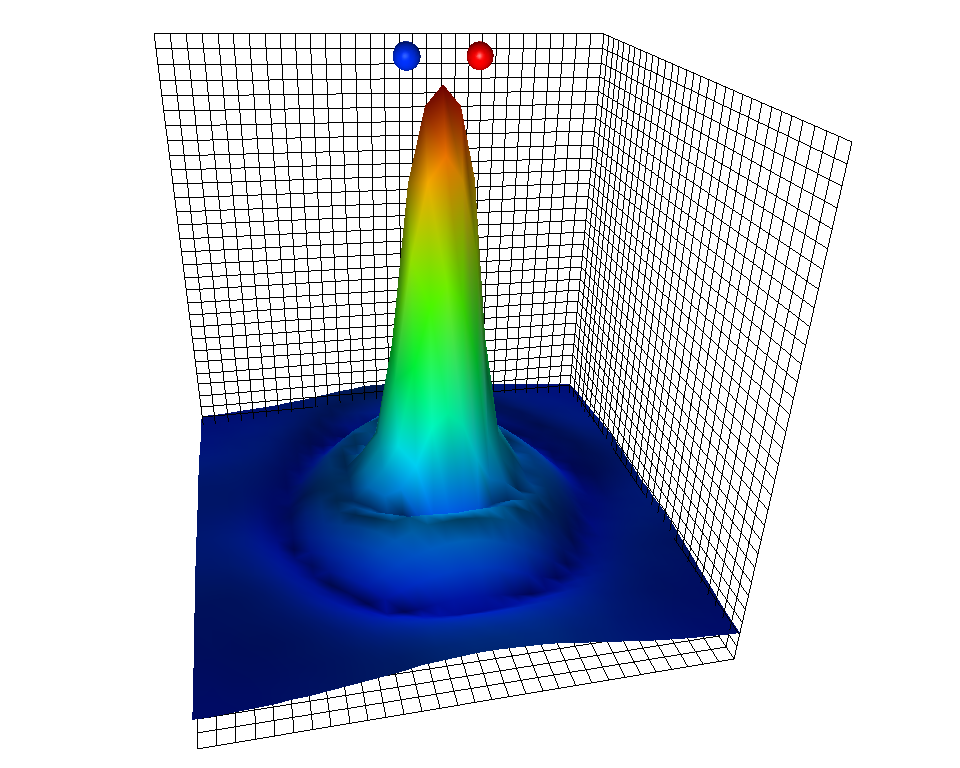}
\includegraphics[clip=true,trim=2.5cm 0.0cm 2.5cm 0.0cm,width=0.24\linewidth]{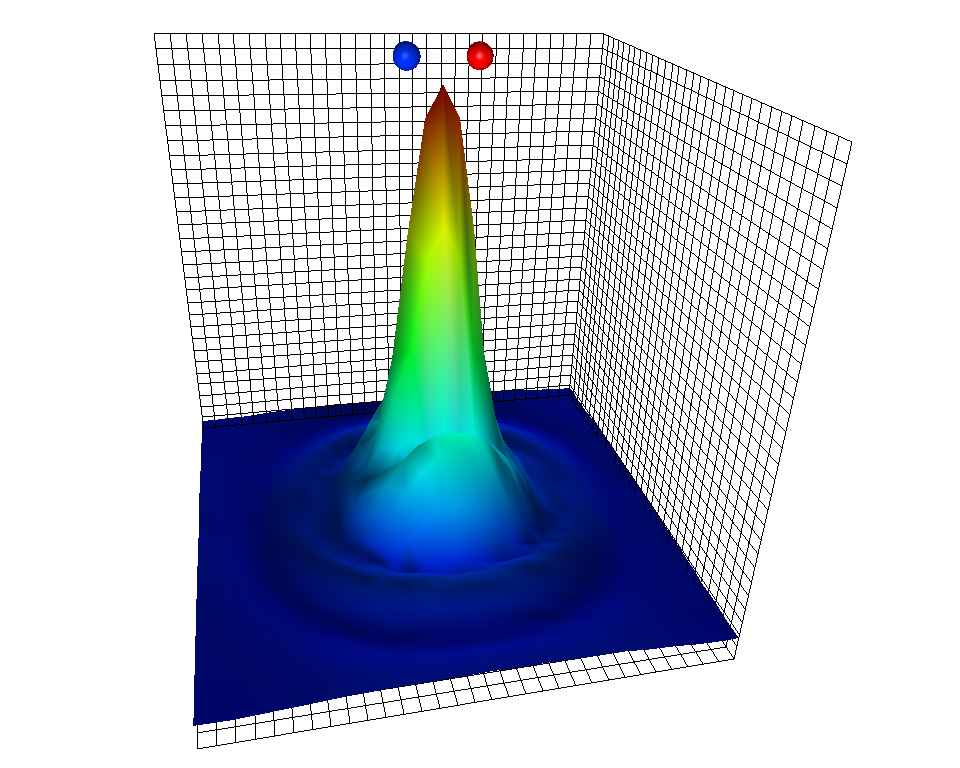} \\[0.1cm]
\includegraphics[clip=true,trim=2.5cm 0.0cm 2.5cm 0.0cm,width=0.24\linewidth]{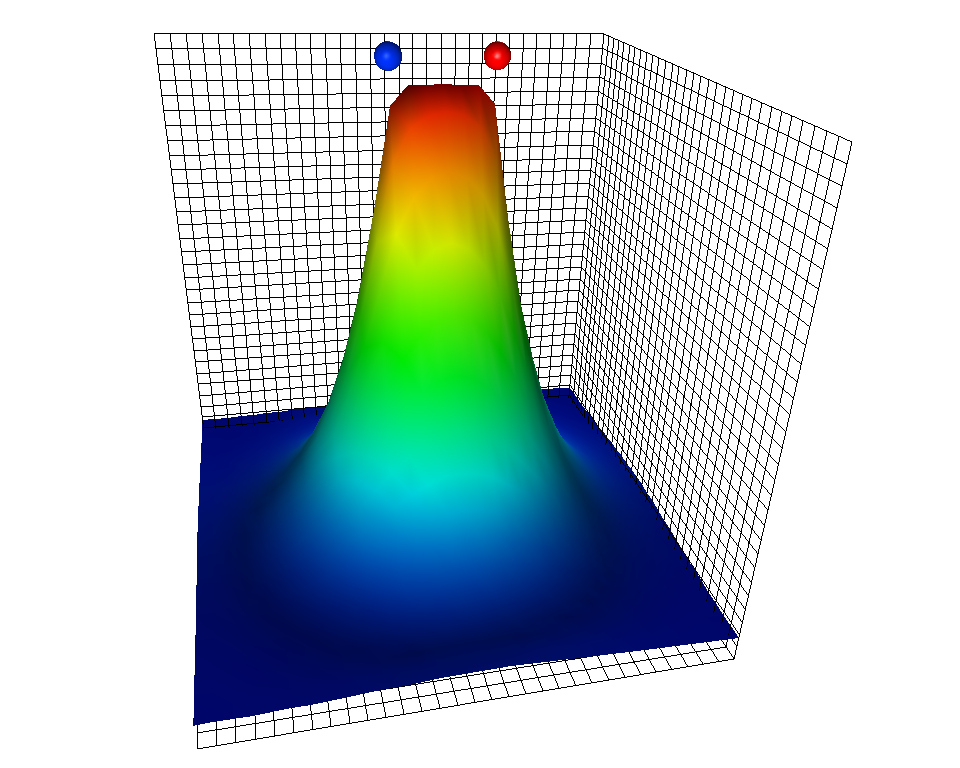}
\includegraphics[clip=true,trim=2.5cm 0.0cm 2.5cm 0.0cm,width=0.24\linewidth]{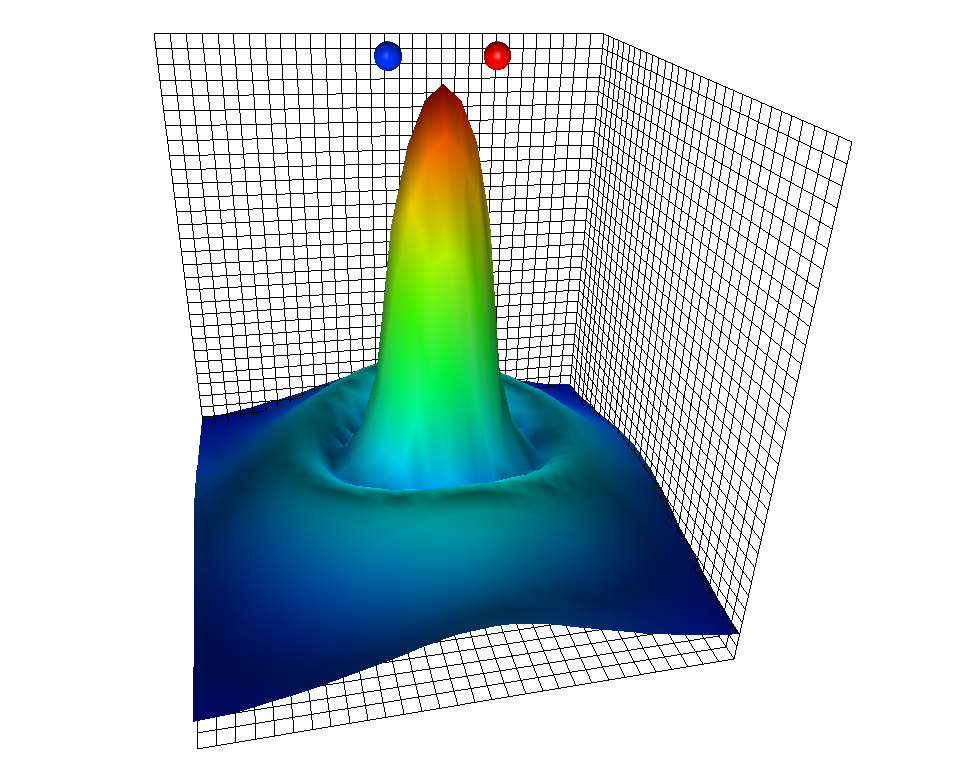}
\includegraphics[clip=true,trim=2.5cm 0.0cm 2.5cm 0.0cm,width=0.24\linewidth]{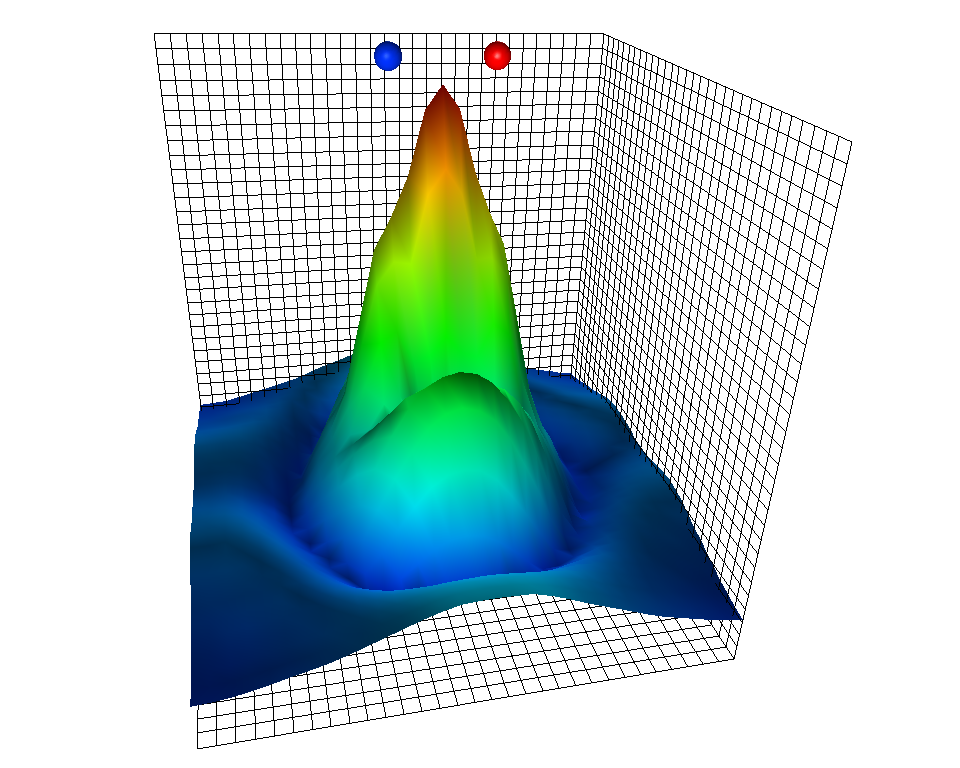}
\includegraphics[clip=true,trim=2.5cm 0.0cm 2.5cm 0.0cm,width=0.24\linewidth]{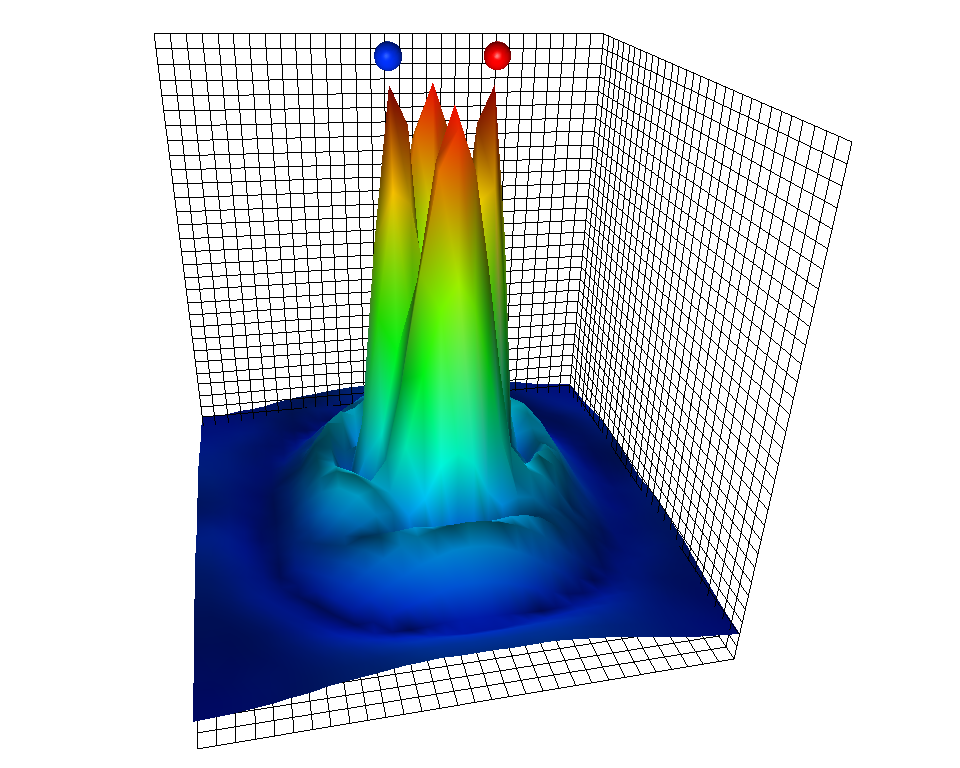} \\[0.1cm]
\includegraphics[clip=true,trim=2.5cm 0.0cm 2.5cm 0.0cm,width=0.24\linewidth]{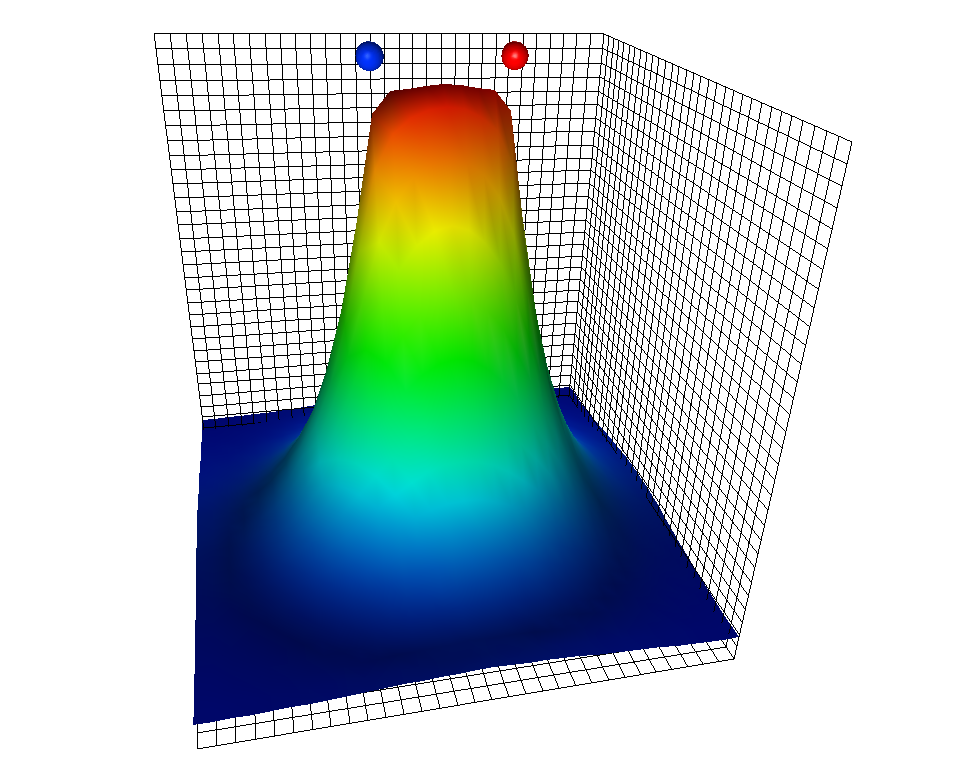}
\includegraphics[clip=true,trim=2.5cm 0.0cm 2.5cm 0.0cm,width=0.24\linewidth]{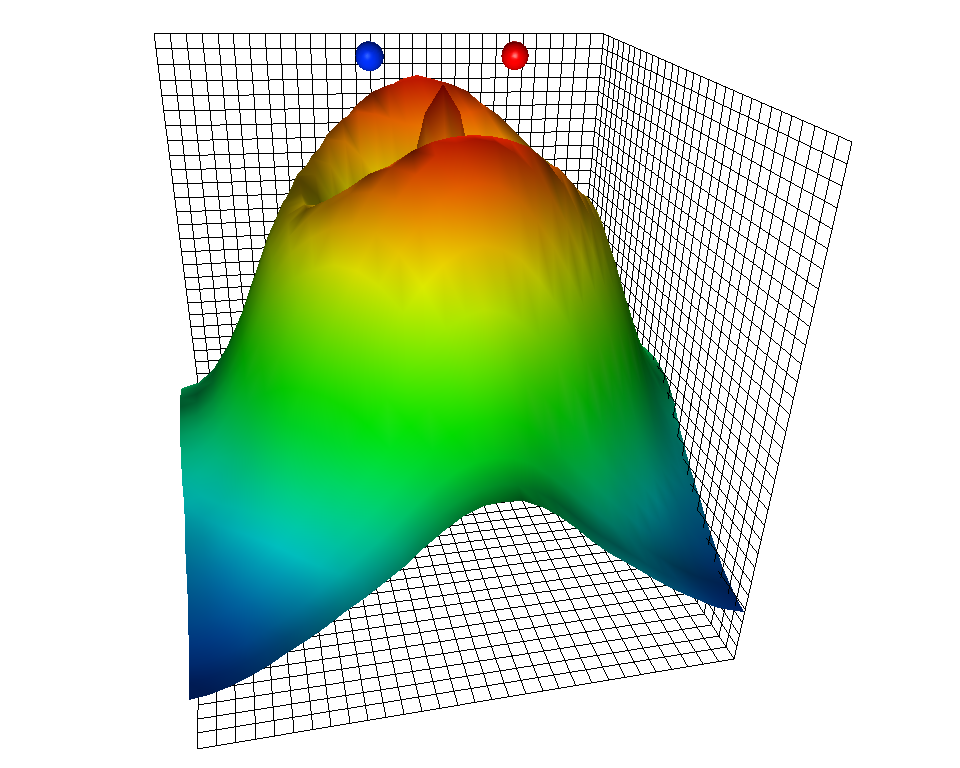}
\includegraphics[clip=true,trim=2.5cm 0.0cm 2.5cm 0.0cm,width=0.24\linewidth]{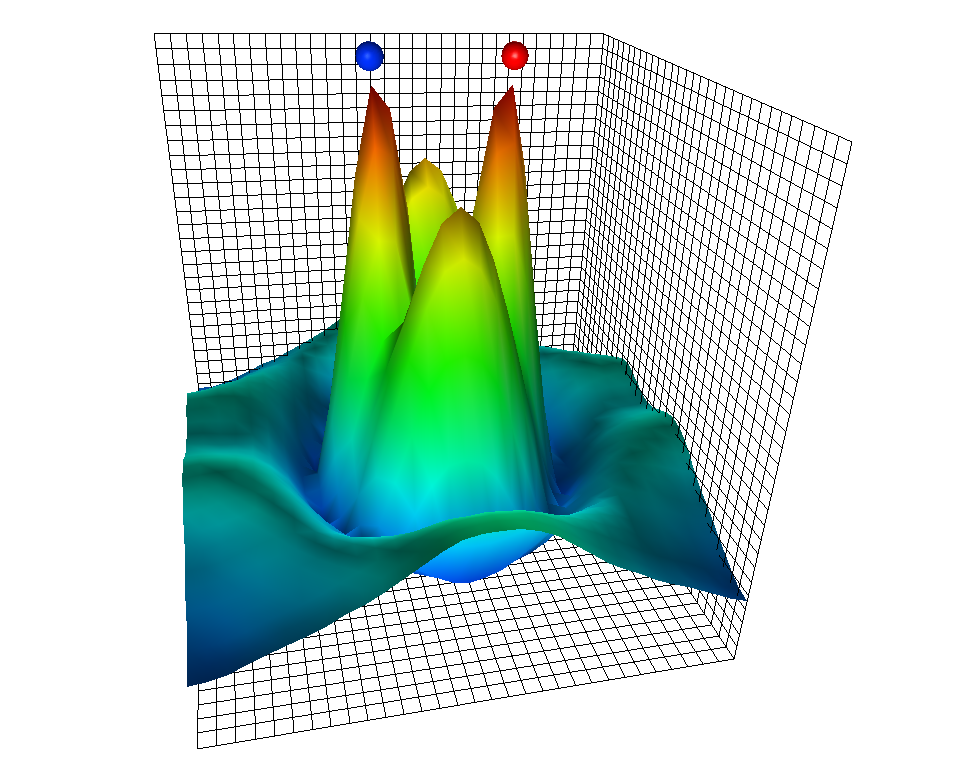}
\includegraphics[clip=true,trim=2.5cm 0.0cm 2.5cm 0.0cm,width=0.24\linewidth]{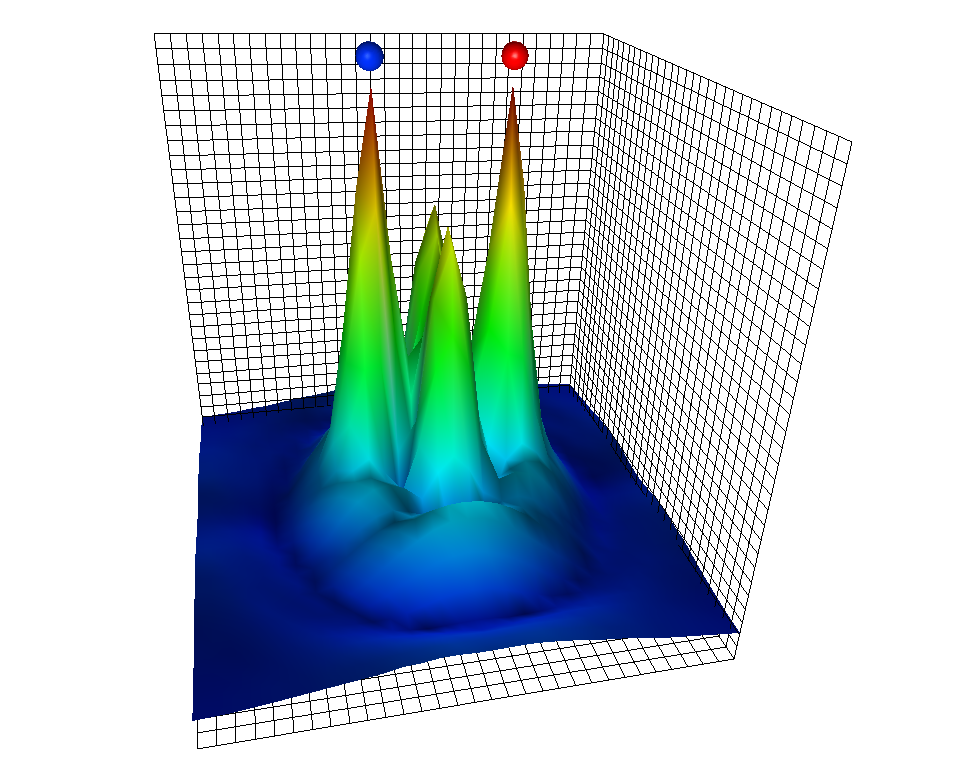}
\caption{The dependence of the $d$-quark probability distribution on
  the positions of the $u$ quarks in the proton and its excited
  states.  From left to right, the columns correspond to the ground,
  first, second and third $S$-wave excitations.  The $u$ quarks are
  fixed on the $x$-axis running from back right through front left
  through the centre of the plot.  The $u$ quarks are fixed a distance
  of $d_1$ and $d_2 = -d_1$ from the origin located at the centre.
  The distance between the quarks, $d = d_1 - d_2$, increases
  from the top row through to the bottom row, taking values 0, 2, 4, 6
  and 8 times the lattice spacing $a=0.0907$ fm.  }
\label{QuarkSepA}
\end{figure*}

\begin{figure*}[p]
\includegraphics[clip=true,trim=2.5cm 0.0cm 2.5cm 0.0cm,width=0.24\linewidth]{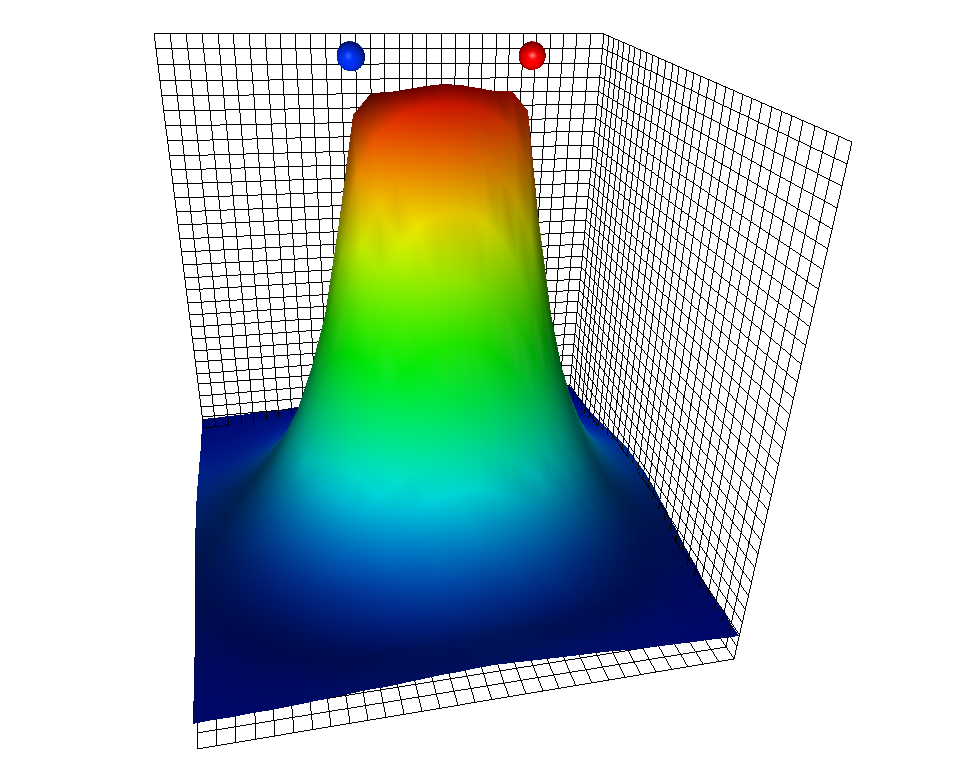}
\includegraphics[clip=true,trim=2.5cm 0.0cm 2.5cm 0.0cm,width=0.24\linewidth]{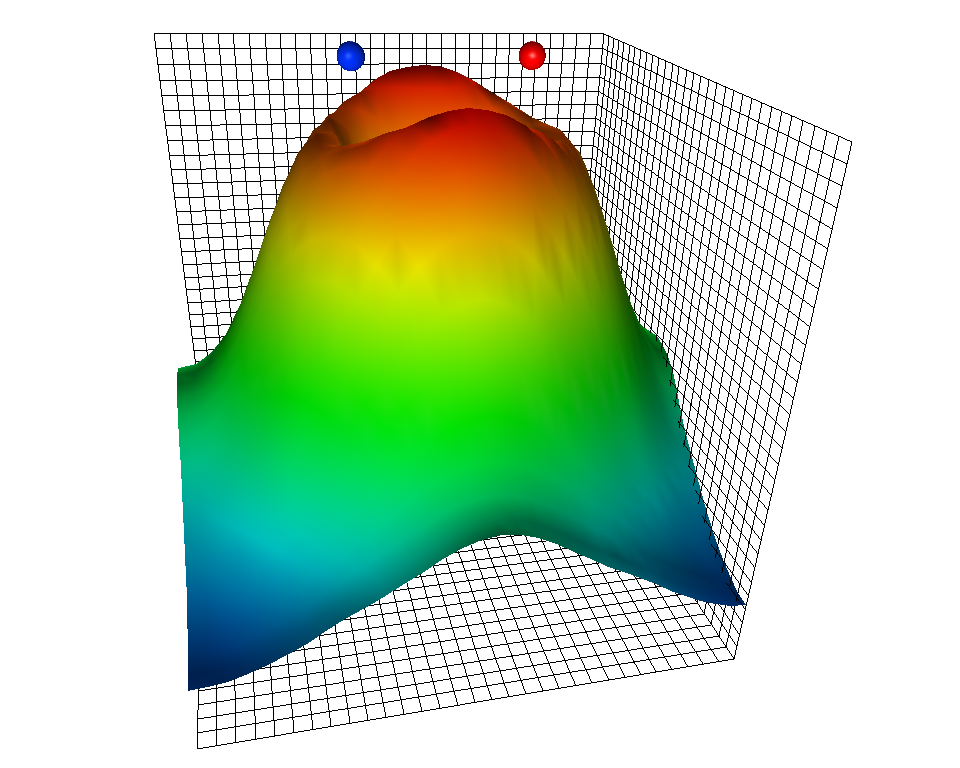}
\includegraphics[clip=true,trim=2.5cm 0.0cm 2.5cm 0.0cm,width=0.24\linewidth]{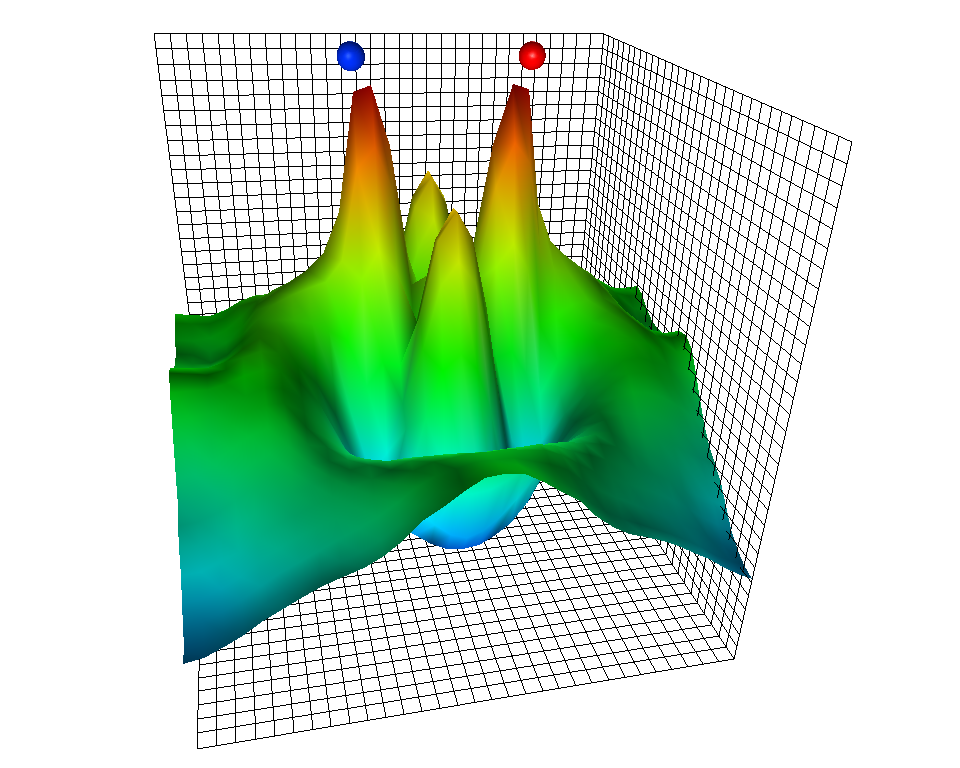}
\includegraphics[clip=true,trim=2.5cm 0.0cm 2.5cm 0.0cm,width=0.24\linewidth]{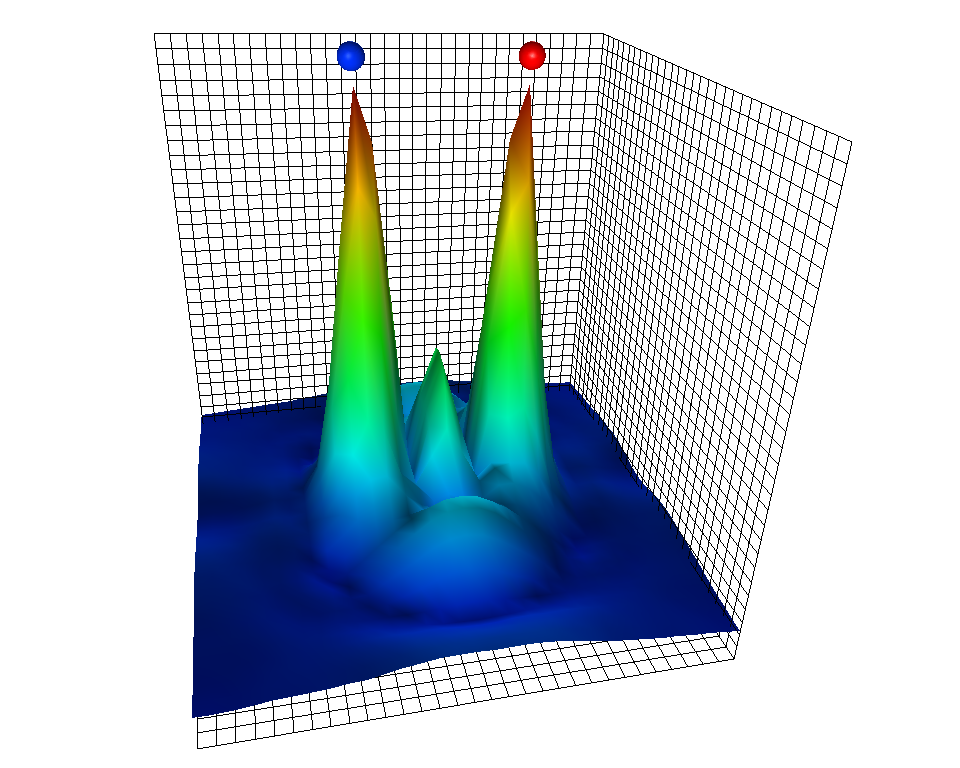} \\[0.1cm]
\includegraphics[clip=true,trim=2.5cm 0.0cm 2.5cm 0.0cm,width=0.24\linewidth]{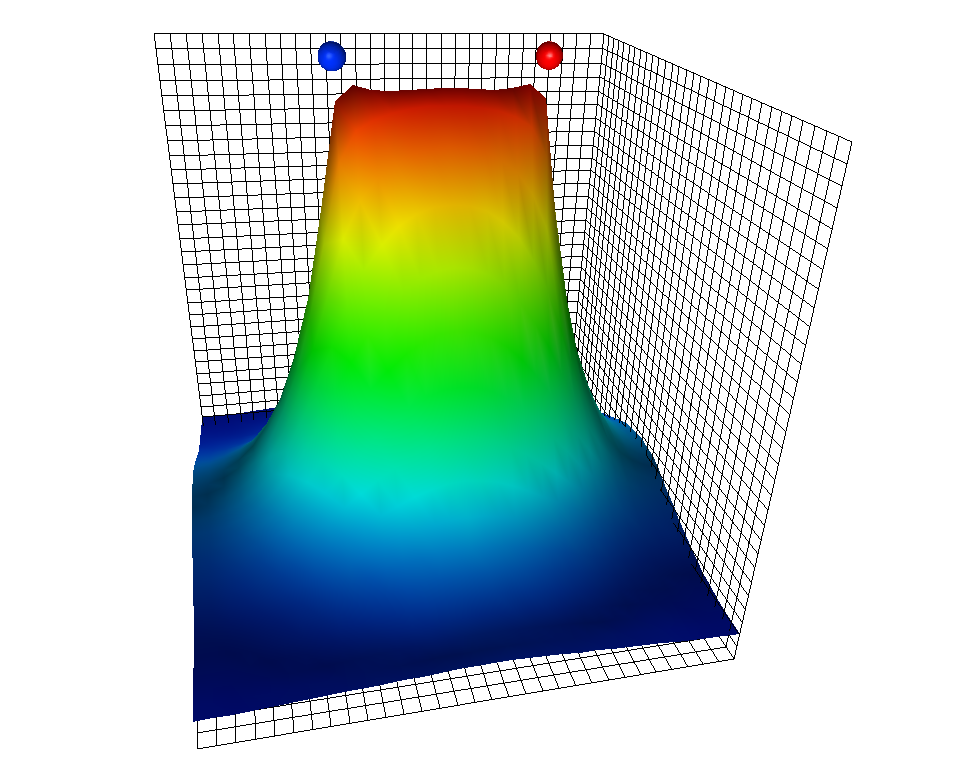}
\includegraphics[clip=true,trim=2.5cm 0.0cm 2.5cm 0.0cm,width=0.24\linewidth]{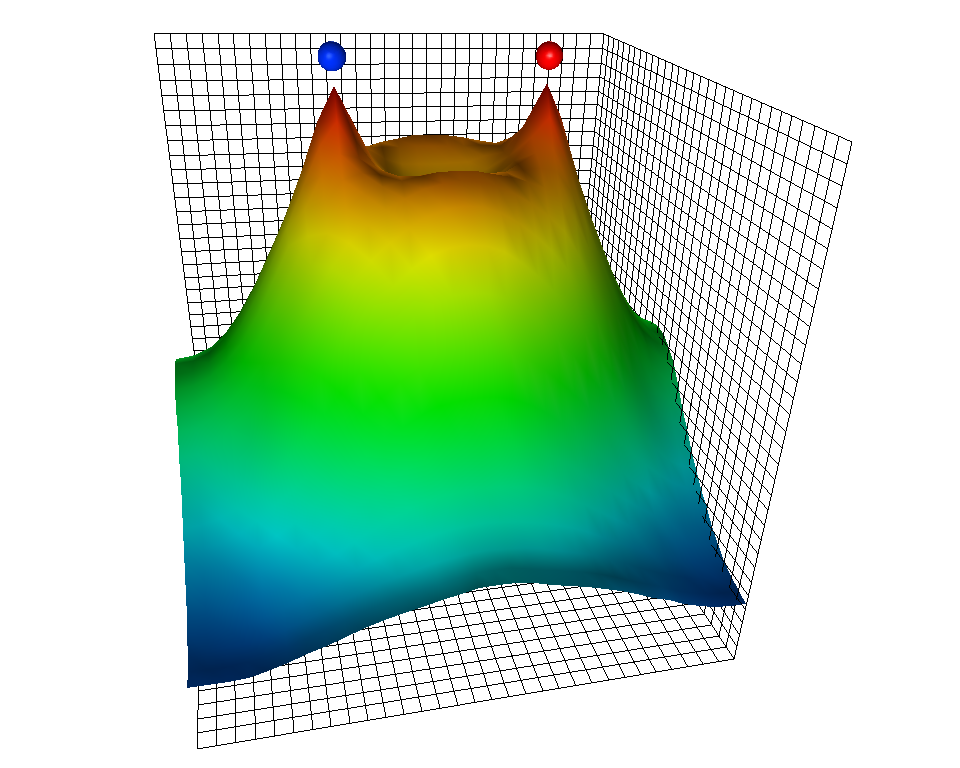}
\includegraphics[clip=true,trim=2.5cm 0.0cm 2.5cm 0.0cm,width=0.24\linewidth]{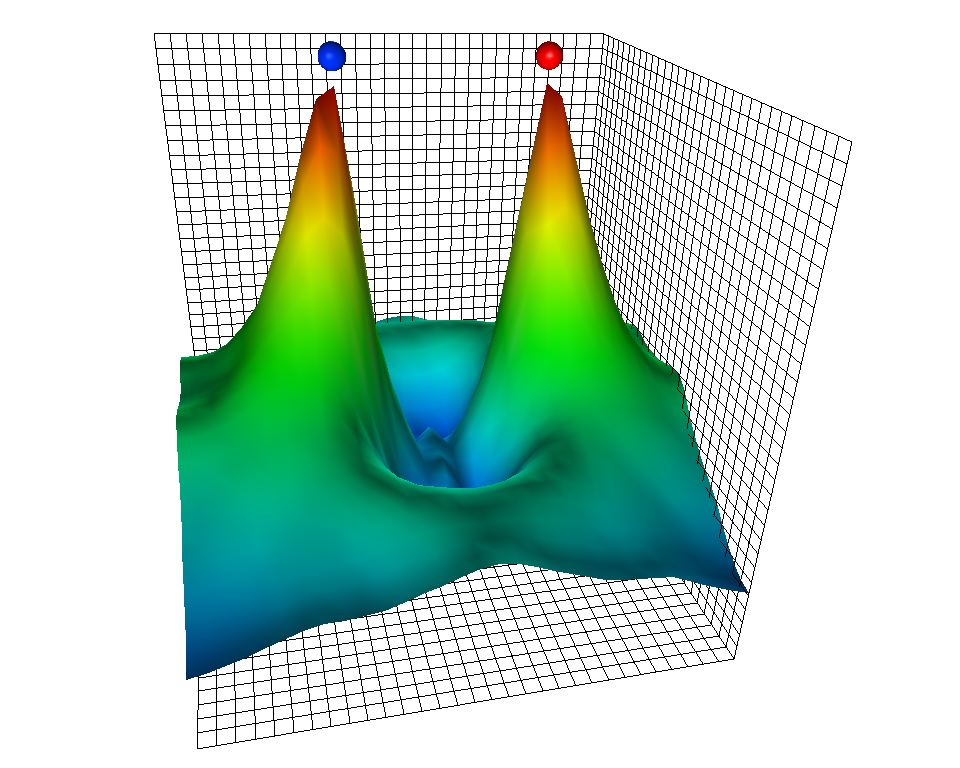}
\includegraphics[clip=true,trim=2.5cm 0.0cm 2.5cm 0.0cm,width=0.24\linewidth]{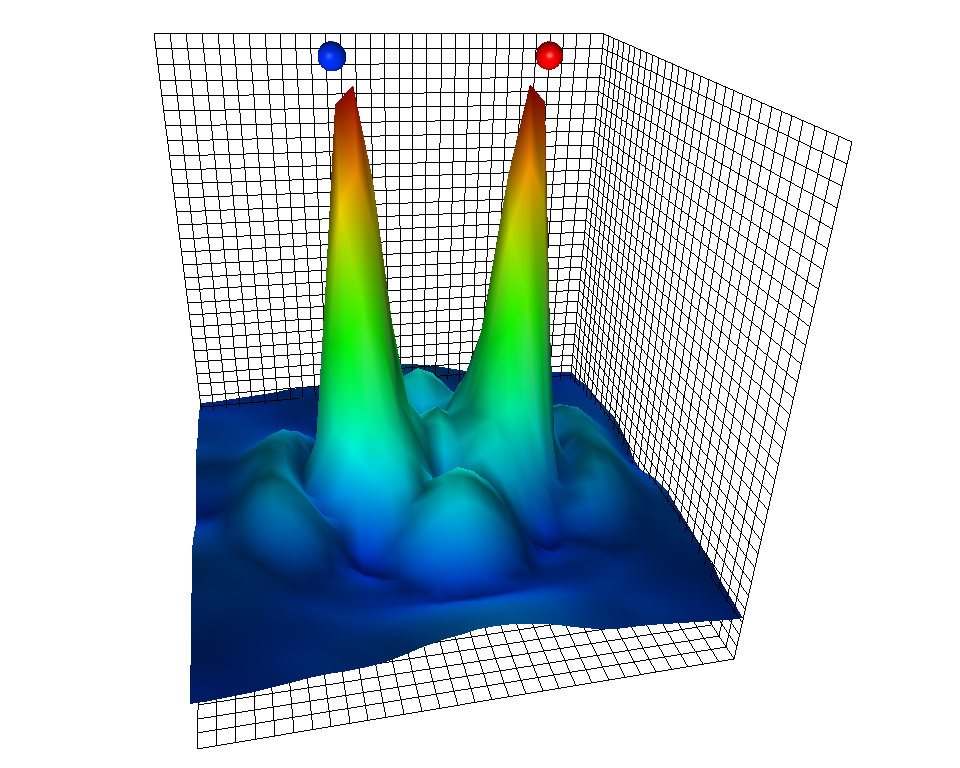} \\[0.1cm]
\includegraphics[clip=true,trim=2.5cm 0.0cm 2.5cm 0.0cm,width=0.24\linewidth]{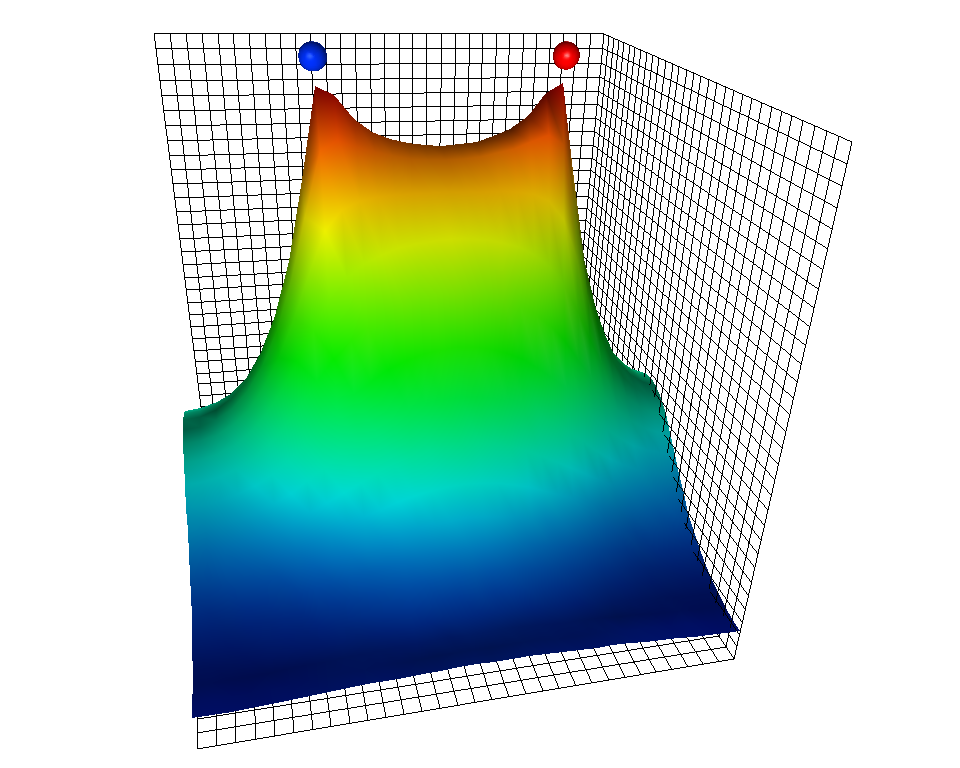}
\includegraphics[clip=true,trim=2.5cm 0.0cm 2.5cm 0.0cm,width=0.24\linewidth]{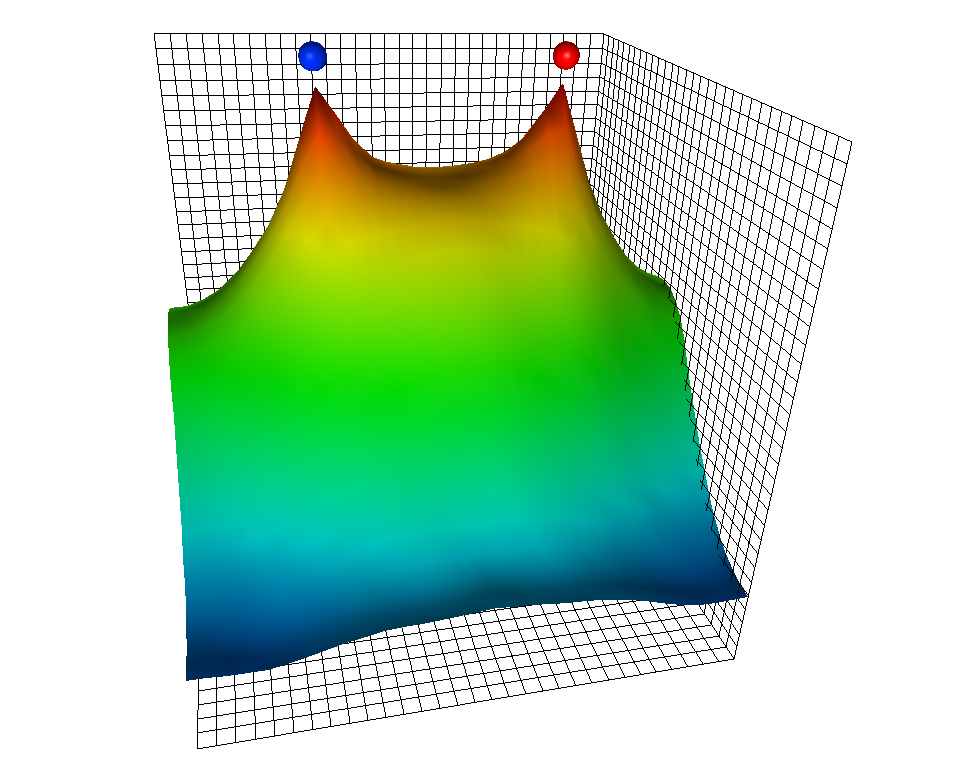}
\includegraphics[clip=true,trim=2.5cm 0.0cm 2.5cm 0.0cm,width=0.24\linewidth]{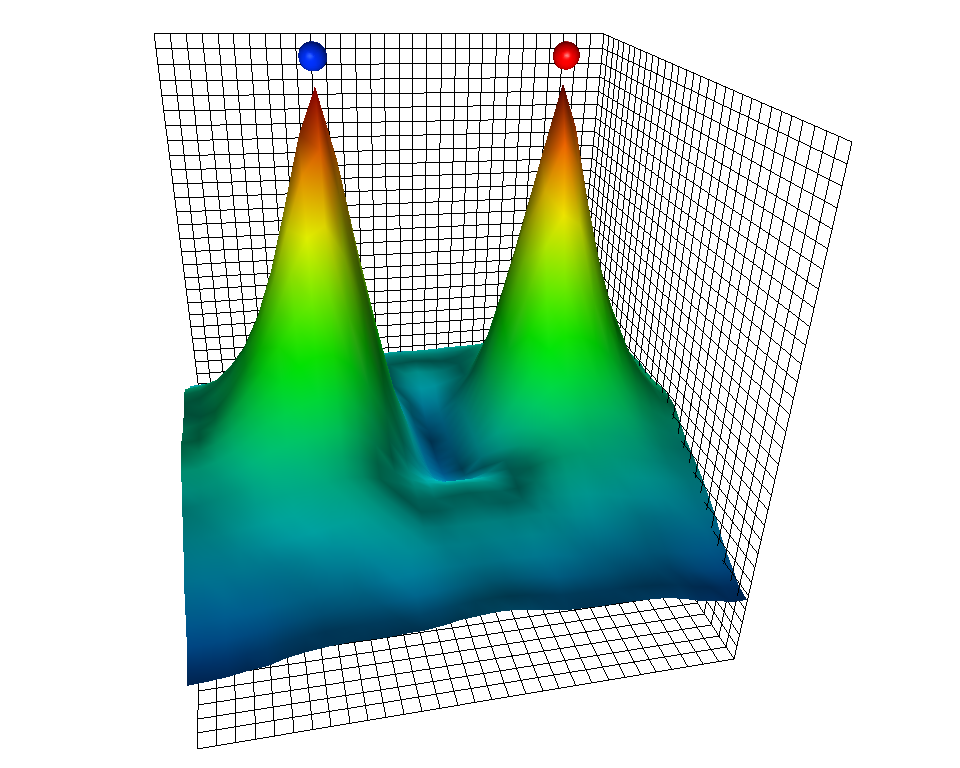}
\includegraphics[clip=true,trim=2.5cm 0.0cm 2.5cm 0.0cm,width=0.24\linewidth]{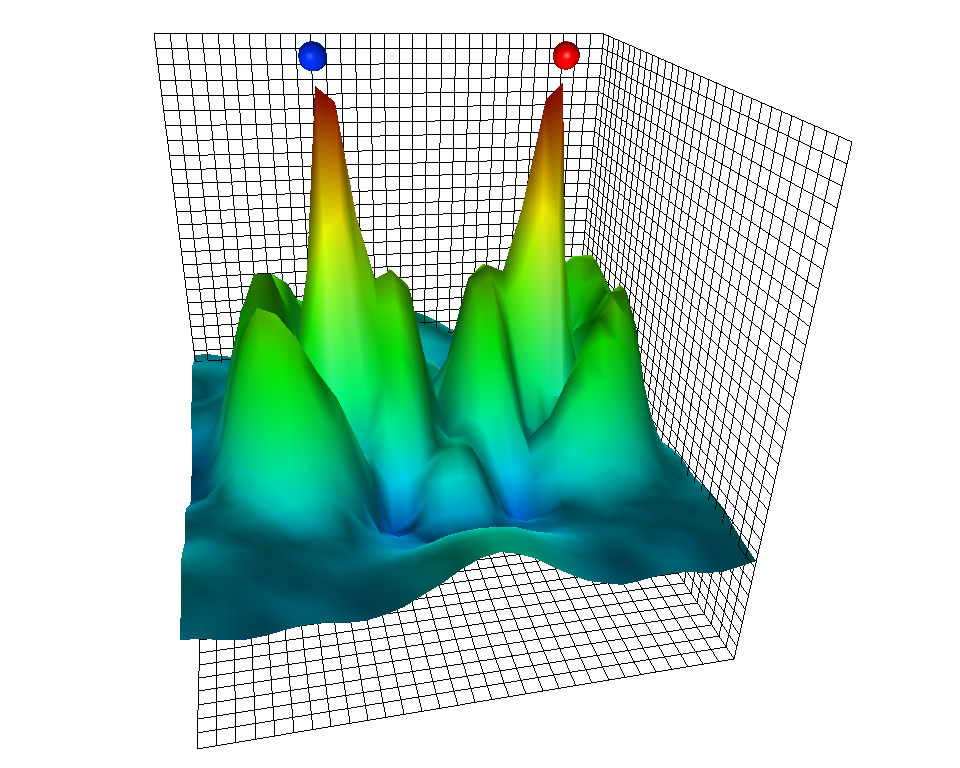} \\[0.1cm]
\includegraphics[clip=true,trim=2.5cm 0.0cm 2.5cm 0.0cm,width=0.24\linewidth]{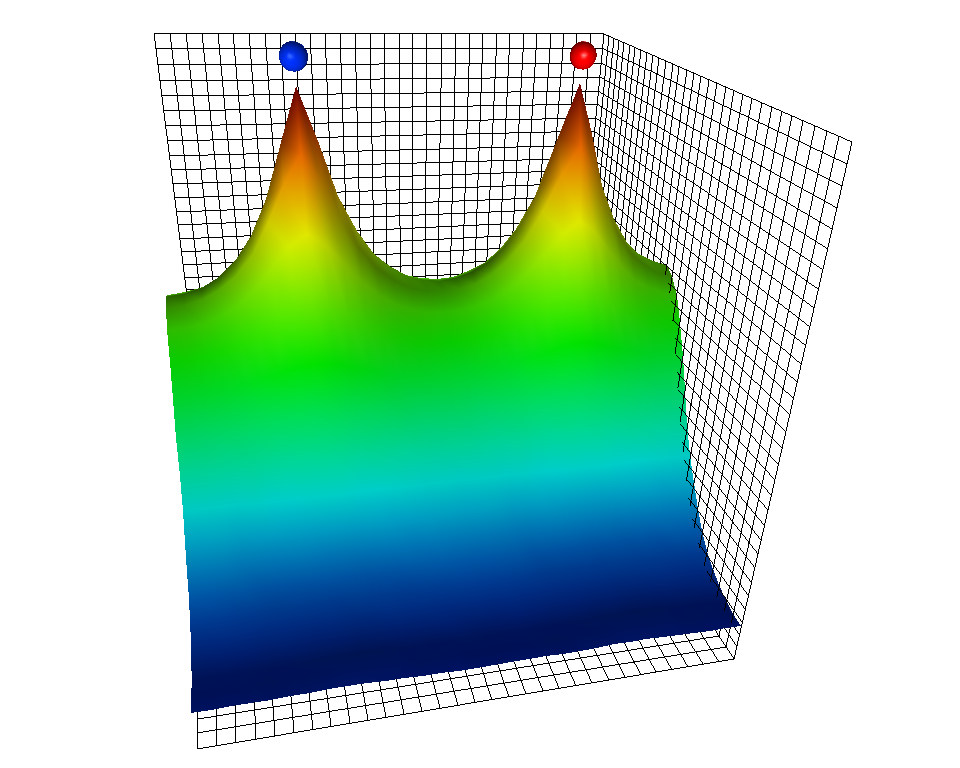}
\includegraphics[clip=true,trim=2.5cm 0.0cm 2.5cm 0.0cm,width=0.24\linewidth]{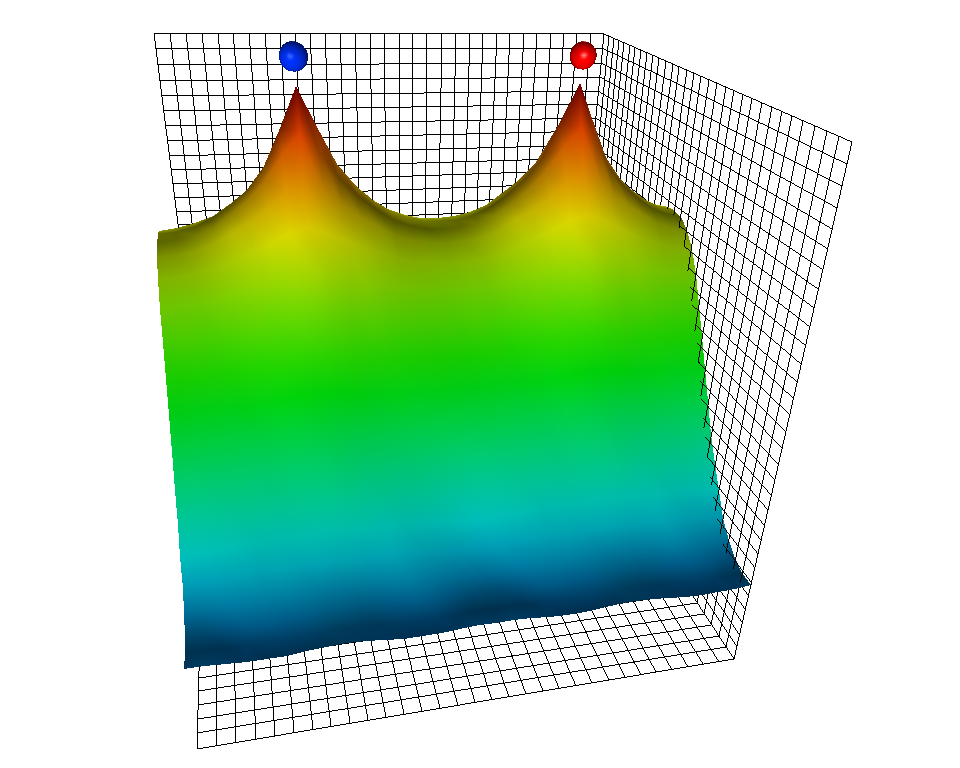}
\includegraphics[clip=true,trim=2.5cm 0.0cm 2.5cm 0.0cm,width=0.24\linewidth]{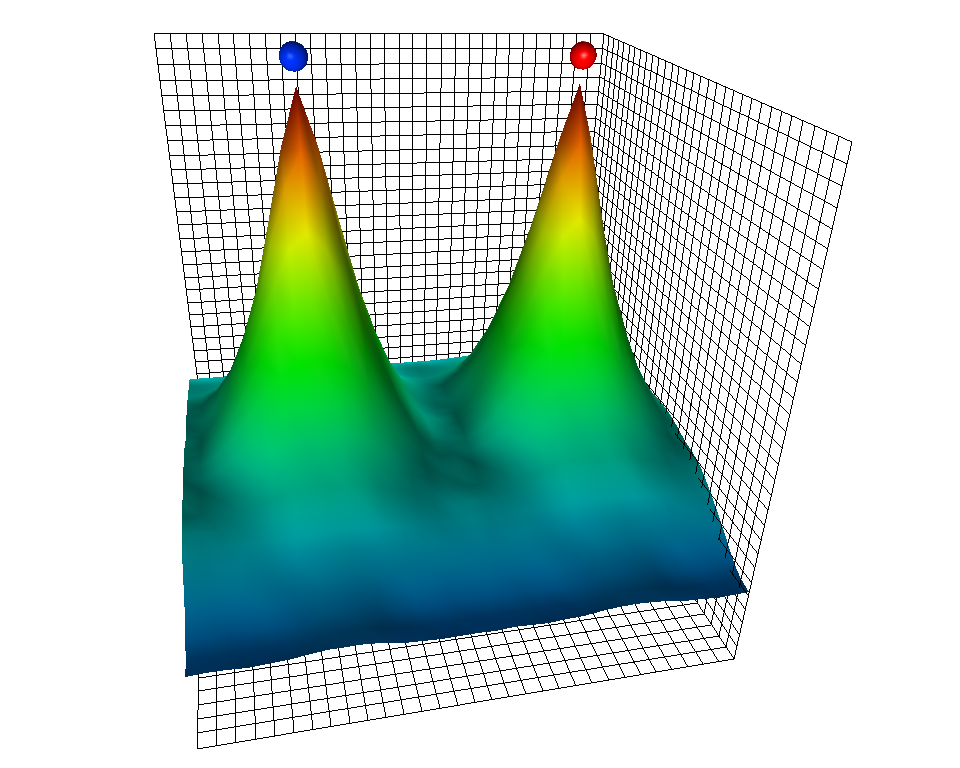}
\includegraphics[clip=true,trim=2.5cm 0.0cm 2.5cm 0.0cm,width=0.24\linewidth]{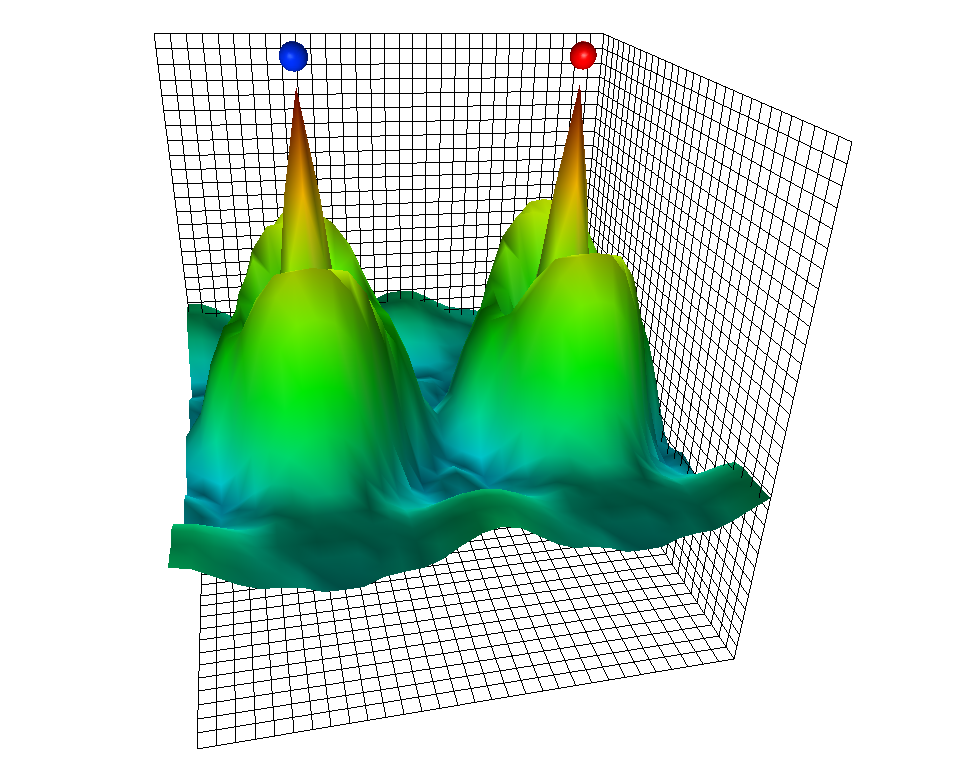}
\caption{The dependence of the $d$-quark probability distribution on
  the positions of the $u$ quarks in the proton and its excited
  states.  From left to right, the columns correspond to the ground,
  first, second and third $S$-wave excitations.  The $u$ quarks are
  fixed on the $x$-axis running from back right through front left
  through the centre of the plot.  The $u$ quarks are fixed a distance
  of $d_1$ and $d_2 = -d_1$ from the origin located at the centre.
  The distance between the quarks, $d = d_1 - d_2$, increases
  from the top row through to the bottom row, taking values 10, 12,
  14, and 16 times the lattice spacing $a=0.0907$ fm.  }
\label{QuarkSepB}
\end{figure*}

Figure~\ref{QuarkSepA} presents the $d$-quark probability
distributions for $u$-quark separations of $d = 0$, 2, 4, 6 and 8
times the lattice spacing $a = 0.0907$ fm and Fig.~\ref{QuarkSepB}
completes the study, illustrating $u$-quark separations of 10, 12, 14
and 16 times the lattice spacing.  Each column corresponds to a
different state with the ground, first-, second-, and
third-excitations illustrated from left to right.  The two small
spheres above the isosurface indicate the positions of the two $u$
quarks.

% @@@@@@@@@@@@@@@@@@@@@@@@@@@@@@@@@@@@@@@@@@@@@@@@@@@@@@@@@@@@@@@@@@@@@

\subsubsection{Ground-State Distribution}

Focusing first on the ground state, on separation of the $u$ quarks
the probability distribution of the $d$ quark forms a single broad
peak.  The structure is slightly rounded until $d = d_1 - d_2 = 12\, a
= 1.09$ fm, with small peaks at the $u$-quark positions.  At a
separation of $d = 13\, a = 1.18$ fm the wave function takes on a
double peak structure associated with scalar-diquark clustering
similar to that illustrated in the third row of Fig.~\ref{QuarkSepB}
for $d/a = 14$.  These results are similar to the earlier quenched
wave function results of Refs.~\cite{Roberts:2010cz,Hecht:1992uq}.

\subsubsection{First Excited-State Distribution}

As the $u$ quarks are separated in the first excited state associated
with the Roper, the central peak of the $d$-quark distribution
broadens in a manner similar to that for the ground state.  However by
$d = 4\, a = 0.36$ fm strength in the wave function is seen to move
from the centre into the outer shell of the $2S$ state.  This
transition continues to $d = 10\, a = 0.91$ fm where the $u$ quarks
are still well inside the original node position of the $2S$
distribution.  At this point, the central peak has been suppressed
entirely leaving a hole inside the ring or shell in three dimensions.
In other words, the node of the wave function has shrunk to the origin.
It's interesting how the ring-like probability density is enhanced in
the direction perpendicular to the separation of the $u$ quarks.

At $d = 11\, a = 1.00$ fm, small peaks form in the
probability distributions at the positions of the $u$ quarks revealing
the first onset of scalar diquark clustering similar to that in the
second row of Fig.~\ref{QuarkSepB}.  At $d = 12\, a = 1.09$ fm, the
$u$ quarks are still within the node of the original wave function.
but the radius of the outer shell of the wave function illustrated by
the ring in the probability density has reduced slightly.  At $d =
14\, a = 1.27$ fm, the $u$ quarks sit in the node of the original wave
function and there is little memory of the original $2S$ structure.
Only a slight swelling at the centre of the distribution remains.  The
central probability density reduces at $d = 15\, a = 1.36$ fm such
that scalar-diquark clustering dominates the probability distribution
at $d= 16\, a = 1.45$ fm.

\subsubsection{Second Excitation}

For the second excited state observed herein, we again see a shift of
the probability density from the central peak to the next shell of the
original $3S$-like wave function.  At $d = 6\, a = 0.54$ fm in the
fourth row of Fig.~\ref{QuarkSepA}, a similar enhancement in the first
shell is observed as for the Roper at $d = 8\, a = 0.73$ fm with
strength in the probability density enhanced in the direction
perpendicular to the separation of the $u$ quarks.

The radius of the first shell about the central peak of the original
distribution shrinks as the $u$ quarks are pulled apart and at $d =
8\, a = 0.73$ fm corresponding to the bottom row of
Fig.~\ref{QuarkSepA}, the $u$ quarks are now in the first shell where
four peaks are apparent.  The original first node has shrunk to the
centre and may have emerged, centered about each of the $u$ quarks.
Further evidence of this process is discussed in the analysis of the
third excitation below.  The second node of the original wave function
now surrounds the four peaks.

The $u$ quarks approach the position of the second node of the wave
function at $d = 10\, a = .91$ fm.  The node is still apparent in the
front and back of the distribution, orthogonal to the $u$-quark
separation axis.  The peaks in the probability distribution are still
associated with the first shell surrounding the central peak of the
original distribution with zero $u$-quark separation.

By $d= 12\, a = 1.09$ fm the quarks have moved beyond the second node.
The radius of node has reduced and can be seen in the dark-blue
regions at the centre of the plot.  At $d = 13\, a = 1.18$ fm the
second node has collapsed to the origin and explains the strong
separation of the two peaks observed at $d = 14\, a = 1.27$ fm in the
third row of Fig.~\ref{QuarkSepB}.  Even at the largest quark
separations examined, the node structure of this state is apparent,
suppressing the probability density between the two peaks once again
governed by scalar-diquark dynamics.

\subsubsection{Third Excitation}

The third excitation displays a wonderfully complex structure that
mirrors the transitions observed for the first and second excitations
for the first few separations.  For example at $d = 4\, a = 0.36$ fm, one
can see the enhancement of the first shell in a direction orthogonal
to the $u$ quark separation axis.

At $d = 6\, a = 0.54$ fm, a four-peak structure emerges as the $u$ quarks
enter the first shell of the wave function.  Remarkably, a new nodal
structure has emerged.  Upon shrinking to the origin, the original
first node emerged surrounding each of the peaks at the $u$ quark
positions.  This node now cuts through the first shell strength of the
underlying $4S$ configuration and divides what would normally be a
ring shape into four peaks.

By $d = 10\, a = 0.91$ fm the third node surrounds all significant
structure in the distribution.  The second node has shrunk to surround
the small peak in the centre and the first surrounds the $u$ quark
peaks.  

At $d = 12\, a = 1.09$ fm the third node now surrounds both of the major
peaks and the fore and aft humps near the centre.  The first node
continues to surround each of the peaks at the $u$ quark positions,
The emergence of a second node around each of the $u$ quarks is
becoming apparent at the left- and right-hand edges of the plot.

At $d = 14\, a = 1.27$ fm the third node maintains a circular structure
centred about the origin and cuts through the ring like structures
forming around each of the $u$ quarks.  The rings clearly reveal the
shifting of the first and second nodes to surround each of the $u$
quarks.

At the largest $u$-quark separation of $d = 16\, a = 1.45$ fm the third
node has shrunk further to just touch the inside edges of the rings
which have formed though the first and second nodes shrinking to the
origin and emerging around the two $u$ quarks.

\section{Conclusions}

In this first study of the quark probability distribution within
excited states of the nucleon, we have shown that all the states
accessed in our correlation matrix analysis display the node structure
associated with radial excitations of the quarks.  For example the
first excited state associated with the Roper resonance displays a
node in the $d$ quark wave function consistent with a radial
excitation of the $d$ quark.  The second and third excitations display
two and three nodes respectively.

It is beautiful to observe the emergence of these corner stones of
quantum mechanics from the complex many body theory of quantum field
theory.  The few-body projection of the underlying physics can be
connected with models, shedding light on the essential effective
phenomena emerging from the complex dynamics of QCD.

On comparing these probability distributions to those predicted by the
constituent quark model, we find good qualitative similarity with
interesting differences.  The core of the states is described very
well by the model and the amplitudes of the $S$-wave shells between
the nodes are predicted very accurately by the constituent quark
model.  The discovery of a node structure provides a deep
understanding of the success of the smeared-source/sink correlation
matrix methods of Ref.~\cite{Mahbub:2010rm}.

Finite volume effects are shown to be particularly significant for
the excited states explored herein at relatively light quark mass.  As
these excited states have a multi-particle component, the interplay
between the lattice volume, the wave function and the associated
energy are key to extracting the resonance parameters of the states. 

Fascinating structure in the $d$-quark probability distributions of
the nucleon excited states is revealed when separating the $u$ quarks
from the origin.  As the $u$ quarks are separated the original node
structure of the wave function shrinks in size.  For example, the
Roper reveals a ringed structure in the surface plots corresponding
to an empty shell in three dimensions as the node collapses to the
origin.  The second excited state reveals a four-peaked structure at
mid-range quark separations.  At large separations these states all
display diquark clustering with the $d$ quark most likely found near
one of the $u$ quarks.  The third state reveals the most exotic
structure with new nodes centred about the $u$ quarks appearing after
the original nodes collapsed to the origin.  

Future calculations will explore the structure of these states in more
detail, examining the effect of the introduction of isospin-1/2
spin-3/2 interpolating fields \cite{Zanotti:2003fx,Nozawa:1990gt} to
reveal the role of $D$-wave contributions.  While our use of improved
actions suppresses lattice discretization errors, ultimately
simulations will be done at a variety of lattice spacings directly at
the physical quark masses.  An analysis of finite volume effects will
also be interesting to further reveal the interplay between the finite
volume of the lattice, the structure of the states and the associated
energy of the states; thus connecting the lattice QCD simulation
results to the resonance physics of Nature.

\section*{Acknowledgements}

We thank PACS-CS Collaboration for making their $2+1$ flavor
configurations available and acknowledge the important ongoing support
of the ILDG.  This research was undertaken with the assistance of
resources at the NCI National Facility in Canberra, Australia, and the
iVEC facilities at Murdoch University (iVEC@Murdoch) and the
University of Western Australia (iVEC@UWA). These resources were
provided through the National Computational Merit Allocation Scheme
and the University of Adelaide Partner Share supported by the
Australian Government.  We also acknowledge eResearch SA for their
support of our supercomputers.  This research is supported by the
Australian Research Council.

\end{document}